\documentclass[12pt]{article}

\pdfoutput=1

\usepackage{amsmath,amssymb,amsfonts,amscd,mathrsfs}
\usepackage{xcolor}
\definecolor{darkblue}{rgb}{0.1,0.1,.7}
\hypersetup{
  colorlinks, linkcolor=darkblue, citecolor=darkblue, urlcolor=darkblue, linktocpage}
\usepackage[square, comma, compress,numbers]{natbib}
\usepackage[]{graphicx}
\usepackage{geometry}
\usepackage[margin=10pt,font=small,labelfont=bf]{caption}
\usepackage{ifthen}
\usepackage{tikz}
\usepackage{subcaption}
\usepackage{booktabs,multirow}
\usepackage{hhline}
\usepackage{tablefootnote}
\usepackage{ragged2e}
\usepackage{array}
\newcolumntype{L}[1]{>{\raggedright\let\newline\\\arraybackslash\hspace{0pt}}m{#1}}
\newcolumntype{C}[1]{>{\centering\let\newline\\\arraybackslash\hspace{0pt}}m{#1}}
\newcolumntype{R}[1]{>{\raggedleft\let\newline\\\arraybackslash\hspace{0pt}}m{#1}}

\usepackage{dsfont} 

\usepackage{booktabs}

\usepackage{titlesec}
\titleformat*{\section}{\large\bfseries}
\titleformat*{\subsection}{\normalsize\bfseries}
\titleformat*{\subsubsection}{\normalsize\it}
\titleformat*{\paragraph}{\normalsize\bfseries}
\titleformat*{\subparagraph}{\normalsize\bfseries}

\newcommand{\reef}[1]{(\ref{#1})}

\def\eps{\epsilon}
\def\phys{{\rm ph}}
\newcommand{\beq}{\begin{equation}} 
\newcommand{\eeq}{\end{equation}}
\def\del {\partial} 
\def\nn{\nonumber} 
\def\bZ {\mathbb{Z}} 
\def\bR {\mathbb{R}} 
 
\def\calO {{\cal O}}

\def\calH {{\cal H}}

\def\calE {{\cal E}} 
\def\calT {{\cal T}} 
\def\bZ {\mathbb{Z}} 
 
\def\bP {\mathbb{P}} 
\def\half{{\textstyle\frac 12}}

\def\ge{\geqslant}
\def\le{\leqslant}

\def\leq{\leqslant}
\newcommand{\diffop}[2]{\ifthenelse{\equal{#2}{1}}{\frac{\mrm{d}}{\mrm{d} #1}}{\frac{\mrm{d}^#2}{\mrm{d} #1^#2}}}
\newcommand{\NO}[1]{{:\!#1\!:}}

\newcommand{\ket}[1]{|#1\rangle}
\newcommand{\bra}[1]{\langle #1|}

\newcommand{\mrm}[1]{{\mathrm #1}}

\def\unit{\mathds{1}} % needs dsfont package
\usepackage[normalem]{ulem}

\def\del{\partial}

\def \Emax{E_{\rm max}}

 \def\cE{{\cal E}}

\newcommand{\be}{\begin{equation}}
\newcommand{\ee}{\end{equation}}
\newcommand{\bea}{\begin{eqnarray}}
\newcommand{\eea}{\end{eqnarray}}
\newcommand{\eq}[1]{Eq.~(\ref{#1})}
 \def\om{\omega}

\usepackage{accents}
\newlength{\dhatheight}

\numberwithin{equation}{section}
\usepackage[tocgraduated]{tocstyle}
\begin{document}

\vspace*{-.6in} \thispagestyle{empty}
\begin{flushright}
CERN-TH-2017-124\\
%LPTENS \ldots
\end{flushright}
\vspace{1cm} {\Large
\begin{center}
{\bf NLO Renormalization in the Hamiltonian Truncation}
\end{center}}
\vspace{1cm}
\begin{center}
{\bf Joan Elias-Mir\'o$^a$, Slava Rychkov$^{b,c}$,  Lorenzo G. Vitale$^{d,e}$}\\[2cm] 
{
$^{a}$ SISSA/ISAS and INFN, I-34136 Trieste, Italy\\
$^b$  CERN, Theoretical Physics Department, 1211 Geneva 23, Switzerland\\
$^c$  Laboratoire de Physique Th\'eorique de l'\'Ecole Normale Sup\'erieure,  \\
PSL Research University,  CNRS,  Sorbonne Universit\'es, UPMC
Univ.\,Paris 06,  \\
24 rue Lhomond, 75231 Paris Cedex 05, France
% Ecole Normale Superieure [Future INSPIRE ID: ENS, Paris, LPT]
% UPMC, Paris (main) [Future INSPIRE ID: UPMC, Paris] 
\\
$^d$ Institut de Th\'eorie des Ph\'enom\`enes Physiques, EPFL, CH-1015 Lausanne, Switzerland\\
$^e$ Department of Physics, Boston University, Boston, MA 02215
}
\vspace{1cm}%\today
\end{center}

\vspace{4mm}

\begin{abstract}
Hamiltonian Truncation (a.k.a. Truncated Spectrum Approach) is a numerical technique for solving strongly coupled QFTs, in which the full Hilbert space 
is truncated to a finite-dimensional low-energy subspace. The accuracy of the method is limited only by the available computational resources. 
The renormalization program improves the accuracy by carefully integrating out the high-energy states, instead of truncating them away. 
In this paper we develop the most accurate ever variant of Hamiltonian Truncation, which implements renormalization at the cubic order in the interaction strength. The novel idea is to interpret the renormalization procedure as a result of
integrating out exactly a certain class of high-energy ``tail states".
We demonstrate the power of the method with high-accuracy computations in the strongly coupled two-dimensional quartic scalar theory, and benchmark it against other existing approaches. Our work will also be useful for the future goal of extending Hamiltonian Truncation to higher spacetime dimensions.

 \end{abstract}
\vspace{.2in}
\vspace{.3in}
\hspace{0.7cm} June 2017

\newpage

\setcounter{tocdepth}{2}

{
\tableofcontents
}
\newpage

\section{Introduction}

Developing reliable and efficient techniques for computations in strongly coupled quantum field theories (QFT) remains one of the critical challenges of modern theoretical physics. In this paper we will be concerned with one such technique---the Hamiltonian Truncation (HT).\footnote{Also known as the TSA---Truncated Space (or Spectrum) Approach.} This method became popular after the work of Yurov and Zamolodchikov \cite{Yurov:1989yu,Yurov:1991my} in the late 80's-early 90's.\footnote{An even earlier paper using the HT \cite{Brooks:1983sb} did not get the attention it deserved.} By now it's an established technique with many nontrivial results (see \cite{Konik-review} for a recent review). 

The HT is applicable to QFTs whose Hamiltonian can be split in the form $H=H_0+V$ where $H_0$ is exactly solvable. $H_0$ may be a free theory or an interacting integrable theory, such as an integrable massive QFT, or a solvable conformal field theory (CFT). $V$ describes additional interactions.\footnote{In what follows we assume that $V$ is a non-gauge interactions. It is a largely open problem how to treat gauge interactions using the HT. The light front quantization \cite{Brodsky:1997de} has long intended to solve this problem, but not many concrete results have been obtained, except in 1+1 dimensions where one can integrate out gauge fields completely, see e.g.~\cite{Bhanot:1993xp,Katz:2013qua,Katz:2014uoa}.} The total Hamiltonian $H$ is in general not exactly solvable and is treated numerically. To set up the calculation, one needs to know the energy eigenstates of $H_0$ in finite volume and the matrix elements of $V$ among them.  Then one represents $H$ as an infinite matrix in the Hilbert space of $H_0$ eigenstates. This matrix is truncated to the subspace of low-energy eigenstates below some energy cutoff $E_T$ and diagonalized numerically. This procedure represents a natural adaptation of the Rayleigh-Ritz method from quantum mechanics to QFT.

The HT method is non-perturbative and a priori works for interactions $V$ of arbitrary strength. It works best if the interaction switches off fast at high energy (in technical language, if $V$ is \emph{strongly relevant}). In this case the method converges rapidly, and accurate results can be obtained with low $E_T$ cutoff and with truncated Hilbert spaces of modest size. If on the other hand $V$ is only \emph{weakly relevant}, then the convergence is poor, as the truncated results exhibit significant $E_T$ cutoff dependence even for the highest numerically affordable $E_T$'s. This is a limitation of the method.

Another, related, limitation is that so far most applications were in $d=2$ spacetime dimensions (although in principle the method can be set up in any $d$ \cite{Hogervorst:2014rta}). The reason is that in $d=2$ there are many physically interesting integrable QFTs and CFTs, which can play the role of $H_0$. Many of these systems possess perturbations $V$ which are strongly relevant---a favorable situation according to the above-mentioned convergence criterion. On the contrary, in $d>2$ the only exactly solvable $H_0$'s are basically free theories, and the available interactions are typically weakly relevant or even marginal, so that the convergence is poor.

Motivated by the need to overcome these limitations, much recent work has focused on improving the convergence of the method. One natural idea is to construct a \emph{renormalized} truncated Hamiltonian, whose couplings are corrected to take into account the effect of states above the cutoff which are truncated away. The renormalized truncated Hamiltonian is still diagonalized numerically, but its eigenvalues exhibit a smaller dependence on the cutoff. This method was developed in \cite{Giokas:2011ix,Hogervorst:2014rta,Lorenzo1} where renormalization corrections of leading (quadratic) order in the interaction $V$ have been considered. Leading-order (LO) renormalization has been successfully used to improve convergence in several HT studies \cite{Hogervorst:2014rta,Lorenzo1,Konik:2015bia,Lencses:2015bpa,Lorenzo2,Rakovszky:2016ugs,Azaria:2016mqb,Horvath:2017wzf}.

A natural hope \cite{Lorenzo1,Konik-review} is that one can improve convergence even further by consider next-to-leading (NLO) order renormalization corrections. Previous work on this problem \cite{Elias-Miro:2015bqk} led to somewhat pessimistic conclusions: it was found that the most straightforward NLO renormalization performs poorly. The goal of our paper will be to present a different implementation of NLO renormalization which overcomes the difficulty found in \cite{Elias-Miro:2015bqk} and improves convergence compared to the LO methods. A short exposition of our results has appeared in \cite{Elias-Miro:2017xxf}.

The paper is structured as follows. In section \ref{sec:general} we review previous work on the renormalized HT and describe our approach to NLO renormalization. Our construction is completely general and is presented as such. In the rest of the paper we apply NLO-renormalized Hamiltonian Truncation (NLO-HT) to one particular strongly coupled QFT---the $\phi^4$ theory in two spacetime dimensions. 
This is a field theory interesting both in its own right, and as a benchmark model for testing the HT method. This theory has been studied by renormalized HT in our prior work \cite{Lorenzo1,Lorenzo2,Elias-Miro:2015bqk},\footnote{It has also been recently studied by Coser et al \cite{Coser:2014lla} using a variant of the Truncated Conformal Space Approach (TCSA) \cite{Yurov:1989yu}, by Bajnok and L\'ajer \cite{Bajnok:2015bgw} using the HT, and in \cite{Chabysheva:2015ynr,Burkardt:2016ffk,Anand:2017yij} via the light front quantization. These papers did not use renormalization improvement.} and so it will be easy to compare the performance.

In section \ref{sec:basic} we remind the setup of the HT method as applied to $(\phi^4)_2$. 
We then explain how our general NLO-HT construction from section \ref{sec:general} can be implemented 
for this theory. In section \ref{sec:numres} we present numerical results. We study the spectrum dependence on the Hilbert space cutoff and show that the convergence is both smoother and more rapid for NLO-HT than for the LO renormalized HT.
We discuss the spectrum dependence on the volume $L$ and the extrapolation to the infinite volume. Finally, we study the dependence of the spectrum on the quartic coupling $g$, and determine the critical coupling where the theory transitions to the phase of spontaneously broken $\bZ_2$ symmetry. Then we conclude.

The interested reader will find much further useful information in the appendices. Appendices \ref{sec:struct},\ref{FE},\ref{sec:other-expansions} are devoted to conceptual issues: general considerations and numerical experiments regarding the structure of interacting eigenstates in finite volume (in particular how the orthogonality catastrophe is avoided), problems with naive implementations of renormalization corrections, and connections of the renormalized HT with the time-honored Brillouin-Wigner and Schrieffer-Wolff constructions of effective Hamiltonians. The rest of the appendices are more technical (see the table of contents).

\section{General theory of the renormalized Hamiltonian Truncation}
\label{sec:general}
\subsection{Review of prior work}

\subsubsection{Raw HT}

Consider a QFT in a finite spatial volume $L$, quantized on surfaces of constant time.\footnote{In relativistic QFTs one can also quantize on surfaces of constant light-cone coordinate. This light front quantization \cite{Brodsky:1997de} is also used in numerical solutions of strongly coupled QFTs via a version of HT; some recent work is \cite{Katz:2013qua,Katz:2014uoa,Chabysheva:2015ynr,Burkardt:2016ffk,Katz:2016hxp,Anand:2017yij}. The structure of the unperturbed Hilbert space is different from the equal time case, which leads to important differences in the numerical procedure. All technical claims in this work will refer exclusively to the equal time quantization.} The Hamiltonian has the form
\beq
H=H_0+V\,.
\eeq
The Hamiltonian $H_0$ is assumed to have an exactly solvable discrete spectrum of eigenstates, which form a basis in the Hilbert space $\calH$. 
The matrix elements of $V$ among $H_0$ eigenstates are assumed known, so that we can view $H$ as an infinite matrix acting in $\calH$. In many applications $V$ is an integral of a local operator:
\beq
V=\int_B \calO \,.
\eeq
For a concrete example, think of $H_0$ describing a free massive scalar field $\phi$ in $1+1$ dimensions, $\calH$ the Fock space, and $\calO=\NO{\phi^4}$ the quartic interaction. This example will be considered in detail below. For the moment we would like to stay general.

Let us now pick an energy cutoff $E_T$ and divide the Hilbert space into the low- and high-energy subspaces:
\beq
\calH=\calH_l\oplus \calH_h\,,
\eeq
where $\calH_l$ is spanned by basis states with $H_0$-eigenvalue $E\le E_T$.\footnote{$E_T=E_{\rm max}$ in the notation of \cite{Lorenzo1}.} Notice that one could in principle consider different types of cutoff, which depend not only on $E_T$ but on other conserved quantum numbers which may be present in the integrable Hamiltonian $H_0$, for example, occupation numbers of individual momentum modes for free $H_0$. It's a tantalizing but little-explored possibility that significant improvement can be achieved by considering alternative cutoffs (see appendix \ref{sec:struct}). 

The HT method constructs the ``truncated Hamiltonian", which is the Hamiltonian $H$ restricted to the finite-dimensional subspace $\calH_l$. The truncated Hamiltonian is diagonalized numerically, producing ``raw" \cite{Hogervorst:2014rta} spectrum. {We will assume that the scaling dimension of the perturbing operator $\calO$ is below $d/2$. In this case the raw spectrum converges to the exact finite volume spectrum for $E_T\to\infty$ \cite{Klassen:1991ze,Hogervorst:2014rta}.} However, in practice one cannot push to very high $E_T$ as the dimension of $\calH_l$ grows exponentially (see appendix \ref{sec:struct}). 
In many practically interesting cases one finds that the convergence error is still non-negligible at the maximal numerically accessible cutoffs. This calls for improvements.

\subsubsection{Integrating out versus truncating}
\label{sec:Heff}

A natural way to reduce the convergence error is to integrate out the high energy states rather than to simply truncate them away. This can be done rigorously as follows. The eigenvalue equation for the full Hamiltonian in the full Hilbert space is:
\beq
H.c=\calE c,\quad c\in \calH\,.
\label{p1}
\eeq
Let $c=(c_l,c_h)$ be the low- and high-energy components of the eigenvector $c$. We have:\footnote{For any operator $A$ acting on $\calH$ we denote $  A_{\alpha\beta}=P_\alpha A P_\beta\,,$
where $P_\alpha$ ($\alpha=l,h$) is the orthogonal projector on $\calH_\alpha$. In this notation $H_{ll}$ is the truncated Hamiltonian.}
\begin{gather}
H_{ll}.c_l + V_{lh}.c_h=\calE c_l\,, \label{loeig}\\
V_{hl}.c_l + H_{hh}.c_h=\calE c_h \label{hieig}\,.
\end{gather}
From the second equation we have
\beq
\label{eq:tail}
c_h=(\calE-H_{hh})^{-1}.V_{hl}.c_l\, .
\eeq
    Substituting this into the first equation we obtain
\beq
\label{eq:eigequiv}
H_{\rm eff}.c_l=\calE c_l,\quad c_l\in \calH_l\,,
\eeq
where 
\begin{gather}
H_{\rm eff} = H_{ll}+\Delta H(\calE)\,,\\
\label{eq:DeltaH}
\Delta H(\calE)=V_{lh}.(\calE-H_{hh})^{-1}.V_{hl}\,.
\end{gather}
The eigenvalue equation \reef{eq:eigequiv} in the truncated Hilbert space is exactly equivalent to the original eigenvalue equation \reef{p1} in the full Hilbert space. 
The term $\Delta H$ takes into account the removal of the high energy states. Needless to say, $\Delta H$ cannot be found exactly in any situation of interest, because $\cE-H_{hh}$ is impossible to invert exactly. However, one can hope that it can be found approximately, and that using these approximations and diagonalizing $H_{\rm eff}$ one can reduce the convergence error compared to the raw truncation at the same cutoff value. This will be discussed below.

{\bf Historical comment}

The above effective Hamiltonian construction was first brought to bear on the problem of renormalized HT in \cite{Hogervorst:2014rta,Lorenzo1}. However, in the general quantum mechanics context, it goes back at least as far as the work of Feshbach \cite{Feshbach1,Feshbach2} and L\"owdin \cite{Lowdin} around 1960. It is also used in quantum chemistry, see e.g.~\cite{Malrieu, Neese}. There, the procedure of dividing the Hilbert space is
called `partitioning', $\calH_l$ and $\calH_h$ the `model' and the `outer' space, and $H_{ll}+\Delta H(\calE)$ the `intermediate' Hamiltonian. The approximation \reef{firstDeltaH2}, see below, is also commonly used.

See also appendix \ref{sec:other-expansions} for parallels between the renormalized HT and two other expansions used previously in quantum physics (the Brillouin-Wigner series and the Schrieffer-Wolff transformation). 

\subsubsection{Leading-order renormalized HT}
\label{sec:leading-order}

The simplest method to reduce cutoff effects and improve convergence of the HT is the local LO renormalization, first argued in \cite{Giokas:2011ix}. It is easy to implement in practice and it has been used in several recent HT studies \cite{Hogervorst:2014rta,Lorenzo1,Konik:2015bia,Lencses:2015bpa,Lorenzo2,Rakovszky:2016ugs,Azaria:2016mqb,Horvath:2017wzf}.

The method is best justified by viewing it as a particular approximation to $\Delta H$ \cite{Hogervorst:2014rta,Lorenzo1}. Earlier work on the renormalized HT idea includes \cite{Feverati:2006ni,Watts:2011cr, Lencses:2014tba}. We disagree with these papers and with \cite{Giokas:2011ix} on several conceptual points, and especially on the treatment of subleading effects, as discussed in \cite{Hogervorst:2014rta}, section 5.4. 

Consider a formal expansion of $\Delta H$ in powers of $V_{hh}$
\begin{gather}
\label{eq:formal}
\Delta H(\calE)=\sum_{n=2}^\infty \Delta H_n(\calE)\,,\qquad
\Delta H_n(\calE)=V_{lh}\frac 1{\calE-H_{0\,hh}}\left(V_{hh}\frac 1{\calE-H_{0\,hh}}\right)^{n-2} V_{hl}\, .
\end{gather}
Let us keep only the first term in this expansion (thus we approximate $H_{hh}\approx H_{0\,hh}$ in \reef{eq:DeltaH}):
\beq
\Delta H(\calE)\approx \Delta H_2(\calE) =  V_{lh}.(\calE-H_{0\, hh})^{-1}.V_{hl}\,.\ \label{firstDeltaH2}
\eeq
Although the matrix in the denominator is now diagonal and easy to invert, the definition still involves an infinite sum over all high energy states, and some approximation is required in order to compute it. The simplest and the most widely used is the local approximation \cite{Hogervorst:2014rta,Lorenzo1}, which adds small corrections to local couplings:
\beq
\label{eq:loc}
\Delta H_2 \approx \Delta H_{2}^{\rm loc} = \sum_{i} \kappa_i(E_T) \int_B \calO_i\,.
\eeq
Here $\calO_i$ are some local operators of the theory (the original interaction $\calO$ will be typically one of them). 
Coefficients $\kappa_i(E_T)$ can be given analytically, using the operator product expansion (OPE) \cite{Giokas:2011ix,Hogervorst:2014rta,Lorenzo1}. The $(\phi^4)_2$ theory case will be treated in detail below. 

Eq.~\reef{eq:loc} can be motivated as follows. By the effective field theory intuition, the local approximation can be expected to work well for the matrix elements $(\Delta H_2)_{ij}$ if the energies of the external states $E_{i,j}$ are much below $E_T$, the lowest intermediate energy summed over in \eq{firstDeltaH2}. These are the most important matrix elements, because the states with $E_i\ll E_T$ dominate the lower energy interacting eigenstates (see appendix \ref{sec:struct}). The matrix elements among states close to the cutoff are not well reproduced by the local approximation, but those states are unimportant.

Replacing $\Delta H$ by $\Delta H_{2}^{\rm loc}$ in $H_{\rm eff}$ gives the local LO renormalized truncated Hamiltonian. Solving the eigenvalue equation \reef{eq:eigequiv} numerically, we obtain the ``local renormalized" \cite{Lorenzo1}  spectrum. {Empirically, this spectrum does show} a smaller $E_T$ cutoff dependence than the raw spectrum, obtained by direct diagonalization of the truncated Hamiltonian $H_{ll}$.

\subsubsection{Beyond local leading-order approximation?}
\label{sec:beyond}

One modest improvement of the local LO approximation is the ``local subleading" approximation discussed in \cite{Hogervorst:2014rta,Lorenzo1}. For states well below $E_T$, it partially takes into account subleading dependence of the matrix elements $(\Delta H_2)_{ij}$ on their energy. It performs slightly but not dramatically better than the local one. So it is important to look for further improvements. 

Our goal will be to develop an NLO approximation, taking the cubic term $\Delta H_3$ into account. Naive NLO would be to use the first two terms in
\reef{eq:formal}
\begin{gather}
\label{eq:uptocubic}
\Delta H\approx \Delta H_2+\Delta H_3\qquad\text{(naive NLO)}\,.
\end{gather}
However, there is a difficulty in following this route \cite{Elias-Miro:2015bqk}. To recognize it, let us go back to the LO approximation \reef{firstDeltaH2} and mention a subtlety glossed over in that discussion.
 
Notice first of all that while the local approximation \reef{eq:loc} is convenient and natural, technically we are not forced to use it.
The local approximation is good for $E_i,E_j\ll E_T$, but if we really wanted, we could actually compute $\Delta H_2$ with reasonable accuracy for all energies below the cutoff, by splitting the infinite sum into two parts, treating one of them exactly, and the other approximately \cite{Elias-Miro:2015bqk} (see section \ref{sec:PI-dh2}). 
Suppose we did it. \emph{Would we get better results for the spectrum using $\Delta H_2$ instead of $\Delta H_2^{\rm loc}$?} 

Surprisingly, the answer is no. The explanation is as follows. When we {replace $\Delta H$ by $\Delta H_2$}, we already make an error. This error is small for $E_i,E_j\ll E_T$, but it turns out that it is very large for energies close to the cutoff. There, $\Delta H_2$ overestimates certain matrix elements by many orders of magnitude.
As we said, states close to the cutoff appear with tiny coefficients in the interacting low-energy eigenstates. So a moderate error involving the matrix elements among those states would not be important. However, the behavior of $\Delta H_2$ near the cutoff turns out to be so bad that it ruins the spectrum. In this respect, {using $\Delta H_2^{\rm loc}$ instead of $\Delta H_2$ is a blessing.} While it adds another small error for $E_i,E_j\ll E_T$, it also regularizes the extremely bad behavior of $\Delta H_2$ near the cutoff. Of course $\Delta H_2^{\rm loc}$ remains inaccurate near the cutoff, but this inaccuracy is order one and does not affect the spectrum appreciably.

Now consider the naive NLO proposal \reef{eq:uptocubic}. The described problem with $\Delta H_2$ is just the first sign that the series expansion \reef{eq:formal} is inadequate for the matrix elements of $\Delta H$ involving states close to the cutoff $E_T$ (see appendix~\ref{FE}). Given this problem, what can we do? To mitigate the bad behavior near the cutoff, we could try to treat $\Delta H_3$ in \reef{eq:uptocubic} via a local approximation. However, to match the expected increase in accuracy, we would have to treat $\Delta H_2$ better than in the local or the local subleading approximation, and at the same time regularize the bad behavior near the cutoff. It's not obvious what such an approximation might be.
 
In the next section we will present a modified approach to NLO renormalization, which neatly avoids all mentioned difficulties. Another approach, to be explored in the future, is outlined in appendix \ref{sec:alternative}.

\subsection{NLO renormalization which works: NLO-HT}
\label{sec:ren-tails}
 
We will now describe our modified approach to NLO renormalization. Let us revisit the effective Hamiltonian construction in section \ref{sec:Heff}. Let's focus on the key equation \reef{eq:tail}, which expresses the ``tail", i.e.~the high energy part $c_h$ of the eigenvector, in terms of its low-energy part $c_l$. If we simply diagonalize $H_{ll}$, we forget about these tails. On the other hand, the correction $\Delta H$ in the effective Hamiltonian takes the tails into account.

Our approach will take the tails into account in a slightly different way, motivated by the already mentioned connection between the Hamiltonian Truncation and the Rayleigh-Ritz (RR) method. In the RR method, one diagonalizes the Hamiltonian truncated to a subspace $\calH_{\rm RR}$ of the full Hilbert space. For example, the raw HT method corresponds to $\calH_{\rm RR}=\calH_l$. The cornerstone of the RR method is the variational characterization of the truncated eigenvalues provided by the min-max principle. It implies, in particular, that as the subspace $\calH_{\rm RR}$ is enlarged, the truncated eigenvalues approach the exact eigenvalues monotonically \emph{from above}.

The raw HT enlarges $\calH_{\rm RR}$ by raising the energy cutoff $E_T$, but this is exponentially expensive.  A more efficient way to enlarge $\calH_{\rm RR}$ would be to add new basis elements capable of reproducing the entire tails \reef{eq:tail}. This is the idea of our approach. Formally, we will proceed as follows. We will be applying the RR method in the subspace $\calH_{\rm RR}$ of the form:
\beq
\label{eq:RR}
\calH_{\rm RR}=\calH_{l}\oplus \calH_{t}\,,
\eeq
where $\calH_{l}$ is the same as above with a certain cutoff $E_T$, and $\calH_{t}$ is a finite-dimensional subspace of $\calH_h$ spanned by ``tail states" defined below. Since this $\calH_{\rm RR}$ is strictly larger than $\calH_{l}$, we are guaranteed to do better than the raw truncation. How much better will depend on the choice of tail states. 

Let $|i\rangle$ be the Fock state basis of $\calH_l$, $i=1\ldots D={\rm dim} \calH_l$. The tail states $\ket{\Psi_i}$ will be vectors in the high-energy Hilbert space $\calH_h$. The ``optimal'' choice for $\ket{\Psi_i}$ would be
\begin{equation}
\label{eq:wouldbe}
 (\cE-H_{hh})^{-1}. V_{hl} \ket{i}\,\quad\text{(would-be optimal tails)}.
\end{equation}
Since $c_l$ in \reef{eq:tail} is a linear combination of $|i\rangle$, using these optimal tail states we could reproduce $c_h$ exactly, and so the RR eigenvalues would be equal to the exact eigenvalues. 

The optimal tails cannot be found and manipulated exactly, for the same reason that $\Delta H$ in \reef{eq:DeltaH} cannot be found exactly. Instead, we will use a simple approximation to the optimal tail states:
\begin{equation}
\label{eq:tailstate}
\ket{\Psi_i} =  (\cE_*-H_{0\,hh})^{-1}. V_{hl} \ket{i}\,\quad\text{(simpler tails used here)}\,.
\end{equation}
Here we replaced the exact eigenvalue $\calE$ by some reference energy ${\cal E}_*$ which will be eventually chosen close to a given eigenvalue of interest. We also replaced $H_{hh}$ by $H_{0\,hh}$. We will see that these simpler tail states are tractable. We will also see that the RR method using the simpler tails performs significantly better than both the raw truncation and the LO renormalization procedures. This is a sign that the simpler tails do approximate the optimal tails reasonably well. 

So, subspace $\calH_{t}$ in \reef{eq:RR} will be spanned by $\ket{\Psi_i}$ defined in \reef{eq:tailstate}. 
In the numerical calculations of this work we will always include the full set of tails $\calT = \{1\ldots D\}$.
However, a priori we can include tail states corresponding to any subset $i\in \mathcal{T} \subset 
\{1\ldots D\}$ of low-energy states. In this section we will develop the theory for such a general case.\footnote{\label{note:tails}One sensible way for selecting $\mathcal{T}$ would be to include only states $\ket{i}$
having a big overlap with the low-energy part $c_l$, so that $c_h$ can still be 
reproduced with a good approximation. As it will become clear later, by doing
so one would reduce the computational cost of the numerical procedure.
In the future, it is worth investigating more carefully the trade-off between the number of included tails
and the accuracy of the method. See also Fig.~\ref{fig:FirstImp} in appendix \ref{sec:struct}. Another way to take advantage of an incomplete set of tails is mentioned
in section \ref{sec:diag}.}

The reader may be wondering what all this has to do with the NLO renormalization. This will become clear later, once we formalize the procedure. Consider the eigenvalue equation \reef{p1} truncated to the $\calH_{\rm RR}$ subspace \reef{eq:RR}. In operator form we have
\beq
P_{\rm RR}H P_{\rm RR}\ket{\psi}=\calE_{\rm RR} \ket{\psi},
\eeq
where $\ket{\psi}\in \calH_{\rm RR}$, $P_{\rm RR}$ is the corresponding projector, and $\calE_{\rm RR}$ is the RR eigenvalue. We will call it $\calE$ from now on, although it's only an approximation to the exact eigenvalue appearing in \reef{p1} and \reef{eq:eigequiv}. In matrix form the equation becomes
\beq
\label{eq:matrixRR}
H_{\rm RR}.c= \cE G_{\rm RR}.c\,,
\eeq
where $c=(c_l,c_t)$ are the components of $\ket{\psi}$ when expanded in the basis of $\calH_{RR}$:
\beq
\label{eq:psi+tail}
\ket{\psi}=\sum_{i=1}^D (c_l)_i\ket{i}+\sum_{j\in\calT}(c_{t})_j\ket{\Psi_j},
\eeq
$H_{\rm RR}$ is the matrix of $H$ in the same basis, and $G_{\rm RR}$ is the Gram matrix. Since the tail states live in $\calH_h$, the Gram matrix has the block-diagonal form:
\begin{equation}
G_{\rm RR}=   \left(\begin{array}{cc}   \mathds{1}  &  \\    & G_{tt}    \end{array}\right)     \, . 
\end{equation}
The part $G_{tt}=G_{tt}(\calE_*)$ is nontrivial because the tail states are not orthogonal; it is given by:
\beq
\label{eq:Gtt}
(G_{tt})_{ij}=\bra{\Psi_i}\Psi_j\rangle = 
\bra{i} V_{l h} \frac{1}{(\cE_*-H_{0 hh})^2} V_{h l} \ket{j}
\,\qquad (i,j\in\calT)\,.
\eeq
Consider now the block structure of $H_{\rm RR}$:
\begin{equation}
H_{\rm RR}= \left(\begin{array}{cc}H_{ll} & H_{lt} \\ H_{tl} & H_{tt} \end{array}\right)  \, .  \label{Htails1}
\end{equation}
Here $H_{ll}$ is the usual Hamiltonian truncated to $\calH_l$. The other blocks must be worked out using the definition of tail states. It turns out that they can be conveniently expressed in terms of $\Delta H_2$ and $\Delta H_3$ discussed in the previous section:
\begin{gather}
\label{eq:Hlt}
(H_{lt})_{ij}= \bra{i}H\ket{\Psi_j} =  \Delta H_2(\cE_*)_{ij}\qquad(i\in\{1\ldots D\},j\in\calT)\, ,\\[2pt]
\label{eq:Htt}
(H_{tt})_{ij} = \bra{\Psi_i}H\ket{\Psi_j}  = [-\Delta H_2(\cE_*)+ \Delta H_3(\cE_*)+\cE_* \, G_{tt}(\calE_*)]_{ij} \ \qquad (i,j\in\calT).
\end{gather}
Eq.~\reef{eq:Hlt} 
is immediate, and \reef{eq:Htt} requires a one-line calculation. We also have
$H_{tl}=H_{lt}^\dagger$.

Let us rewrite the generalized eigenvalue problem \reef{eq:matrixRR} in a form 
analogous to \reef{loeig}, \reef{hieig},
\begin{align}
H_{ll}.c_l + H_{lt}.c_t&=\calE c_l\,, \\
H_{tl}.c_l + H_{tt}.c_t&=\calE G_{tt}.c_t\,.
\label{eq:eigeq}
\end{align}
See section \ref{sec:idea} for a discussion of how one could proceed to find the spectrum directly from these equations and of computational advantages it could bring (in the context of the $\phi^4$ theory). In this paper we will instead transform the problem to an equivalent form by eliminating the tail components $c_t$ and deriving an effective equation involving only $c_l$. While this step is not strictly speaking necessary, it will bring additional physical insight on the method. 
So, expressing $c_t$ from the second equation and substituting into the first, we get an analogue of \reef{eq:eigequiv}:
\begin{gather}
(H_{ll}+\Delta \widetilde H).c_l=\calE c_l\,, \label{Htails2}\\
\Delta \widetilde H= H_{lt}.(\calE G_{tt}-H_{tt})^{-1}.H_{tl}\,.
\end{gather}
Using \reef{eq:Hlt}, \reef{eq:Htt} we obtain
\beq
\Delta \widetilde H= \Delta H_2(\cE_*)_{lt} \frac{1}{\Delta H_2(\cE_*)_{tt}- \Delta H_3(\cE_*)_{tt}+(\cE-\cE_*)\, G(\cE_*)_{tt}}\Delta H_2(\cE_*)_{tl}   \label{neff}  \, .
\ee
We emphasize the notation: every time a matrix has a subscript $l$ (resp. $t$) it means that the corresponding index runs over the full $\{1\ldots D\}$ (resp. over the subset $\calT$).

In our computations we will always choose $\calE_*$ sufficiently close to $\calE$ 
for the states of interest (which will be the lowest energy states in both 
parity sectors), and neglect the last term in the denominator.\footnote{The correction proportional to $G$ could be comparable to $\Delta H_3$ for the excited states, 
for which $\cE-\cE_*$ is order one. We could add this correction exactly or perturbatively as in \cite{Lorenzo1}, but we will not do it in this work.} 
Also let us specialize to the case when $\calT$ is the full set of tails, as will be in all numerical computations below.
In this case we obtain a simplified expression:
\be 
\label{eq:mainresult}
\boxed{\Delta \widetilde H= \Delta H_2(\cE_*)\frac{1}{\Delta H_2(\cE_*)- \Delta H_3(\cE_*)}\Delta H_2(\cE_*)}   \, ,
\ee
where all matrices have indices running over the full basis of $\calH_l$. This is our main theoretical result. In the rest of the paper we will test how this correction performs, in the context of the two dimensional $\phi^4$ theory.

Finally let us clarify the relation with NLO. Performing a formal power series expansion of $\Delta \widetilde H$ in $\Delta H_3$ up to the first order, we obtain:
 \be
 \label{eq:dhtformal}
\Delta \widetilde H = \Delta H_2(\cE_*) + \Delta H_3(\cE_*) +\ldots\,.
\ee
For $\calE\approx \calE_*$, these are the same two terms as in the naive NLO correction \reef{eq:uptocubic}. We see that our approach based on $\Delta \widetilde H$ will capture $O(V^3)$ corrections, unlike the studies in \cite{Hogervorst:2014rta, Lorenzo1,Lorenzo2,Elias-Miro:2015bqk} based on $\Delta H_2$. For this reason we will refer to $\Delta \widetilde H$ as ``NLO renormalization correction". In practice we will of course use the full expression \reef{eq:mainresult} without expanding. 

Of course, $\Delta \widetilde H$ is not identical to the naive NLO correction, differing by the higher order \ldots terms in \reef{eq:dhtformal}. That's good because naive NLO fails, as discussed in section \ref{sec:beyond}. On the other hand our NLO approach is guaranteed not to fail. This is because we arrived at our $\Delta \widetilde H$ via a variational route. Since the Hilbert space \reef{eq:RR} is strictly larger than the raw truncated Hilbert space $\calH_l$, our NLO renormalization is guaranteed to perform better than the raw truncation. As we will see, it also performs better than the local LO renormalization from section \ref{sec:leading-order}.

\section{NLO-HT for $(\phi^4)_2$ theory}
\label{sec:basic}

In the previous section we gave a general description of NLO renormalized Hamiltonian Truncation (NLO-HT). In the rest of the paper we will apply this method to one particular strongly coupled QFT: the $\phi^4$ theory in $d=2$ spacetime dimensions. 
In this section we describe implementation of the method, and in the next one the numerical results. As we have already studied the $(\phi^4)_2$ theory in \cite{Lorenzo1,Lorenzo2,Elias-Miro:2015bqk} using the LO renormalization, it will be very instructive to compare.

\subsection{The $(\phi^4)_2$ theory}
 
We give here only the minimal information, see \cite{Lorenzo1} for the details. The theory is defined by the normal-ordered Euclidean action 
\beq
S=\half \int d^2 x\,[ \NO{(\del\phi)^2+m^2\phi^2}+g\, \NO{\phi^4}]\,.
\eeq
We quantize it canonically on a cylinder with periodic boundary conditions, expanding the field into creation and annihilation operators:
\begin{gather}
\phi(x,\tau=0)=\sum_k \frac{1}{\sqrt{2L\omega_k}}(a_k e^{ikx}+a_k^\dagger e^{-ikx})\,,\\
k=2\pi n/L\ (n\in \bZ),\quad \omega_k=\sqrt{m^2+k^2}\,,\qquad [a_k,a_{k'}]=0,\quad [a_k,a^\dagger_{k'}]=\delta_{kk'}\,.
\end{gather}
Here $x$ is the coordinate along the spacial circle of length $L$, while $\tau\in\bR$ is the Euclidean time along the cylinder.

In terms of normal-ordered operators, the Hamiltonian is 
a sum of the free piece and the quartic interaction, plus finite-volume corrections,
\begin{gather}
H=H_0 + g \left[ V_4 + 6 z(L) V_2 \right] + \left[ E_0(L) + 3 z(L)^2 g L \right]
\,,\qquad H_0=\sum_k \omega_k a^\dagger_k  a_k\,,\nn\\
V_2 = L \sum_{k} \frac 1{2L\omega_k}
\Big[ a_{k}a_{-k} + 2 a^\dagger_{k}a_{k} +a^\dagger_{k}a^\dagger_{-k}  \Big]\,,
\label{eq:Hdef}
\\ V_4 = L \sum_{\sum k_i=0} \frac 1{\prod\sqrt{2L\omega_i}}
\Big[ (a_{k_1}a_{k_2}a_{k_3}a_{k_4} +
4 a^\dagger_{-k_1}a_{k_2}a_{k_3}a_{k_4} +\text{h.c.}) +
6 a^\dagger_{-k_1}a^\dagger_{-k_2}a_{k_3}a_{k_4} 
 \Big]\,.\nn
\end{gather}
The $E_0(L)$ and $z(L)$ terms are exponentially suppressed in the limit $L m \gg 1$. They are discussed in \cite{Lorenzo1} and defined in Eqs.~(2.10), (2.18) of that paper, which we do not reproduce here. Introduction of these terms is necessary for putting the theory correctly in finite volume. For example, $E_0(L)$ can be understood as 
the Casimir energy. In \cite{Lorenzo1} these contributions
were described, but then neglected in the numerical analysis. In this work they will be kept, as the numerical error will be sometimes
smaller in comparison, allowing us to analyze these exponentially suppressed effects.

The Hamiltonian $H$ acts in the free theory Fock space $\calH_{\rm Fock}$ in finite volume $L$ (we will consider volumes up to $10 m^{-1}$). There are three conserved quantum numbers:  total momentum $P$, spatial parity $\bP$ ($x\to-x)$, and field parity $\bZ_2$ ($\phi\to -\phi$). As in \cite{Lorenzo1,Lorenzo2,Elias-Miro:2015bqk}, we will focus on the invariant subspaces $\calH^{\pm}$ consisting of states with $P=0$, $\bP=+$, $\bZ_2=\pm $. The states in $\calH^{+}$ (resp.~$\calH^{-}$) contain even (resp.~odd) number of free quanta.
The basic problem is to find eigenstates of $H$ belonging to $\calH^{\pm}$.  The two subspaces don't mix and the diagonalization can be done separately.

The lowest eigenstate in $\calH^{+}$ is the ground state in finite volume (the interacting vacuum). The interpretation of the lowest eigenstate in $\calH^{-}$ depends on the phase of the theory, namely if the $\bZ_2$ symmetry is spontaneously broken in infinite volume or not. 
The $\bZ_2$-preserving phase is realized for moderate quartic couplings $g/m^2< g_c$, where the critical coupling was measured as $g_c=2.97(13)$ in \cite{Lorenzo1}, while here we will find a smaller but compatible value $g_c\approx 2.8$. In the $\bZ_2$-preserving phase, the lowest $\calH^{-}$ eigenstate is the one-particle excitation at zero momentum. Excitation energy over the ground state then measures the physical particle mass $m_{\rm ph}$. In the $\bZ_2$-broken phase at $g/m^2>  g_c$, the lowest $\calH^{-}$ eigenstate is the second vacuum, exponentially degenerate with the first one at finite $L$ \cite{Lorenzo2,Bajnok:2015bgw}. 

In this paper we will focus on the $\bZ_2$-preserving phase, below $g_c$. We will use the NLO-HT method to measure the physical mass $m_{\rm ph}$ as a function of the quartic coupling. We will also measure $g_c$, as the point where $m_{\rm ph}$ goes to zero. It will be instructive to compare with \cite{Lorenzo1} where these measurements were done using the LO renormalized HT.

\subsection{NLO-HT implementation outline}
\label{PI}

Here and below we will fix the units of energy by setting the mass to $m=1$. 

In our {\tt python} code, we first build the Fock state basis of $\calH=\calH^{\pm}$ up to a fixed energy cutoff $E_T$. For example, we will use $E_T=20$ for $L=10$, corresponding to order $10^4$ states. We then evaluate the matrix elements of $H$ between these states (i.e.~the matrix $H_{ll}$) directly from the definition \reef{eq:Hdef}. This matrix is sparse, and it is important to organize this computation exploiting this sparsity maximally efficiently. Our current algorithm improves on \cite{Lorenzo1}; it is described in appendix \ref{sec:compdetails}.
The subsequent steps are the computation of $\Delta \widetilde H$ and the numerical diagonalization; they are discussed below. 

\subsubsection{$\Delta H_2$}
\label{sec:PI-dh2}

We need to evaluate the matrix element $(\Delta H_2)_{ij}$ between any two $\calH_l$ states. 
Recall that $\Delta H_2$ is defined by \reef{firstDeltaH2} which is an infinite sum over intermediate states in $\calH_h$. The choice of $\calE_*$ will be described below; for now let's keep it as a free parameter. 

We introduce a new cutoff  
$E_L>E_T$ {('$L$' for `local approximation')} and split this sum into ``moderately high" states in the range $E_T< E_k\leq E_L$ and ``ultrahigh" ones of energy $E_k>E_L$~\cite{Elias-Miro:2015bqk}:
\begin{align}
\Delta H_2(\calE_*) &= \Delta H_2^< + \Delta H_2^>\, ,\label{eq:splitDH2}\\
(\Delta H_2^<)_{ij} &=  \sum_{k: E_T< E_k\leq E_L} V_{ik}\frac{1}{\cE_*-E_k}V_{kj} \, ,  \label{nonl1} \\
(\Delta H_2^>)_{ij} &= \sum_{k: E_k>E_L} \text{(same)}\,. 
\label{l1} 
\end{align}

The number of ``moderately high'' states, which contribute to $\Delta H_2^<$, is large but finite. 
We will choose $E_L$ not excessively large, so that this finite sum can be done exactly; see appendix \ref{sec:compdetails} for the algorithmic details. 
On the other hand, while the number of ultrahigh states contributing to $\Delta H_2^>$ is infinite, all of these states have energy significantly higher than the external energies $E_{i,j}$. 
For this reason we will be able to approximate the matrix $\Delta H_2^>$ by a sum of local operators:
\beq
\label{eq:loc>}
(\Delta H_2^>)_{ij} \approx   \sum_{N=0,2,4} \kappa_N(E_L) (V_N)_{ij} \,,\qquad
V_N = \int_0^Ldx\,\NO{\phi(x)^N}\,.
\eeq
This is similar in spirit to the local approximation which we already encountered in Eq.~\reef{eq:loc}, with $E_T$ replaced by $E_L$. The operators $\NO{\phi(x)^N}$ are the particular examples of operators $\calO_i$ in that formula, as appropriate for the $\phi^4$ theory under consideration. For an explanation why only operators up to $V_4$ occur at this order, see \cite{Lorenzo1} and appendix \ref{la}.

The point of introducing the intermediate scale $E_L$ is that we want \reef{eq:loc>} to be a good approximation for all $E_{i,j}\le E_T$. Without the intermediate scale the approximation would break down close to the cutoff, as is the case for Eq.~\reef{eq:loc} that is true only for $E_{i,j}\ll E_T$. 

The expected accuracy of the local approximation \reef{eq:loc>} is $(E_T/E_L)^2$. In principle we need $E_L\gg E_T$, but in practice we will choose $E_L\approx 3 E_T$ and we will check that it already gives a reasonable approximation (see appendix~\ref{convEL}).
The local approximation can be justified using the operator product expansion (OPE) as in \cite{Lorenzo1}; it can also be connected with the diagram technique (see appendix \ref{la}). The coefficients $\kappa_{N}$ are given by~\cite{Lorenzo1}:
\be
\label{eq:kappamu}
\kappa_N(E_L)=g^2\int_{E_L}^\infty dE \, \frac{ \mu_N(E) }{\cE_*-E} \, ,
\ee
where $\mu_N$ can be conveniently expressed as the relativistic phase-space integrals~\cite{Elias-Miro:2015bqk}. They can be computed in an $m/E$ expansion and the leading terms are \cite{Lorenzo1}:\footnote{\label{note-infL}$\mu_{N}=\mu_{44N}$ in the notation of \cite{Lorenzo1}. In obtaining \eq{mus}, the infinite length limit $L\rightarrow\infty$ was taken. This is a good approximation for the volumes that we consider later in the numerical study. 
 While we will keep exponentially suppressed term in the zeroth-order Hamiltonian \reef{eq:Hdef}, keeping such terms in renormalization corrections is unimportant at the current level of accuracy.}
\be
\mu_0(E) = \frac{1}{E^2}\left\{ \frac{18}{\pi^3}(\log E/m)^2 -\frac{3}{2\pi}   \right\} \ , \quad \mu_2(E)= \frac{72 \log E/m}{\pi^2 E^2}\ , \quad \mu_4(E)= \frac{36}{\pi E^2}\, .\label{mus}
\ee

\subsubsection{$\Delta H_3$}
\label{sec:dh3-322}
The evaluation of $\Delta H_3$ follows the same strategy as for $\Delta H_2$. We introduce an intermediate cutoff $E_L'$ (in general different from $E_L$) and split the definition into four sums depending if the exchanged states $k,k'$ are moderately high or ultrahigh with respect to $E_L'$:
\begin{align}
\Delta H_3(\calE_*) &=\Delta H_3^{<<}  + \Delta H_3^{>>} + (\Delta H_3^{<>} + \text {h.c.}) \, , \label{d3org}\\[2pt]
(\Delta H_3^{<<})_{ij}  &=  \sum_{k, k^\prime: E_T< E_{k,k^\prime }\leq E'_L} V_{ik}\frac{1}{\cE_*-E_k}V_{kk^\prime}\frac{1}{\cE_*-E_{k^\prime}} V_{k^\prime j}   \label{nonl2}  \ , \\ 
(\Delta H_3^{>>})_{ij}   &=  \sum_{k, k^\prime:  E_{k,k^\prime }> E'_L}   \text{(same)}\,, \label{l2} \\
(\Delta H_3^{<>})_{ij}&=\sum_{\substack{k: E_T< E_k\leq E'_L\\ k^\prime :  E_{k^\prime }> E'_L}} \text{(same)},\label{mix}
\end{align}
We compute $\Delta H_3^{<<}$ by evaluating and multiplying the involved finite matrices; see appendix \ref{sec:compdetails}. For $\Delta H_3^{>>}$ we use a local approximation: 
\be
\Delta H_3^{>>} \approx  \sum_{N=0,2,4,6} \lambda_{N}V_N+\lambda_{2|4}\,\NO{V_{2}V_{4}}+\lambda_{4|4}\,\NO{V_{4}V_{4}}  \,. \label{localexp2}
\ee
This involves local operators up to $V_6$ as well as bilocal operators with up to eight fields, whose appearance is a novelty first observed here (see section \ref{sdh3} and appendix \ref{detailsDH3} for details).\footnote{Ref.~\cite{Konik-review} briefly discussed the local approximation at the cubic order, for the TCSA case when $H_0$ describes a CFT. Their Eq.~(321) {appears} incomplete, as it does not allow for bilocal operators. See also appendix \ref{sec:lessons}.\label{note:cubicTCSA}}

Concerning $\Delta H_3^{<>}$, its definition can be rewritten as a finite sum over moderately high $k$:
\be
(\Delta H_3^{<>})_{ij}= \sum_{ k: E_T< E_k\leq E'_L} V_{ik}\frac{1}{\cE_*-E_k} (\Delta H_2^>)_{k j}\, .  \label{mix2}
\ee
The $\Delta H_2^>$ here is the piece of $\Delta H_2$ receiving the contribution from the ultrahigh states; it is given by \reef{l1} with $E_L\to E_L'$. For \reef{l1} we could use a local approximation since both external energies were much below the cutoff, but here we cannot do this right away, since $E_k$ may be close to the cutoff $E'_L$. To deal with this nuisance, we introduce a further cutoff $E_L''> E'_L$. Then, in the sum defining $(\Delta H_2^>)_{k j}$, the part over the intermediate 
states below $E_L''$ is performed explicitly, and for the part above $E_L''$ the local approximation \reef{eq:loc>} is used (with $E_L\to E_L''$).

Let us now make some remarks on the computational cost of evaluating $\Delta H_3$, which is the most expensive step in the procedure. 
For the choice of parameters $L$, $E_T$, $E_L'$, $E_L''$ which we will use in section \ref{sec:numres}, the expressions 
\eqref{nonl2} and \eqref{mix} involve double sums over tens of millions of 
high-energy states. We are able to take advantage of the sparsity 
of the matrices to perform these sums relatively efficiently (see
appendix \ref{sec:compdetails} for the details). 
Still, this step limits the value of the local cutoffs and/or the number of tails 
that can be included. In the future, one may have to devise more efficient
approximate procedure to evaluating the matrix $\Delta H_3$. One simple option would be to discard tail states that are not important for modeling the 
high-energy part of the eigenvectors $c_h$ in \eqref{eq:tail} (see note \ref{note:tails}).
Alternatively, one could consider varying degrees of approximation for the 
different matrix entries of $(\Delta H_3)_{i j}$. For instance, if for a pair of states
$i, j$ it happens that $(\Delta H_2)_{i j} \gg (\Delta H_3)_{i j}$, one might be 
justified in discarding altogether the smaller contribution for this matrix element. 
It would be very interesting to explore these and other possibilities. We leave this for future work, 
while here we will stick to the simple prescription described so far.

This finishes a rough outline of how the needed matrices will be evaluated. We would like to emphasize one feature of the proposed 
algorithm: the systematic split of all sums into moderately high and ultrahigh parts. The moderately high sums are done by simply evaluating and multiplying the needed finite matrices, while in the ultrahigh parts the local approximation can be used. In appendix \ref{DT} we will review a diagrammatic technique of \cite{Elias-Miro:2015bqk}, which in principle provides a different way of organizing the computation of $\Delta H_n$. Since that technique is not easily automatizable, we will not use it here for the moderately high region computations. However, it will be instrumental for analyzing the local approximation for the ultrahigh parts (appendices \ref{la}, \ref{detailsDH3}).

\subsubsection{Idea for the future}
\label{sec:idea}

We would like to record here a promising idea which occurred to us late in this 
project, so that we had not had the chance to test it in detail. 
In the setup outlined in section \ref{sec:basic}, in which the full set of tails
is added to the variational ansatz, we can raise the cutoff up to $E_T=20$, 
beyond which computing the matrix $\Delta H_3$ becomes too expensive.
On the other hand the raw truncation can be implemented up to $E_T=35$.
We could further increase the accuracy of our procedure combining the two, 
i.e.~by considering $E_T=35$ but introducing an incomplete set of tails for states
below $E_t=20<E_T$ (see note \ref{note:tails}).
We think this combination may be affordable if we analyze this problem directly via 
\reef{eq:eigeq}, without integrating out the tails, as we explain in section 
\ref{sec:diag}.

\subsubsection{Diagonalization}
\label{sec:diag}
We used an iterative Lanczos method diagonalization routine {\tt scipy.sparse.linalg.eigsh} (based on ARPACK), with the parameter {\tt which=`SA'}, intended for computing algebraically smallest eigenvalues. With this choice of parameter the matrix is not inverted and diagonalization times are smallest.
Notice that this routine works both for sparse and non-sparse matrices. In our problem, the matrices $H_{ll}$, $\Delta H_2$, $\Delta H_3$ are sparse, but the matrix $ \Delta \tilde H$ is not sparse because of the matrix inversion involved in its definition.

With the same routine one could implement the idea outlined in section 
\ref{sec:idea}, by passing the inverse of the Gram matrix $G_{\rm RR}$
and solving directly \reef{eq:eigeq}. 
In that case every large matrix needed for the numerical algorithm 
will be kept in the sparse format, while the only non sparse matrix will
be $G_{tt}^{-1}$, of modest size.
 
Below we will also compare NLO-HT to the raw truncation at much higher cutoff, up to $E_T=35$ when the Hilbert space contains millions of states.  
In this work, the full needed matrix is always evaluated and saved in memory, and then the diagonalization routine is called. When the involved matrices are sparse, it might be possible to use the option of evaluating the needed matrix elements `on the fly', as opposed to prior evaluation and storage of the whole matrix. We have not explored this option in this work.

\section{Numerical results}
\label{sec:numres}

In the previous section we described how to set up the NLO-renormalized HT method for the $(\phi^4)_2$ theory in finite volume. 
In this section we will present the numerical results which come out of this implementation. Recall that we are working in the units in which $m=1$.

Our code is written in {\tt python} and was run on a cluster with 100 Gb RAM nodes. As an example of required computational resources, one NLO-HT data point in figure \ref{fig:specvsET1} for $L=10$ and $E_T=20$ requires 40 CPU hours and about 80 Gb RAM. Running time and memory requirements quickly decrease with $E_T$. The whole scan for $E_T=10$ - 20 in steps of 0.5 for a given $g$ takes about 140 CPU hours. For the raw and leading LO renormalized HT, the maximal attainable $E_T$ was limited by available RAM, while the running time was faster than for the NLO-HT.

\subsection{$E_T$ dependence}
\label{ETplots}

The numerical accuracy of the NLO-HT method is determined by the cutoff $E_T$ of the low-energy
Hilbert space, and by the auxiliary ``local'' cutoffs $E_L, E_L', E_L''$ 
introduced in section \ref{PI}. The latter cutoffs are used in the computation of $\Delta \tilde H$;
here we will fix them relative to $E_T$ as
$E_L=3 E_T$, $E_L'=2 E_T$, $E_L''=3 E_T$. This is high enough so that
$\Delta \tilde H$ is approximated sufficiently well (see also the checks in appendix \ref{convEL}).
The $\cE_*$ parameter in \reef{eq:mainresult} will be fixed as follows. At each $E_T$, we will choose $\cE_*$ 
equal to the energy of the lowest state in each $\bZ_2$ parity sector, as computed for the same $E_T$ in the local LO renormalized approximation
(section \ref{sec:leading-order}).

With the auxiliary cutoffs fixed as above, $E_T$ remains the only free parameter.
Therefore, the numerical error will be estimated just by varying $E_T$.

\begin{figure}[h!tbp]
\begin{center}
\includegraphics[scale=.48]{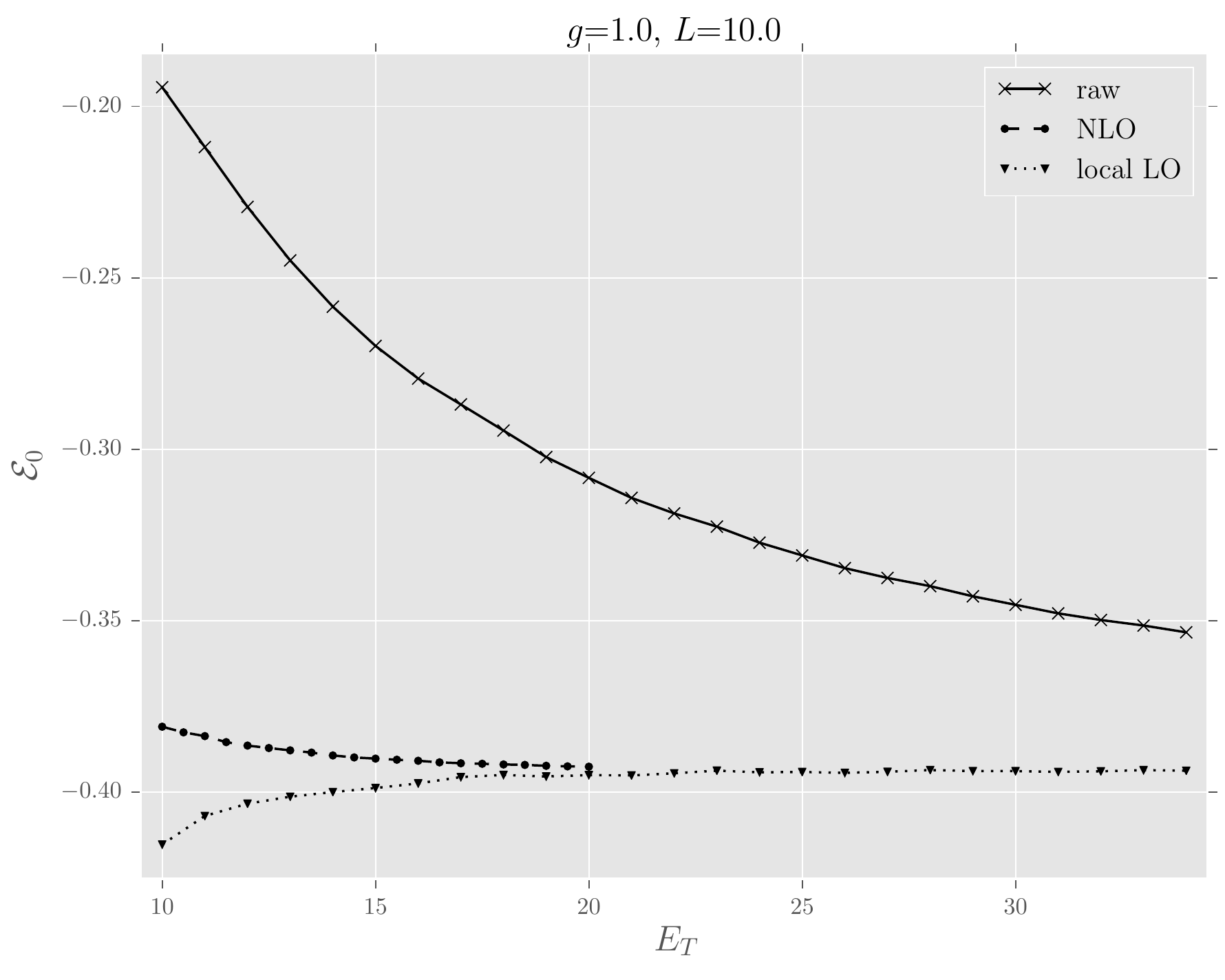}
\includegraphics[scale=.48]{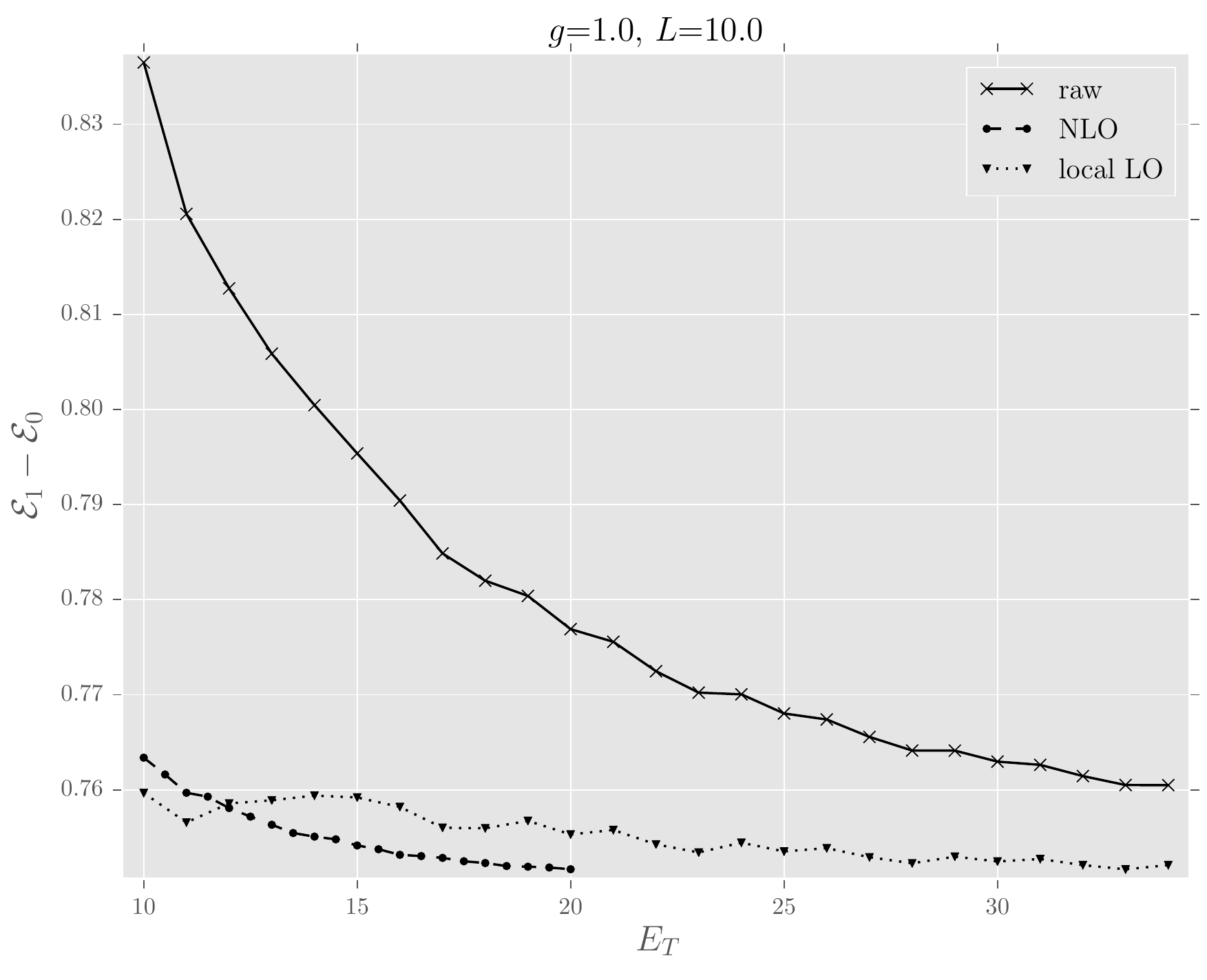}
\includegraphics[scale=.48]{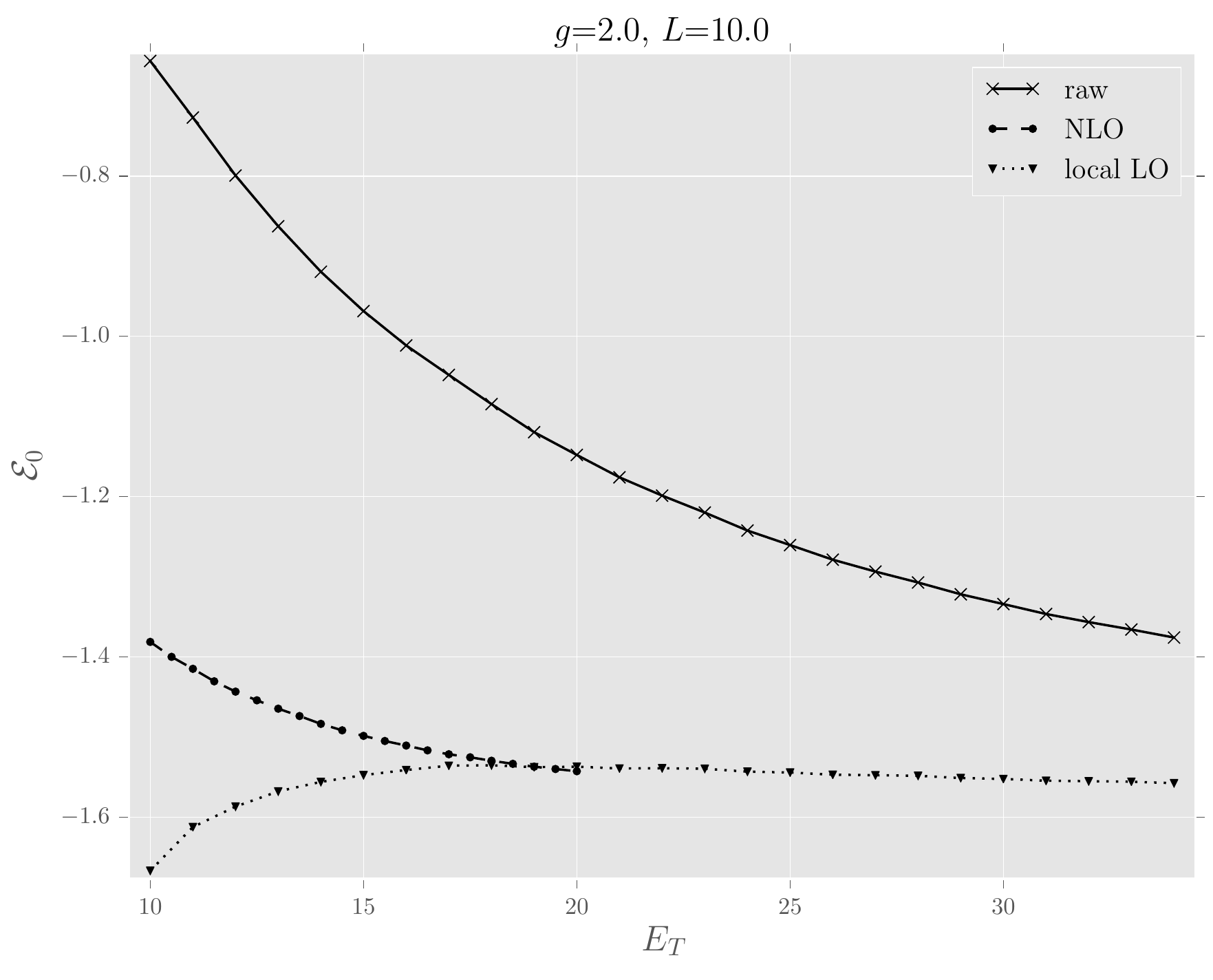}
\includegraphics[scale=.48]{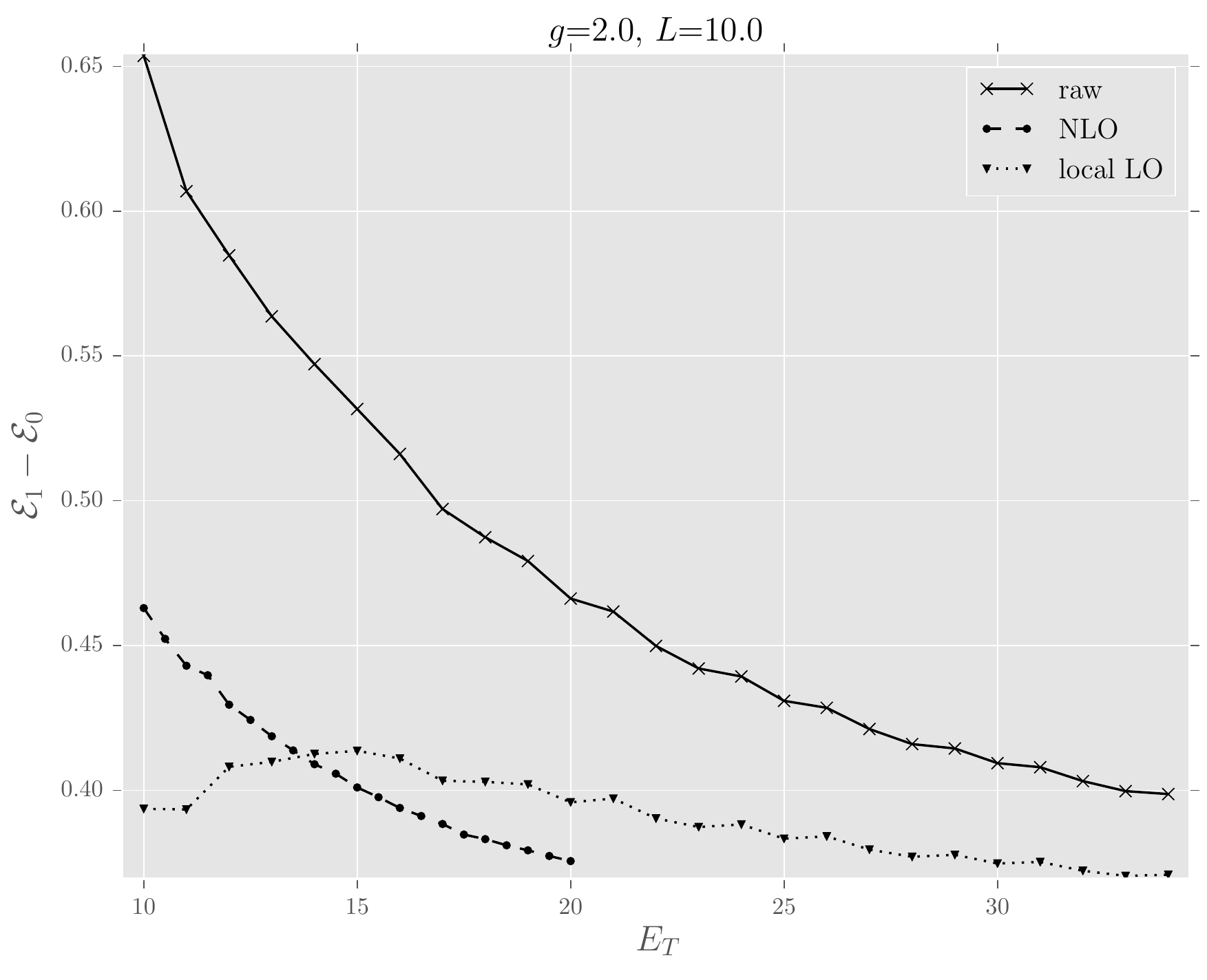}
\caption{The vacuum energy (left) and the physical mass (right) for $L=10$, plotted as a function of $E_T$ for the three methods: raw HT, local LO renormalized HT, and NLO-HT. The top (bottom) plots refer to $g=1$ ($g=2$). }
\label{fig:specvsET1}
\end{center}
\end{figure}

In Fig.~\ref{fig:specvsET1} 
we plot the NLO-HT vacuum 
energy $\cE_0$ and the physical mass $\cE_1 - \cE_0$ as a function of $E_T$, for $g=1$ and $2$ and $L=10$.
Notice that $g=1,2$, while being smaller than the critical coupling $g_c \approx 2.8$,
are well above the window $g \lesssim 0.2$ where perturbation theory is 
accurate.\footnote{See appendix B of \cite{Lorenzo1}.}
We could push the NLO-HT cutoff up to $E_T=20$ for $L = 10$, 
corresponding to $\sim 10^4$ states. The main numerical bottleneck which prevents us from going higher is the evaluation of $\Delta H_3$.

For the sake of comparison, in the same figure we overlay the 
numerical results obtained by two of the methods described in \cite{Lorenzo1}.
These are the raw truncation, in which the correction term $\Delta H (\cE)$ 
in \eqref{eq:eigequiv} is simply thrown away,
and the local LO renormalized (referred to as simply ``local" below) procedure, in which $\Delta H (\cE_*)$ is replaced by the
simpler correction term $\Delta H^{\rm loc}_2(\cE_*)$ computed in a fully local fashion, 
as discussed in section \ref{sec:leading-order}.\footnote{In this case $\calE_*$ is taken from raw HT. We do not include
the results obtained by the ``local subleading'' method of \cite{Lorenzo1}, 
which are only marginally more accurate than the local ones.}
As one can see, we are able
to push the cutoff $E_T$ much higher for these simpler methods, 
up to $E_T = 34$ for $L = 10$, corresponding to $\sim 10^7$ 
states. 

The first observation is that the raw HT and the NLO-HT are variational procedures, and hence always provide upper bounds on the eigenvalues, which become monotonically more accurate with increasing $E_T$. This is visible in the figure. 
On the contrary local renormalization is not variational and does not have to be monotonic.

From Fig.~\ref{fig:specvsET1} 
it is evident that the raw HT is by far the 
least accurate, therefore we will not report results of this method in the rest of the discussion. We will keep showing local results as a baseline to judge the relative 
advantages of the NLO-HT, and to justify its additional complexity.

\begin{figure}[tbp]
\begin{center}
\includegraphics[scale=.75]{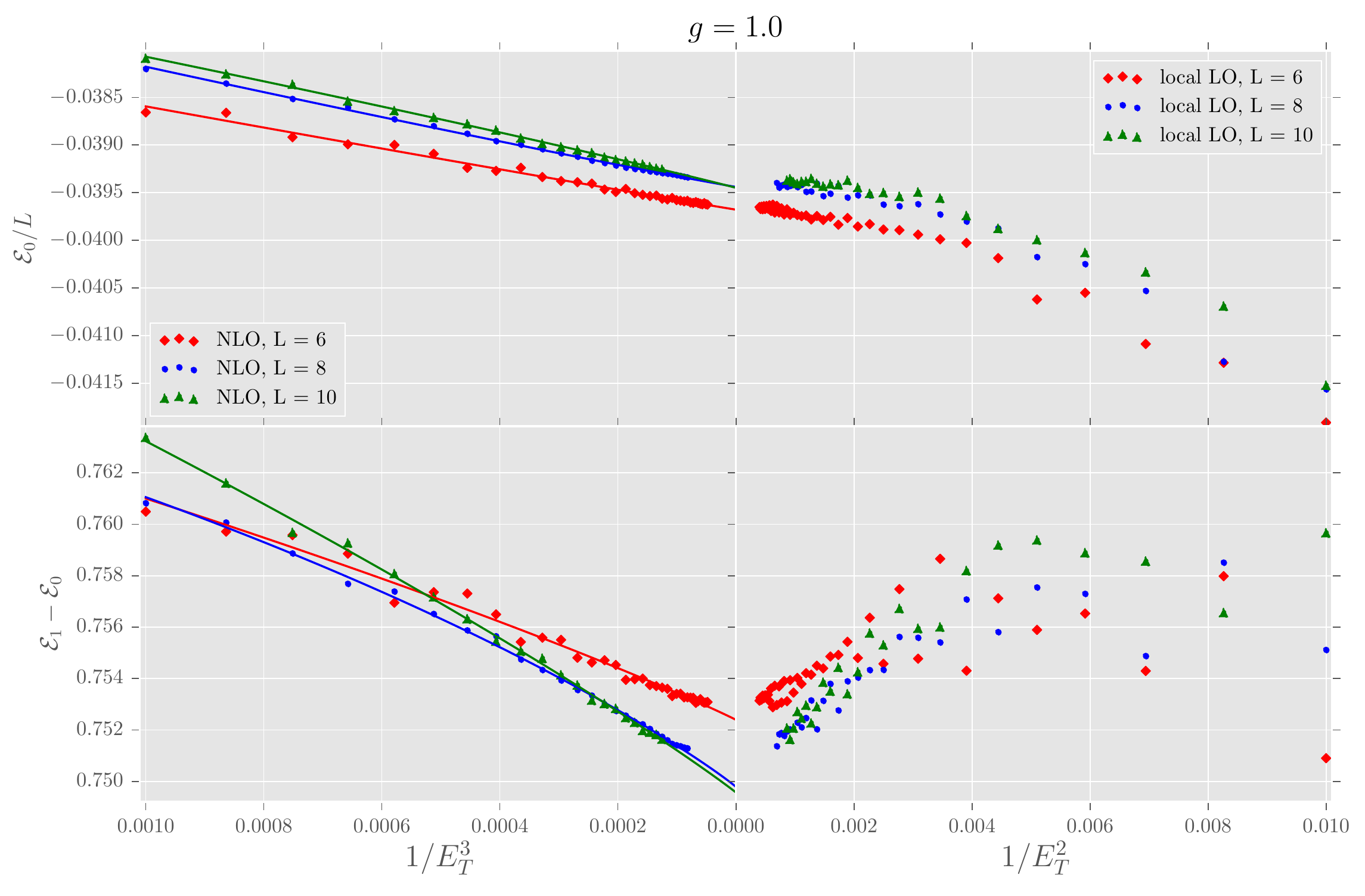}
\includegraphics[scale=.75]{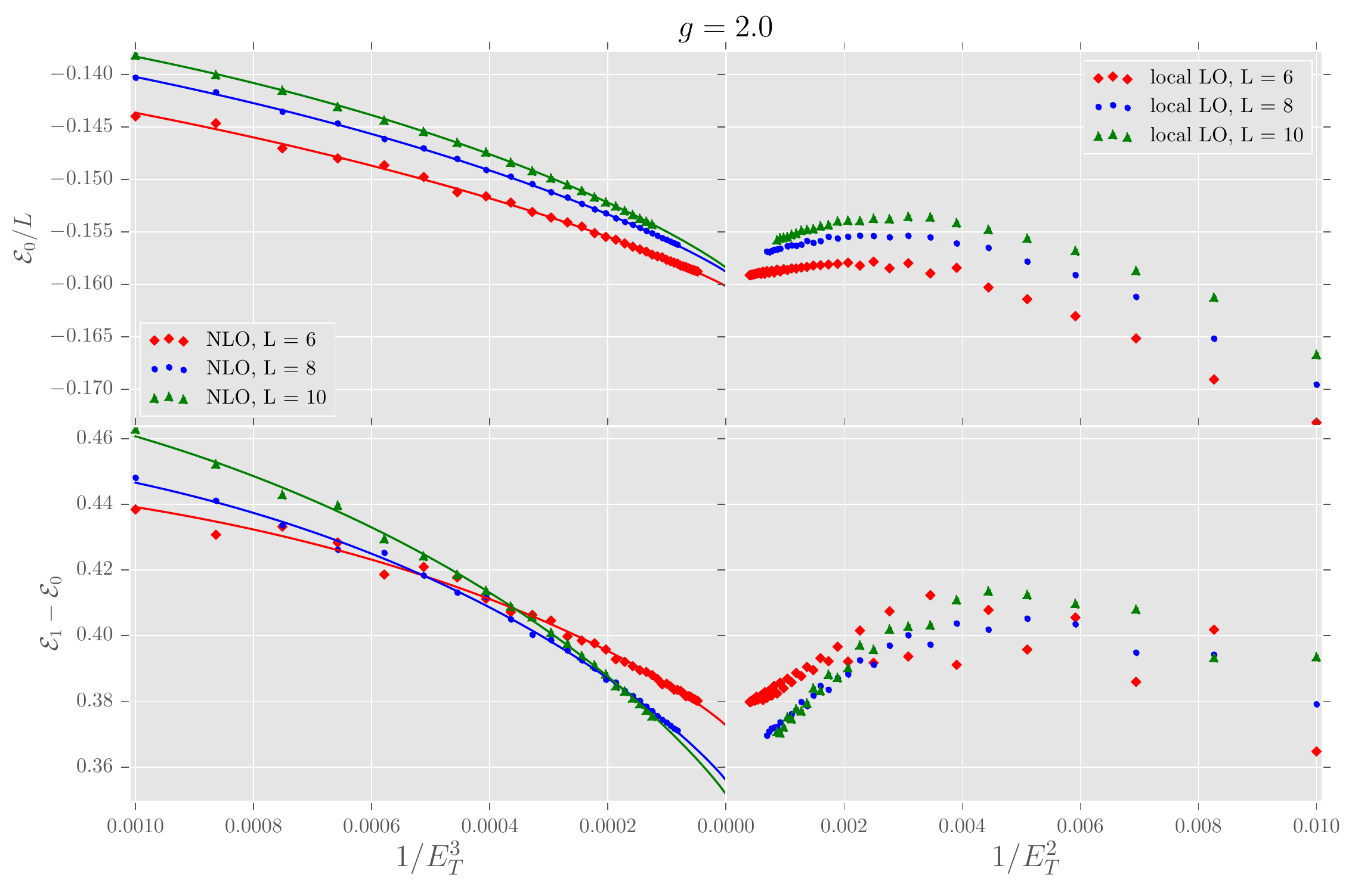}
\caption{Convergence rate of NLO-HT vs local LO renormalized HT. See the text.}
\label{fig:specExtrapolated1}
\end{center}
\end{figure}

Fig.~\ref{fig:specExtrapolated1} compares the rate of convergence of these two methods.
 In this figure the vacuum energy density $\cE_0/L$ and mass $\cE_1-\cE_0$ are shown for $L=6, 8, 10$ and $g=1$ and $g=2$. {These plots are consistent with the expectation that both methods converge to the same} 
asymptotic values as $E_T\to\infty$. Notice that the local data are plotted versus $1/E_T^2$, while the NLO-HT data versus $1/E_T^3$. At asymptotically large $E_T$, both methods appear to have linear convergence with respect to these two variables. Notice that for the smaller values of $L$ we could push the cutoff higher than for $L=10$, due to larger gaps in the free spectrum.

Naively, we may have expected faster convergence with the cutoff: $1/E_T^3$ for the local and $1/E_T^4$ for the NLO-HT. For example, the coefficients of the local correction terms given in \reef{eq:kappamu} behave as $1/E_T^2$ times logarithms. If these coefficients were to correct the $1/E_T^2$ behavior fully, we would have remained with an error decreasing one power of $E_T$ faster. Apparently this does not happen. Similarly, in the NLO-HT case, the largest local coefficient at the cubic order decreases as the cubic power of the cutoff, see Table \ref{tabasym}, and again this does not seem sufficient to fully correct the $1/E_T^3$ behavior of the spectrum. 
While we don't understand why the naive expectations concerning the convergence 
rate fail,\footnote{A possible reason for the NLO-HT might have to do with
the local approximation of $\Delta H_3$, see appendix \ref{convEL}.} 
it remains true that the observed convergence for NLO-HT is much 
faster than for the local (which in turn is much faster than for the raw HT).

The local data in Figs.~\ref{fig:specvsET1} and \ref{fig:specExtrapolated1} show significant fluctuations on top of the $1/E_T^2$ approach, especially pronounced for the mass. The origin of these fluctuations lies in the discreteness of the spectrum. For a continuously increasing $E_T$, the truncated Hilbert space changes discontinuously when the high-energy states fall below the cutoff. At the same time, the local correction term
$\Delta H^{\rm loc}_2(\cE_*)$ varies continuously with the cutoff, and so is unable to compensate the effects of discreteness.\footnote{It should be pointed out that Ref.~\cite{Bajnok:2015bgw} was able to fit the raw HT data by a fitting function inspired by the $E_T$ dependence theoretically predicted in \cite{Lorenzo1}. They used a slightly different definition of Hilbert space cutoff and fitted only a subsequence of cutoff values, which was reducing the fluctuations around a smooth fit.}

On the other hand, the correction term $\Delta \tilde H$ in the NLO-HT method adjusts itself discontinuously with the cutoff, because the sum over states just above the cutoff if performed exactly and not in the local approximation. For this reason the NLO-HT provides a much smoother dependence on $E_T$, as Figs.~\ref{fig:specvsET1} and \ref{fig:specExtrapolated1} demonstrate. This makes the NLO-HT data well amenable to a fit.
We tried various fitting procedures, and the one which seemed to worked best is to fit the NLO-HT points by a polynomial in $1/E_T$ of the form 
\begin{equation}
F(E_T) = \alpha + \beta/E_T^3 + \gamma/E_T^4\,.
\label{eq:etfit}
\end{equation}
From these fits we extract predictions for the eigenvalues at $E_T=\infty$, with error estimates, which will be used in the subsequent sections. For more details on the fitting procedure see appendix \ref{sec:fitdetails}.

The reader may notice that some points in the left panels of figure 
\ref{fig:specExtrapolated1} {violate monotonicity in $E_T$ by a small amount,
which is in apparent contradiction} with was what stated earlier about the
variational nature of the NLO-HT procedure. {These fluctuations are numerical artifacts having negligible
impact on the accuracy of the method. Their presence is explained by the following two reasons.} 
First, as explained in section \ref{PI}, the ultrahigh energy contributions to the matrices $\Delta H_2$ and $\Delta H_3$ in \eqref{eq:mainresult} have been computed in the local approximation, rather than
exactly. Second, in our prescription, we choose the parameter $\cE_*$ 
in $\Delta \tilde H$ to depend on $E_T$, as explained above, implying 
that increasing $E_T$ does not strictly correspond to enlarging the variational 
ansatz. 

\subsection{$L$ dependence}
\label{sec:Ldep}

In this section we study the dependence of the numerical eigenvalues on the volume $L$. 
Finite volume effects in quantum field theory are very well understood theoretically \cite{Luscher:1985dn,Luscher:1986pf, Klassen:1990ub}. This will allow us to perform interesting consistency checks of our results, and to devise a procedure for extracting infinite volume predictions.

Let us discuss first the theoretical expectations for the vacuum energy density and for the physical particle mass in finite volume. The vacuum energy at $L\gg 1/ m_{\phys}$ should behave as
\begin{eqnarray}
\calE_0(L)/L &= \Lambda -\frac{m_\phys}{\pi L} K_1(m_\phys L)
+ \frac{a}{4\sqrt{\pi}}  \left(\frac{m_\phys }{L^3}\right)^{1/2}e^{-2 m_\phys L} +\ldots 
\qquad (L\gg 1/m_\phys)\,,
\label{eq:largeL}
\end{eqnarray}
where $\Lambda$ is the infinite volume vacuum energy density (the cosmological constant) and $m_\phys$ is the
physical mass of the lightest particle. This formula is valid in any massive quantum field theory in 1+1 dimensions in absence of bound states (i.e.~particles with mass below $2 m_{\phys}$). See the discussion in \cite{Lorenzo1} after Eq.~(4.4), as well as \cite{Klassen:1990dx}, Eq.~(90) and later. Free bosons/fermions have $a=\pm 1$. For interacting theories we expect $a=O(1)$. This is satisfied by the fits below.

The physical mass in finite volume is defined as $\calE_1-\calE_0$ where $\calE_1$ is the lightest excited energy level at zero momentum. The large $L$ corrections to this quantity can be understood as contributions to the one-particle self-energy arising from virtual particles traveling around the cylinder representing (spatial circle)$\times$(time) \cite{Luscher:1985dn, Klassen:1990ub}. In a 1+1 dimensional theory with unbroken $\bZ_2$ symmetry they can be expressed as:\footnote{The role of the $\bZ_2$ symmetry is to forbid the cubic coupling. With a cubic coupling there would be an extra leading term in the r.h.s.~scaling as $\exp(-\sqrt{3/4}\,m_{\phys}L)$ \cite{Luscher:1985dn,Klassen:1990ub}. The given value of the exponent $\sigma$ is for a generic $1+1$ dimensional QFT with $\bZ_2$ symmetry. 
Generic theories in higher dimensions and/or without $\bZ_2$ symmetry will have smaller $\sigma$ (see \cite{Klassen:1990ub}), while specific theories with restricted interactions may have larger $\sigma$. E.g.~the critical 2d Ising model perturbed by the temperature perturbation has $\sigma=3$.}
\begin{eqnarray}
\label{eq:massfinL}
\cE_1(L) -  \cE_0(L) &=& m_{\phys} + \Delta m(L) + O(e^{- \sigma\, m_{\phys}L}) \,,\qquad \sigma = \sqrt{3}  \,,
\\ \Delta m(L) &=& -\frac{1}{8 \pi m_\phys} \int d\theta \, e^{- m_{\phys} L \,
\cosh \theta} 
F(\theta + i \pi/2)\,,
\\ F(\theta) &=& -4 i m_\phys^2 \sinh(\theta) \left(S(\theta) -1 \right)\,,
\end{eqnarray}
where $S(\theta)$ is the S-matrix for $2 \to 2$ scattering, with 
$\theta$ the rapidity difference. The third term in \eqref{eq:massfinL} is given by contributions in which
virtual particles travel around the cylinder multiple times. 

While the S-matrix can be measured in the HT approach by studying the $L$ dependence of two particle states \cite{Yurov:1991my}, this will not be done in this work. Instead, we will parametrize our ignorance of the S-matrix replacing $S(\theta + i \pi/2)$ 
with a Taylor series expansion around $\theta=0$.
This is reasonable because the integral in $\Delta m$ is dominated by small $\theta$. We obtain:
\begin{eqnarray}
\Delta m(L)/m_{\phys} \approx b K_1(m_\phys L) + \frac{c}{(m_\phys L)^{3/2}} e^{-{L m_{\phys}}}\,.
\label{eq:Lfit}
\end{eqnarray}
The Bessel function here would be the exact answer for a constant
$S(\theta)$, while the second term comes from $\theta^2$ in $S(\theta + i \pi/2)$ 
doing the integral via the steepest descent (the linear term vanishes in the integral). 
Further corrections are suppressed by additional powers of $1/(m_{\phys} L)$.

\begin{figure}[tbp]
\begin{center}
\includegraphics[scale=.47]{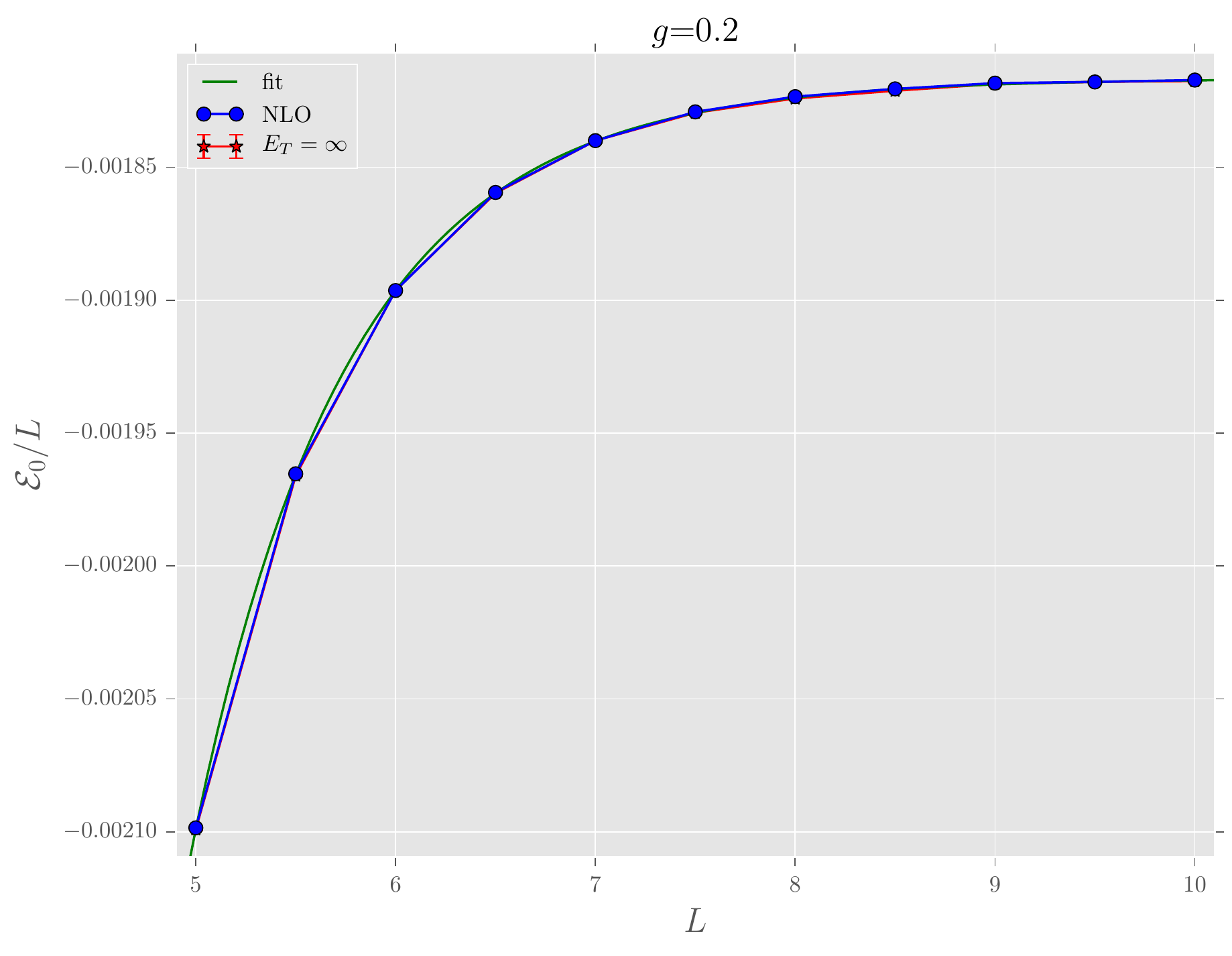} 
\includegraphics[scale=.47]{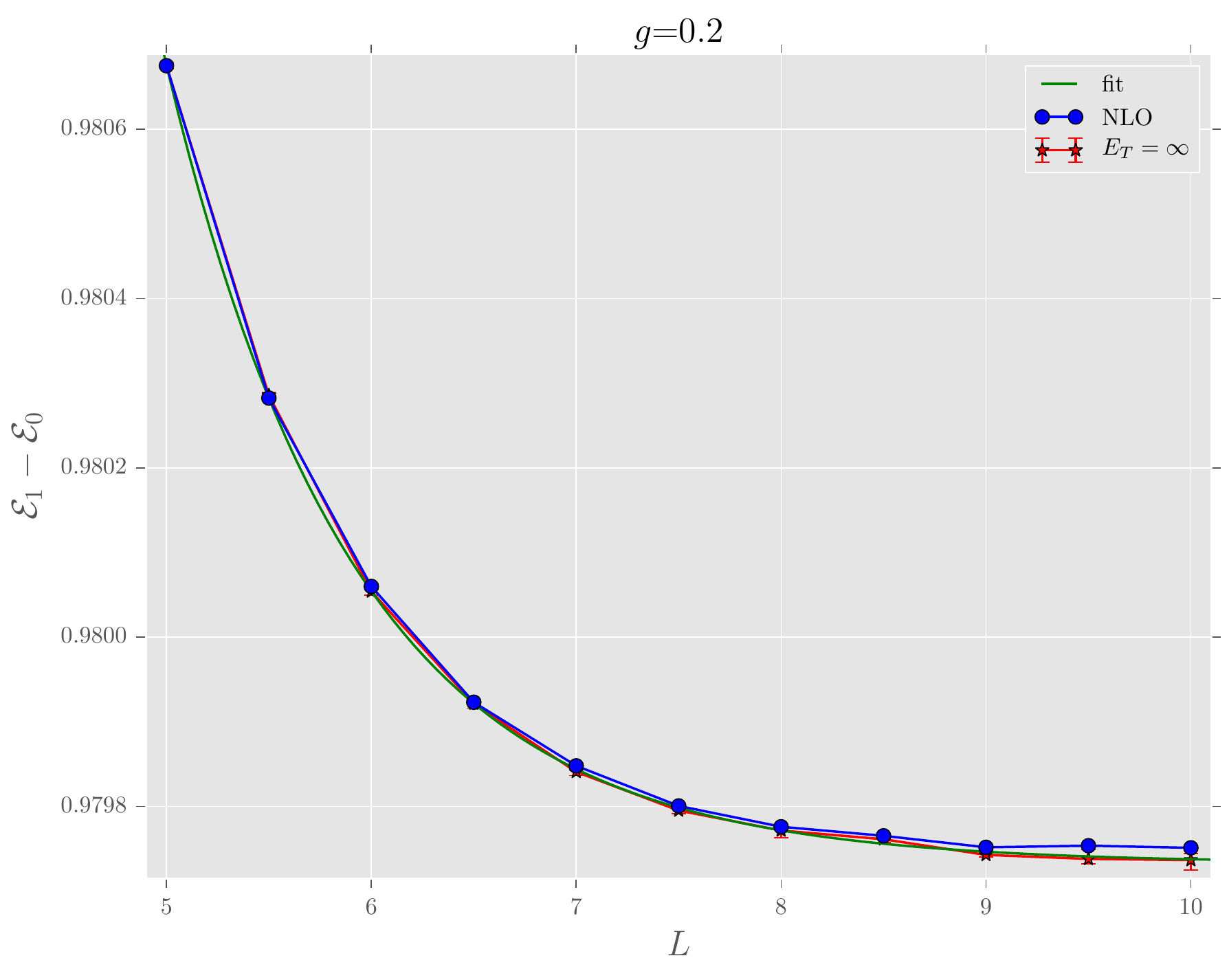}
\includegraphics[scale=.47]{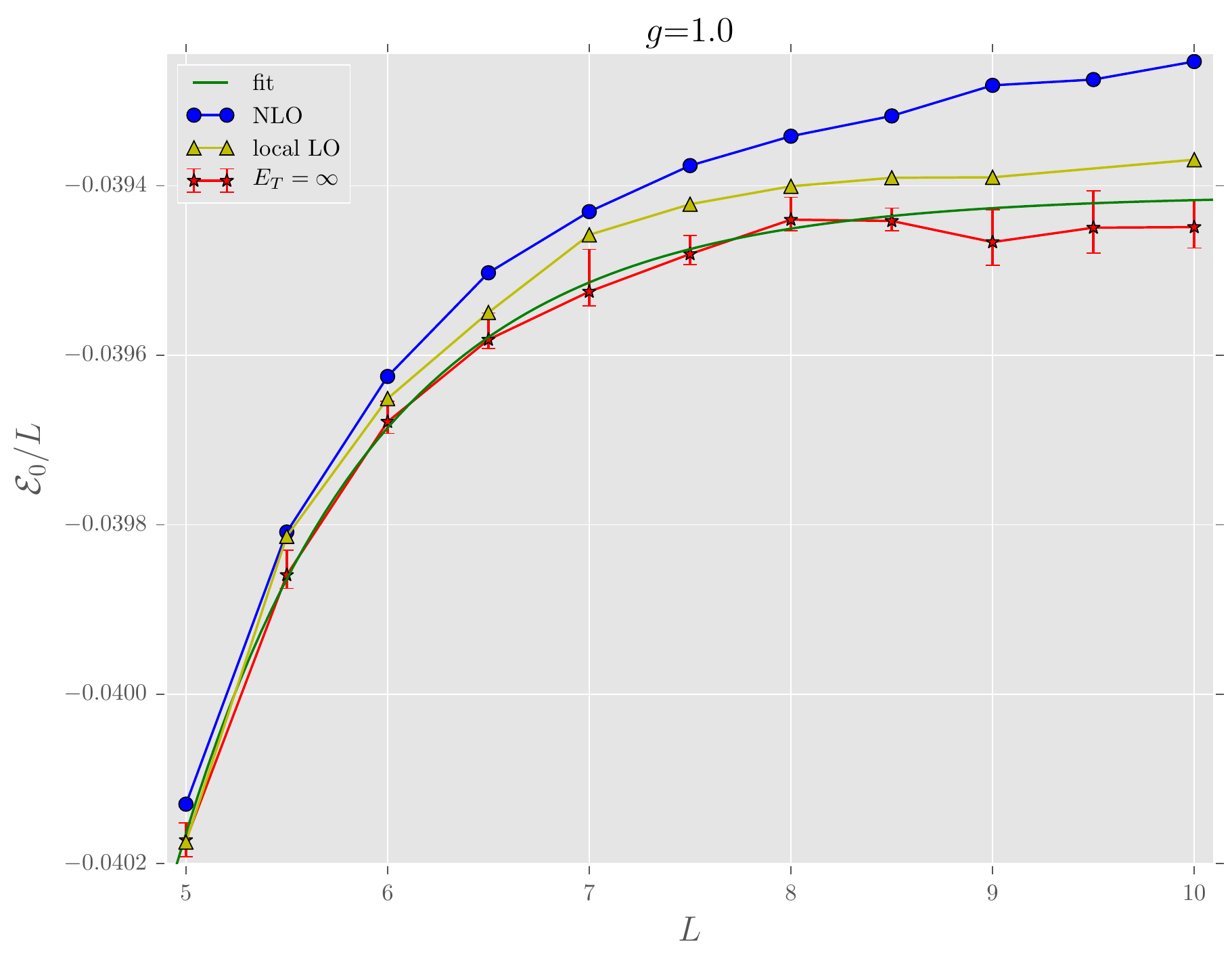} 
\includegraphics[scale=.47]{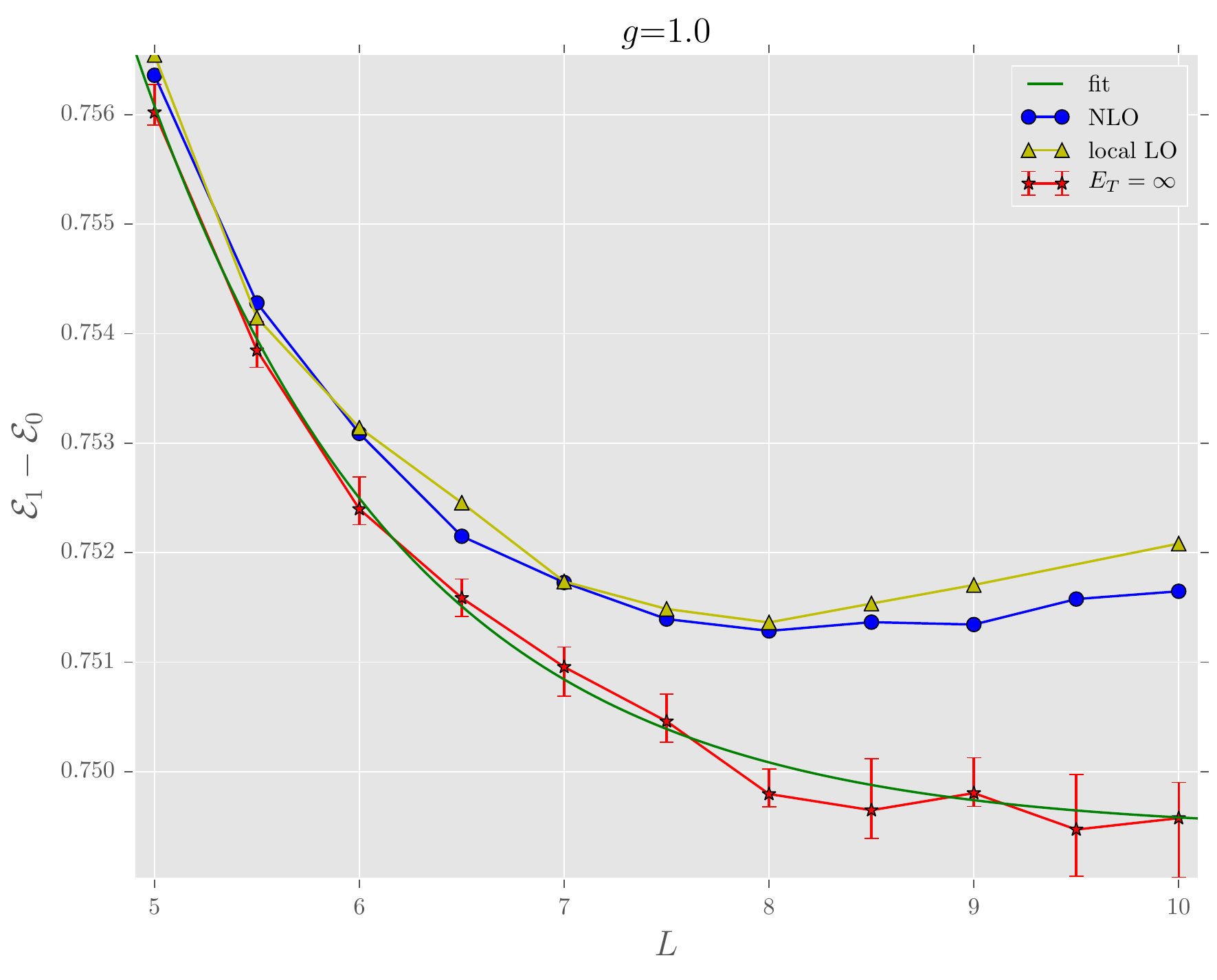}
\includegraphics[scale=.47]{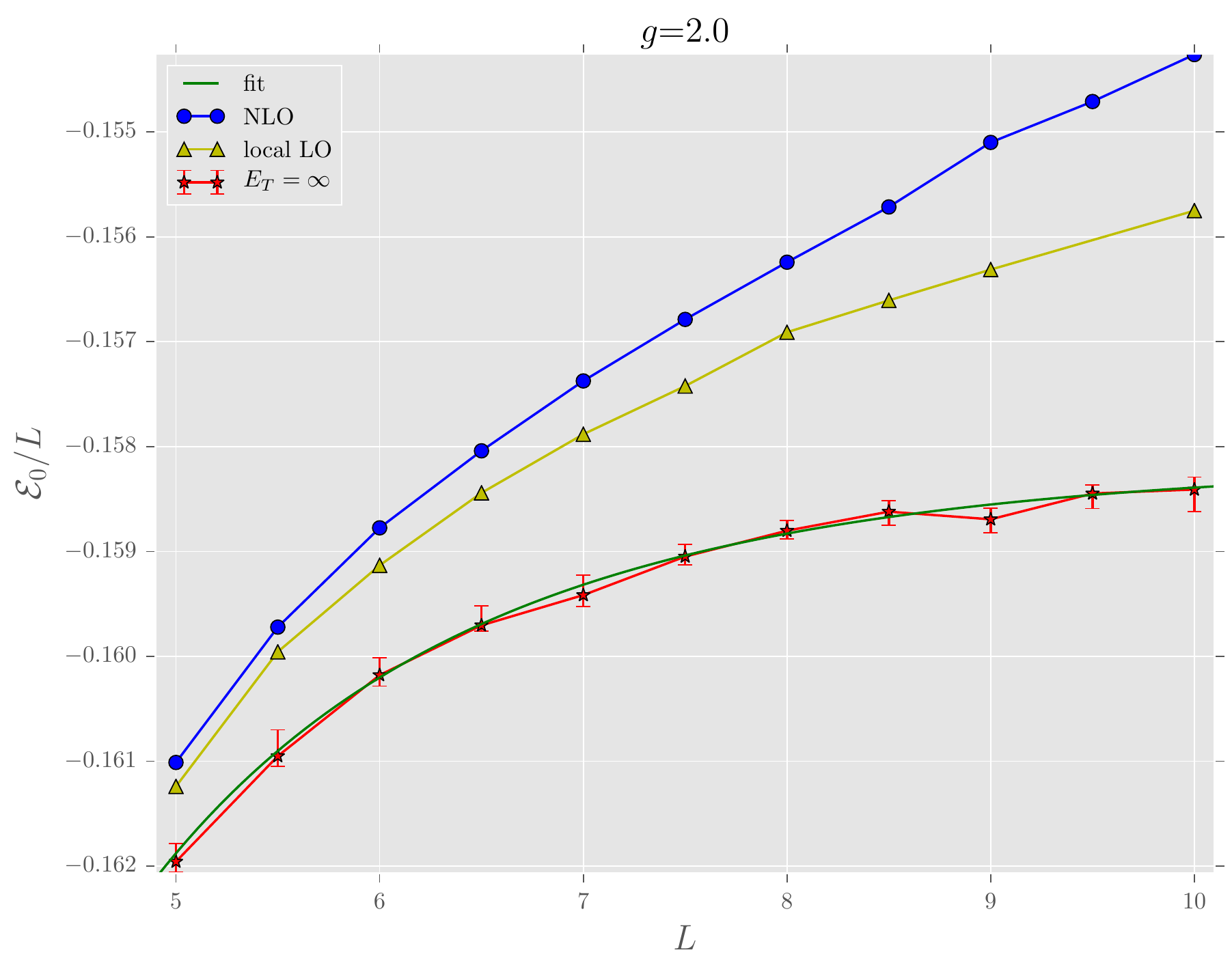} 
\includegraphics[scale=.47]{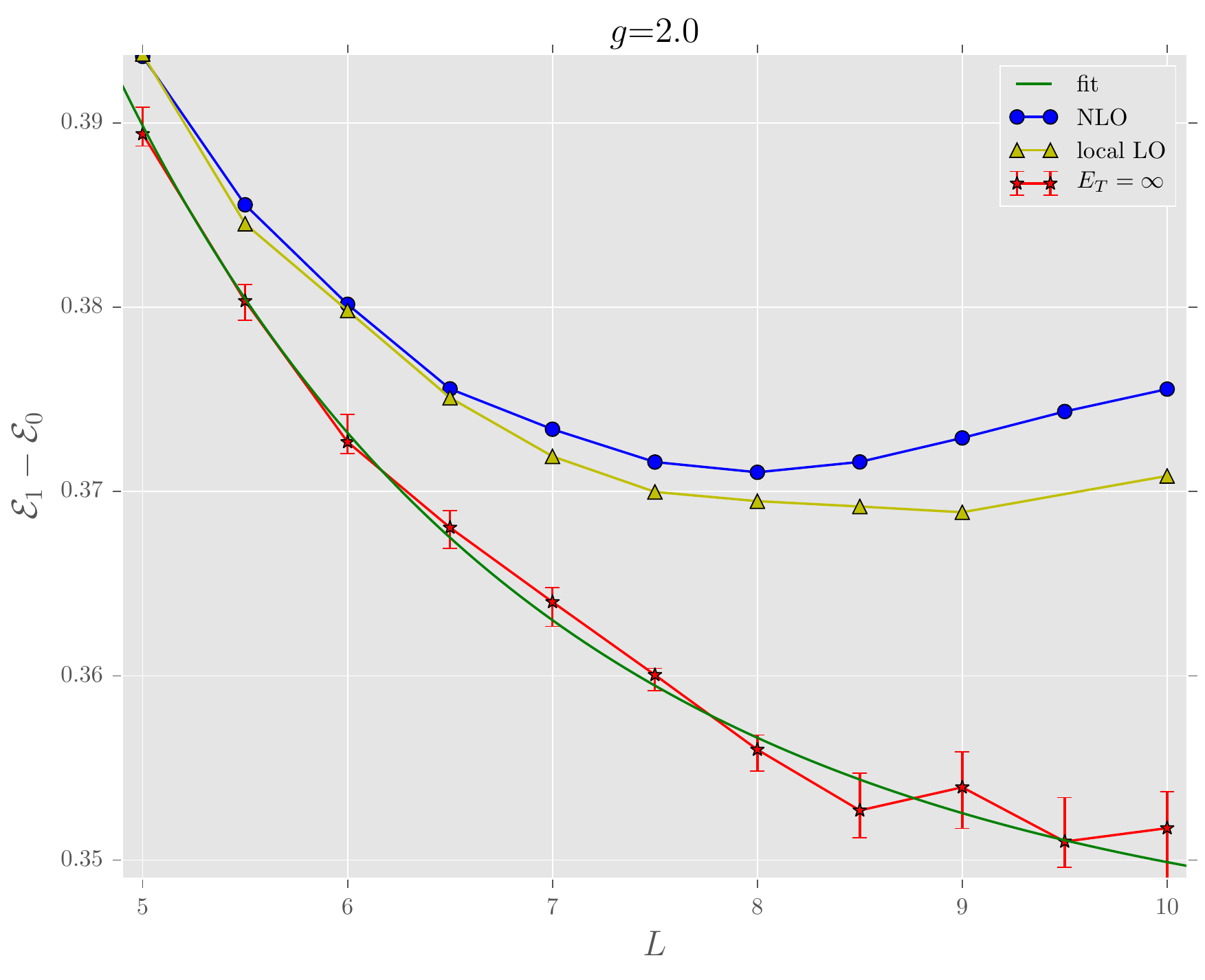}
\caption{The vacuum energy density $\calE_0/L$ and the physical mass $\calE_1-\calE_0$ as functions of $L$ for three representative values of $g$.}
\label{fig:specvsL0}
\end{center}
\end{figure}

In Fig.~\ref{fig:specvsL0} we present the numerical data: the vacuum energy density $\cE_0/L$ and the
physical mass $\cE_1 - \cE_0$ as functions of $L$ for three values of the coupling $g=0.2,1,2$. 
We include the NLO-HT data points at the highest $E_T$ we could reach for the given $L$ (blue), the NLO-HT data fit-extrapolated to $E_T=\infty$ as discussed in the previous section (red error bars), and the local data at its highest $E_T$ {(yellow)}.\footnote{We don't show local data for $g=0.2$ because they are very similar to NLO-HT for this small coupling.} 

Let us interpret this data theoretically, starting with with weak coupling $g=0.2$ which lies at the boundary of the region where fixed order perturbation theory ceases to be reliable \cite{Lorenzo1}. We fit the $E_T=\infty$ data for the physical mass using Eq.~\reef{eq:massfinL} where we neglect the third term and approximate $\Delta m$ by Eq.~\reef{eq:Lfit}.
The fit has three parameters: $m_\phys$, $b$, $c$. The fit works well in the whole range of $L$ and allows us to extract the value of $m_\phys$ reported in Table \ref{table:res0}. The uncertainty on $m_{\phys}$ was determined by fitting the upper and lower ends of the error bars.

We next fit the $E_T=\infty$ data for the vacuum energy using Eq.~\reef{eq:largeL} with $\Lambda$, $m_\phys$, $a$ as fit parameters. Including the error term $\propto a$ is not very important to achieve a good fit for this low value of $g$, but it's important for $g \gtrsim 2$ considered below. We checked that a very good fit can be obtained with $m_\phys$ in the range determined from $\calE_1 - \calE_0$. 
The final determination of $\Lambda$ reported in Table \ref{table:res0} is obtained using a constrained fit\footnote{We use the Trust Region Reflective 
algorithm for the least square optimization with bounds (calling {\tt curve\_fit()}
with the {\tt method$=$"trf"} argument in {\tt python}).} restricting $m_\phys$ to that range. Notice that it is crucial for this test not to neglect the corrections $E_0(L)$, $z(L)$ in the Hamiltonian \reef{eq:Hdef} whose decrease rate $e^{-m L}$ is close to the $e^{-m_\phys L}$ effects we are trying to observe.

\begin{table}[h]
\centering
\begin{tabular}{L{.7cm}  L{2.4cm}  L{2.7cm} }
$g$ & $m_\text{ph}$ & $\Lambda$ \\
\midrule
0.2 & 0.979733(5) & $-0.0018166(5)$\\
1  & 0.7494(2) &  $-0.03941(2)$\\
2 & 0.345(2) & $-0.1581(1)$\\
\end{tabular}
\caption{The values of $m_\phys$ and $\Lambda$ extracted from the NLO-HT data in Fig.~\ref{fig:specvsL0}.}
\label{table:res0}
\end{table}

{Passing to $g=1,2$, for these stronger couplings there is much more difference between the three curves.} The NLO-HT data at the maximal attainable cutoff do not show dependence on $L$ compatible with theoretical expectations. However, the same data extrapolated to $E_T=\infty$ can be {fitted} very well. We use the same fitting procedures as for $g=0.2$.
The fits are good and the physical mass from the two determinations agrees within errors. See Table \ref{table:res0} for the extracted $m_\phys$ and $\Lambda$.

Let us comment on the local data in Fig.~\ref{fig:specvsL0}. For $g=1,2$ they are unsuitable to perform the fit, just as the non-extrapolated NLO-HT data.\footnote{The $g=1$ vacuum energy data could perhaps be fitted. Notice that the fluctuations in this data are much smaller than in Fig. 7 (left) of \cite{Lorenzo1}, because of higher $E_T$ cutoff.} While the local LO data can be pushed to a much higher $E_T$, it is more difficult to extrapolate them to $E_T=\infty$ than the NLO-HT due to pronounced fluctuations within the asymptotic $1/E_T^2$ convergence rate. We tried extrapolating the local data and obtained results largely consistent with NLO-HT but with
larger error bars.
Also, we remark that in higher dimensions, where the Hilbert space
grows more quickly with the cutoff, and the convergence is slower,
we expect the local LO approximation to perform even worse than here with respect 
to the NLO-HT approach, as the cutoff cannot be pushed as high. 

In all the above fits the third term in \reef{eq:massfinL} was neglected. We checked that this assumption gives a reasonable fit up to $g=2.6$. As $g$ is increased further it gets close to the critical coupling $g _c\approx 2.8$.
On the one hand, fitting finite volume data in this region becomes more difficult as the physical mass approaches zero and the neglect of subleading terms suppressed by higher powers of $e^{-m_\phys L}$ is no longer justified. On the other hand, we know that at $g=g_c$ the $\phi^4$ theory should flow to the critical Ising model. So, for $g$ near $g_c$, the flow must lead to the Ising field theory (IFT)---the critical Ising perturbed by the $\eps$ operator, up to irrelevant corrections which go to zero as $g\to g_c$ and which we will neglect in the subsequent discussion.\footnote{For many although not all purposes the IFT can be thought of as the theory of free massive Majorana fermions. Our fits prefer negative values for $a$ in Eq.~\reef{eq:largeL} close to $g=g_c$, as appropriate for fermionic excitations.}
The IFT is integrable and its finite volume partition function is known exactly. In particular, the functional dependence of $\calE_1(L)-\calE_0(L)$ and $\calE_0(L)-\calE_0(\infty)$ on $m_{\phys} L$ is known. One could use this information to improve our fitting procedure for $g$ near $g_c$. For instance, the coefficients $b$ and $c$ in \eqref{eq:Lfit} in that region would be fixed to the values to $2/\pi $ and $0$ \cite{Klassen:1992eq}, rather than being fitted from the data. This reasoning also explains why neglecting the third term in
\eqref{eq:massfinL} works even for relatively small values of $m_{\phys}$, since in the IFT $\sigma=3$, above the generic value $\sqrt{3}$. Using the IFT predictions would lead to more accurate estimates of $\Lambda$ and $m_\phys$ for $g$ close to $g_c$. However, in the present work we will be content with our simplified analysis, 
not using explicitly this additional piece of information.

\subsection{$g$ dependence and the critical coupling}

In the previous sections we explained how NLO-HT data can be extrapolated to $E_T=\infty$ and then to $L=\infty$.
We will now use these procedures to study the spectrum dependence on $g$.

In Fig.~\ref{fig:speccrit} we show the NLO-HT data for the vacuum energy density and 
the physical mass for $g\in[0,3]$ in steps of 0.2. Green error bars refer to $L=10$ NLO-HT data extrapolated to $E_T=\infty$, while red error bars are the infinite volume estimates (we only perform the latter for 
$g \le 2.6$, i.e.~not too close to the critical point).

\begin{figure}[t]
\begin{center}
\includegraphics[scale=.38]{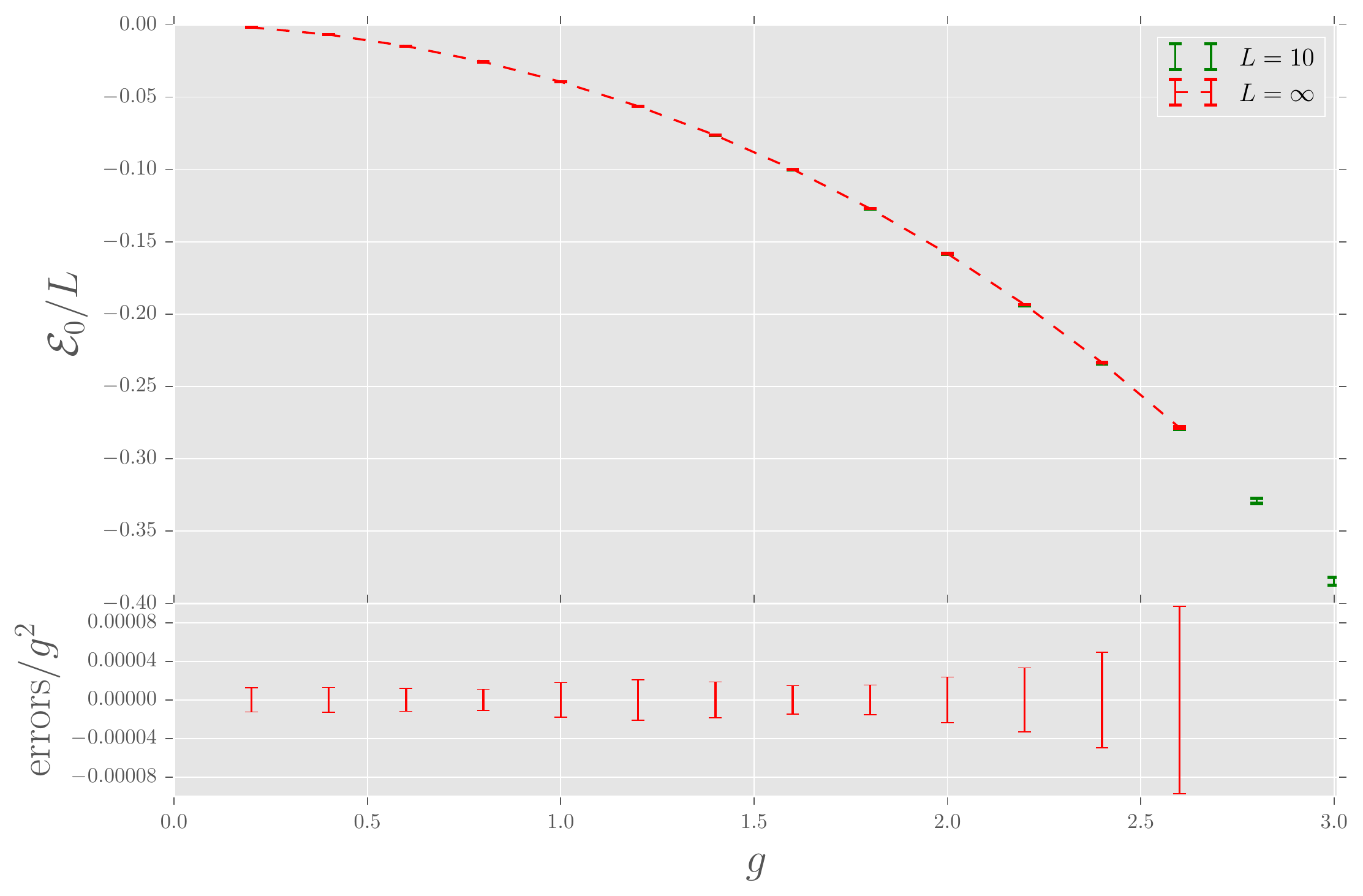} 
\includegraphics[scale=.38]{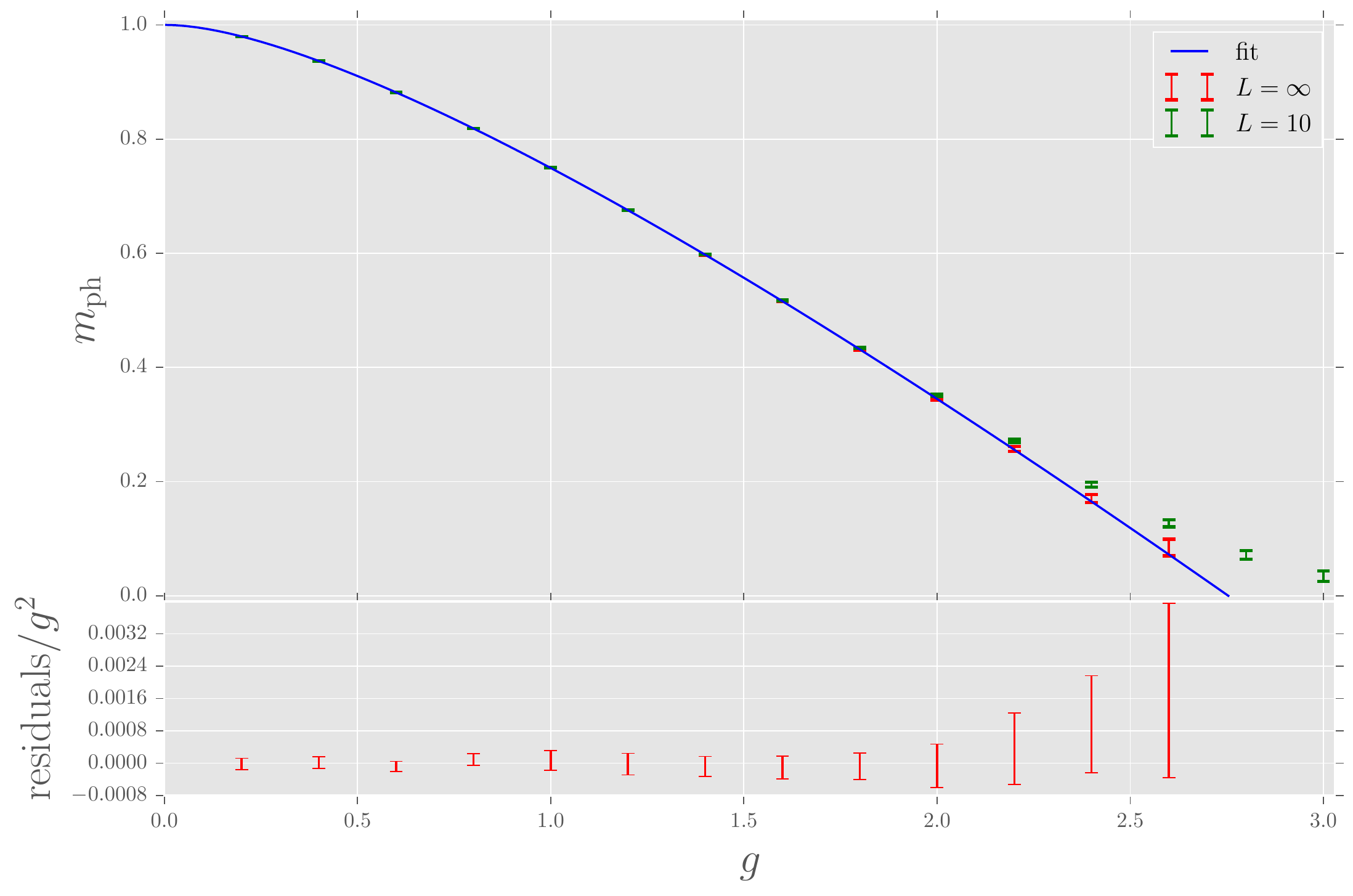} 
\caption{\emph{Left:} the vacuum energy density as a function of $g$. The dashed line joining the points is not a fit; it is included to guide the eye. We also show errors divided by $g^2$. \emph{Right:} the physical mass as a function of $g$. The line is a fit described in the text. We also show fit residuals divided by $g^2$.}
\label{fig:specvsG}
\end{center}
\end{figure}

These plots should be compared to Fig.~5 in \cite{Lorenzo1}, taking into account that in those figures we did not attempt to extrapolate to infinite $E_T$ and $L$ and did not provide error estimates. The current results are clearly superior in that these sources of systematic error are properly taken into account.

There is not much structure in the vacuum energy plot except that it is a monotonically decreasing function of $g$. The physical mass plot is more interesting. We see by eye that the mass gap vanishes somewhere close to $g\approx 2.8$. This is in accord with the previous theoretical \cite{Chang:1976ek} and numerical \cite{Schaich:2009jk,Lorenzo1,Milsted:2013rxa,Bosetti:2015lsa} studies, which found that our theory undergoes a second order phase transition at a critical value
of the coupling. To give a more accurate estimate of $g_c$, we perform a fit of the red data points
 in the range $g \in [0, 2.6]$. We use a rational function: 
\begin{equation}
\label{eq:ansatz}
f(g) = \frac{( 1 + g (\frac{1}{g_1} + \frac{1}{g_2} + \frac{1}{g_3} + \frac{1}{g_c}) 
+ a g^2) (1 - \frac{g}{g_c})^\nu}
{( 1 + \frac{g}{g_1}) (1 + \frac{g}{g_2}) (1 + \frac{g}{g_3}) } \,,
\end{equation}
with fit parameters $a$, $g_1$, $g_2$, $g_3$, $g_c$, and $\nu$. We demand that $g_1,g_2,g_3>0$ so that $m_{\rm fit}(g)$ has poles at the negative real axis. We see that $f(g_c)=0$ by construction. Performing the fit, we get our final estimate for the critical coupling, {reported in} Table \ref{table:gc}.

The $\nu$ parameter in the above fit is a critical exponent, and assuming the Ising model universality class for the phase transition, we expect $\nu = (2 - \Delta_\epsilon)^{-1} = 1$ using  $\Delta_\epsilon = 1$, the dimension of the most relevant non-trivial $\bZ_2$-even operator of the critical Ising model \cite{Lorenzo1}. In our fit we fixed $\nu$ to this exact value. Relaxing this assumption gives the same prediction with somewhat larger error bars.

The rationale behind introducing the poles into the ansatz $f(g)$ is that they are supposed to
approximate the effect of a branch cut along the negative real axis, which the analytically continued function $m_{\phys} (g)$ may be expected to have. In fact it's impossible to get a good fit using a purely polynomial approximation. The number of poles is somewhat arbitrary. {Three poles as in \reef{eq:ansatz} gives a good fit, and we checked} that increasing the number of poles does not change the prediction for $g_c$ appreciably. 

The ansatz $f(g)=1+O(g^2)$ by construction. We checked that the $g^2$ and $g^3$ coefficients of our best fit are roughly consistent with the perturbation theory prediction (appendix B of \cite{Lorenzo1})
\beq
m_{\phys}(g)=1-1.5 g^2+2.86460(20) g^3+\ldots. 
\eeq
Using a slightly more complicated ansatz
\begin{equation}
\frac{( 1 + g (\frac{1}{g_1} + \frac{1}{g_2} + \frac{1}{g_3}+\frac{1}{g_4} + \frac{1}{g_c}) 
+ a g^2 +b g^3+ c g^4) (1 - \frac{g}{g_c})}
{( 1 + \frac{g}{g_1}) (1 + \frac{g}{g_2}) (1 + \frac{g}{g_3}) (1 + \frac{g}{g_4})} \, ,
\end{equation}
we could find a fit which agrees with perturbation theory precisely. {The $g_c$ estimate from such a fit comes out nearly identical with the one provided above.} This is not surprising because most of the constraining power of the fit relevant for determining $g_c$ comes from the region $1\lesssim g\lesssim 2$ where perturbation theory is anyway {not adequate}.

\begin{table}[h!]
\begin{center}
\begin{tabular}{L{2cm} L{2.5cm} L{5.7cm}}
Year, ref. & $g_c$  & Method  \\ \midrule
This work &  2.76(3) & NLO-HT \\ 
\hline
2015 \cite{Lorenzo1}  & 2.97(14)  & LO renormalized HT  \\ %\hline
2016 \cite{Bajnok:2015bgw} & 2.78(6) & raw HT\tablefootnote{The $\bZ_2$-broken phase of the theory was studied, using minisuperspace treatment for the zero mode (as in \cite{Lorenzo2}). Their estimate for the critical coupling has been translated to our convention using the Chang duality \cite{Chang:1976ek,Lorenzo1}.} \\
\hline
2009 \cite{Schaich:2009jk}  &  $2.70^{+0.025}_{-0.013}$  &Lattice Monte Carlo \\ %\hline
2013 \cite{Milsted:2013rxa}  &  2.766(5) & Uniform matrix product states  \\ %\hline
2015 \cite{Bosetti:2015lsa} &  $2.788(15)(8)$ &  Lattice Monte Carlo  \\ %\hline
2015 \cite{Pelissetto:2015yha} & 2.75(1)  & Resummed perturbation theory \\ %\hline
\end{tabular}
\caption{Estimates of $g_c$ from various techniques.}
\label{table:gc}
\end{center}
\end{table}

 In Table \ref{table:gc}, we compare our estimate for $g_c$ with other recent results in the literature.
Our original HT estimate in \cite{Lorenzo1} was a bit high, evidently because the effects of the extrapolating to $E_T\to\infty$, $L\to\infty$ were not taken into account. 
It's reassuring that our current estimate agrees well with the HT estimate from \cite{Bajnok:2015bgw}, obtained approaching the critical point from the other side, i.e.~from within the $\bZ_2$-broken phase. 

The last four results in the table are based on studies of latticized $\phi^4$ models, such as lattice Monte Carlo simulations of the euclidean model \cite{Schaich:2009jk,Bosetti:2015lsa} or matrix product states approach to the latticized Hamiltonian formulation \cite{Milsted:2013rxa}. Lattice considerations also enter \cite{Pelissetto:2015yha} which determines the critical coupling via resummed perturbation theory. It should be pointed out that matching to the continuum limit is particularly subtle in the two dimensional lattice $\phi^4$ theory, because of the presence of an infinite number of relevant and marginal operators \cite{Lorenzo1}. The above lattice studies do not perform careful matching, and use the simplest possible discretization. The agreement with HT is good, and so this simplest discretization seems to have the right continuum limit. It would be interesting to understand why this is so.

Recently, the two dimensional $\phi^4$ theory was also studied using the light front quantization \cite{Chabysheva:2015ynr,Burkardt:2016ffk,Anand:2017yij}\footnote{See 
also \cite{Katz:2016hxp} for an application to the three dimensional $\phi^4$ 
theory at large $N$.}
using a wavefunction basis superior to the old discrete light cone quantization work \cite{Harindranath:1988zt}. The light front quantization scheme is different from the equal-time quantization scheme used here. This difference is apparent already at the perturbative level, since certain diagrams contributing to vacuum energy and mass renormalization are absent in the light front scheme. The vacuum energy cannot be compared between the two schemes as it is set identically zero in the light front scheme. On the other hand, it is believed that the physical mass can be compared between the two schemes, with an appropriate non-perturbative coupling redefinition.\footnote{This is believed to be true at least in the $\bZ_2$-invariant phase. Accessing the $\bZ_2$-broken phase on the light front is a much harder problem, and we are not aware of any concrete computations.} A method to perform such a coupling redefinition was recently proposed in \cite{Burkardt:2016ffk,Chabysheva:2016wvl} (see \cite{Burkardt:1992sz} for previous related work). We refer to those works for the comparison of the critical coupling estimates obtained using the two methods.

\begin{figure}[t]
\begin{center}
\includegraphics[scale=.38]{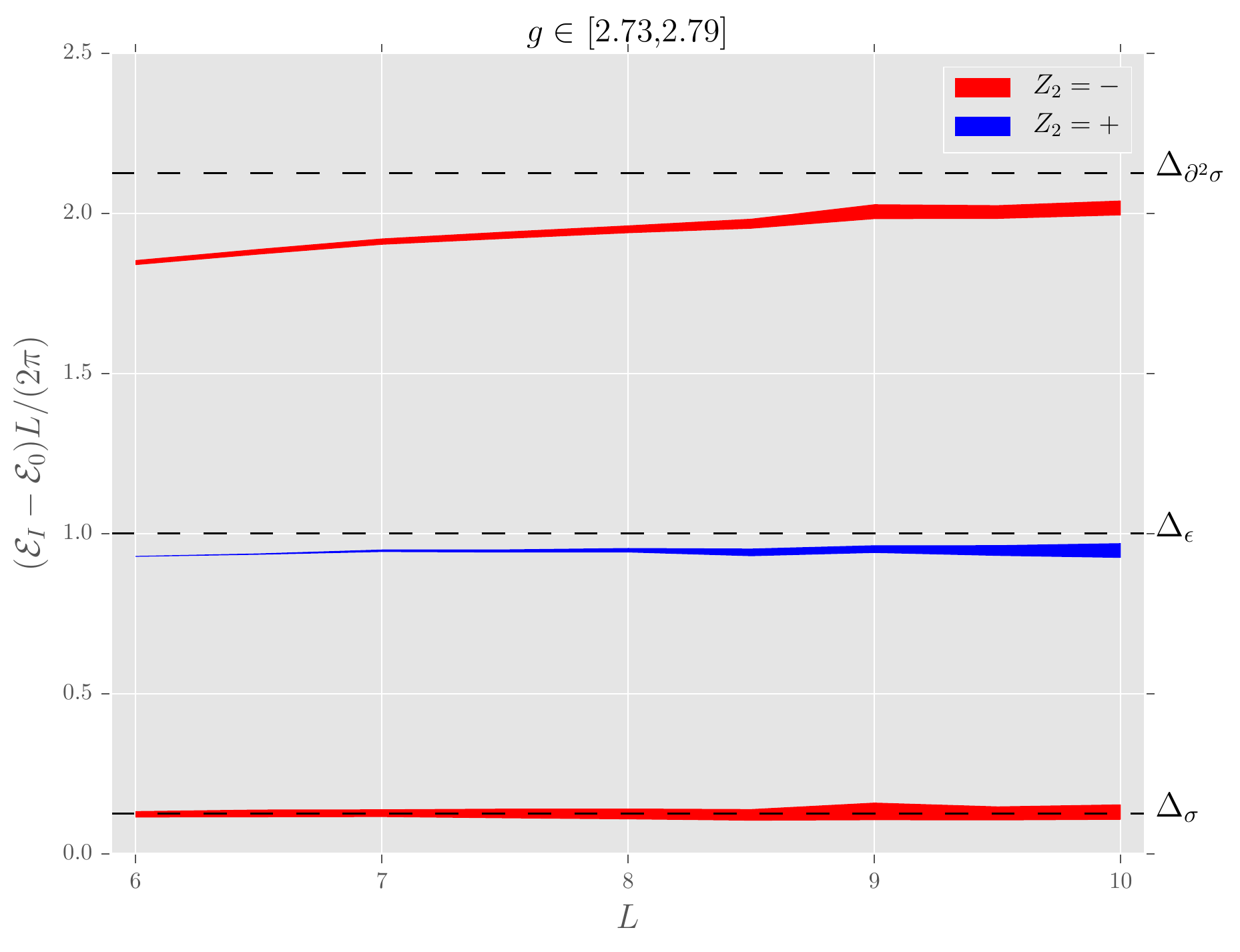} 
\caption{Comparison of energy levels at $g=g_c$ with CFT predictions.}
\label{fig:speccrit}
\end{center}
\end{figure}

We conclude this section with a rough check that our method reproduces the physics of the phase transition 
at criticality. Conformal field theory predicts that at $g=g_c$ the energy levels should vary with $L$ as
\begin{equation}
\cE_I(L) - \cE_0(L) \sim 2\pi \Delta_I/L\,\qquad(L\gg1)\,,
\label{eq:cft}
\end{equation}
where $\Delta_I$ are operator dimensions in the critical Ising model. In Fig.~\ref{fig:speccrit} we test this relation for the first three energy levels above the vacuum, which should correspond to the operators with dimensions
$\Delta_\sigma = 1/8$, $\Delta_\epsilon = 1$, $\Delta_{\partial^2 \sigma} = 2 + 1/8$.
The bands correspond to varying $g$ in the range 2.76(3).
We see reasonable agreement for $\sigma$ and $\epsilon$, 
while it is possible that the agreement for $\partial^2 \sigma$ 
will be reached at higher values of $L$. This figure can be compared to Fig.~6 in \cite{Lorenzo1} and Figs.~22, 23 of \cite{Bajnok:2015bgw}, which show similar behavior.

\section{Conclusions and outlook}

\label{sec:conclusions}

In this work we have addressed several conceptual and practical issues regarding the renormalization improvement of the Hamiltonian Truncation (HT) technique. This led us to propose the NLO-HT, a variant of the HT using a variational correction term to the Hamiltonian, of next-to-leading-order accuracy in the interaction. 
The NLO-HT method puts on a firmer theoretical footing the renormalization theory 
in the context of Hamiltonian Truncation, and at the same time rigorously
improves the numerics with respect to previous work.

In the second part of the paper, we tested the NLO-HT in the context of the two-dimensional $\phi^4$ theory. We also benchmarked
the NLO-HT against the simpler existing versions of the HT---the raw truncation and the local leading-order renormalization. Compared to these, the NLO-HT results exhibit smoother and more rapidly convergent dependence on the Hilbert space cutoff $E_T$. Therefore, they lend themselves to more accurate extrapolations to $E_T = \infty$ and ultimately provide more accurate determinations of the true eigenvalues.

In this work, we focused on the massive region where the $\bZ_2$ symmetry is preserved, and on the critical region, where the mass gap vanishes. We computed 
the mass gap and vacuum energy density over the whole range of couplings, as well as the critical exponents at the critical point.
In the future it will be interesting to use NLO-HT to also study the region beyond the phase transition, where the $\bZ_2$ symmetry breaks spontaneously. That region was previously studied in \cite{Lorenzo2,Bajnok:2015bgw} using the local LO renormalized and raw Hamiltonian Truncation.

The implementation of the NLO-HT method required a refinement of the local approximation of the counterterms, which formed the basis of the previously used local LO renormalization. We have discussed and addressed novel issues arising in the local approximation at the cubic level, such as the presence of bilocal operators. Following \cite{Elias-Miro:2015bqk},  we used the local approximation only to approximate the ``ultrahigh'' energy parts of the correction terms, while the moderately high parts were evaluated exactly. This required additional computational effort, but as a result all matrix elements of the correction terms were accurately taken into account.
At present, the evaluation of counterterms presents a computational bottleneck demanding significant time and memory resources. This step is the main limiting factor in the performance of the method. In this regard, we outlined 
several directions for future development. One promising idea was already mentioned in sections \ref{sec:idea}, others are scattered in the main text, see e.g.~note \ref{note:tails} and section \ref{sec:dh3-322}.
Other interesting questions for developing the method include:
\begin{itemize}
\item
Is it worth it/possible to enrich the variational ansatz to allow for an even 
more accurate reproduction of the would-be optimal tails \reef{eq:wouldbe}?
\item
Can we deal more efficiently with the states with high occupation numbers, 
which at present occupy a fraction of the Hilbert space disproportionally 
large compared to their total weight? See appendix \ref{sec:struct}.
\end{itemize}

However, while further improvements in the method are welcome, they are not strictly speaking necessary. The NLO-HT is already one of the most advanced implementations of Hamiltonian Truncation currently available. It would be great to see it applied in further HT studies of the $\phi^4$ theory or of other strongly coupled QFTs. We will be happy to share our code upon request. One $\phi^4$ application we are currently thinking about is to investigate the analytic structure of $m_{\rm ph}$ and $\Lambda$ for the complexified quartic coupling $g$. The Hamiltonian Truncation seems to be the only non-perturbative technique currently suitable for this task.

Finally, we believe that Hamiltonian Truncation is now in a much better shape 
to attack strongly coupled renormalization group (RG) flows in higher dimensions.
For instance, as the next step one could study models of the Landau-Ginzburg or 
Yukawa type in three dimensions, and their RG flow either to a gapped or to 
a conformal phase. Furthermore, one could 
apply the renormalization procedure described in this
work in the context of TCSA, in order to deform interacting fixed points directly.
For instance, it would be interesting to study the temperature and/or magnetic 
deformation in the 3D Ising model, in which the UV data (OPE coefficients and scaling  
dimensions) for the low-lying primary operators are known to high accuracy 
\cite{El-Showk:2014dwa,Kos:2016ysd,Simmons-Duffin:2016wlq}.

\section*{Acknowledgements} 

We thank Richard Brower, Ami Katz, Robert Konik, Iman Mahyaeh, Marco Serone, Gabor Tak\'acs, Giovanni Villadoro and Matthew Walters for the useful discussions. SR is supported by the National Centre of Competence in Research SwissMAP funded by the Swiss National Science Foundation, and by the Simons Foundation grant 488655 (Simons collaboration on the Non-perturbative bootstrap). 
The work of LV was supported by the Simons Foundation grant on the Nonperturbative 
Bootstrap and by the Swiss National Science Foundation under grant 200020-150060.
The computations were performed on the BU SCC and SISSA Ulysses clusters.

%%%%%%%%%%%%%%%%%%%%%%%%%%%%
%%%%%%%%%%                    %%%%%%%%%%%%
%%%%%%%%%%   Appendix  %%%%%%%%%%%%
%%%%%%%%%%                    %%%%%%%%%%%%
%%%%%%%%%%%%%%%%%%%%%%%%%%%%

\addtocontents{toc}{\protect\contentsline {chapter}{%
\vspace{0.1in}{\hspace{0.5cm}\bf Appendices:}\\
%\newline
\vspace{-0.1in}
}{}{}}

\appendix

\section{Structure of the interacting eigenstates}
\label{sec:struct}

Much of the motivation underlying the HT method is based on the idea of decoupling---that interacting eigenstates in finite volume are dominated by the low-energy non-interacting states. In this appendix we will show some plots demonstrating the validity of this idea, in the context of the $(\phi^4)_2$ theory (see also the related discussion in \cite{Konik-review}, Section VII.B). We will also discuss, and resolve, the apparent contradiction with the ``orthogonality catastrophe".

All plots in this appendix will correspond to $m=1$, $L=10$, and cutoff $E_T=20$. We will be showing data for the raw truncated Hamiltonian eigenstates -- as we are interested here in the qualitative features, it's not crucial to include renormalization corrections. 

\begin{figure}[tbp]
\begin{center}
\includegraphics[width=0.45\textwidth]{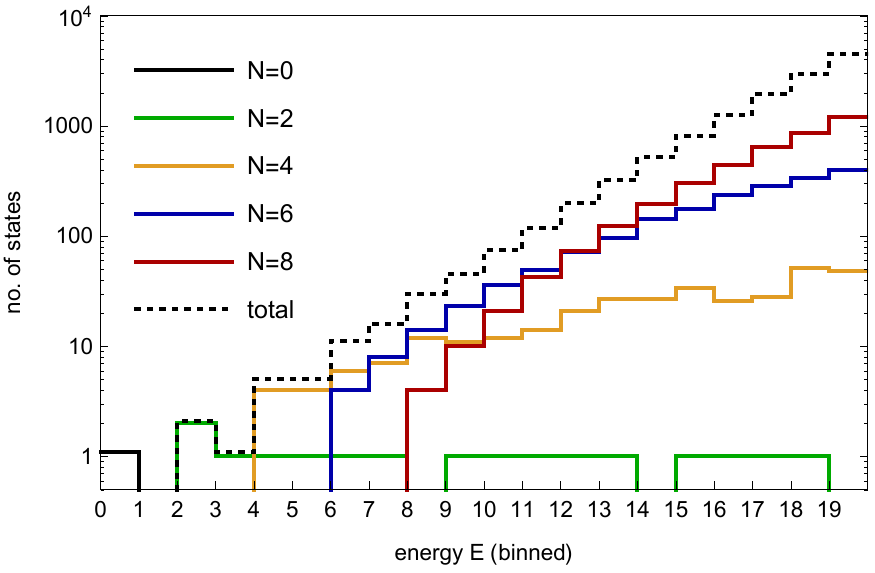}
\caption{The composition of the $\bZ_2$ even Hilbert space for $m=1$, $L=10$. For each occupation number sector (up to $N=8$), we show the number of states per unit interval of energy up to the cutoff $E_T=20$. The dashed line shows the total number of states in the same interval (all allowed occupation numbers).}
\label{fig:HilbertSize} 
\end{center}
\end{figure}

We start by showing the composition of the $\bZ_2$ even truncated Hilbert space subject to the constraints $P=0$, $\bP=0$. In Fig.~\ref{fig:HilbertSize} we plot the distribution of the number of states
per particle number (0,2,4,\ldots) and per interval $[E,E+1)$ of energy. As this plot illustrates, the total Hilbert space dimension (dashed line) grows exponentially with the cutoff. We expect that the leading exponential asymptotics will be the same as in the massless scalar boson CFT, $\sim \exp \sqrt{x}$, $x=(2\pi/3) LE_T$. (Fixing the prefactor would require, among other things, taking into account the zero momentum constraint.) We see that most states have rather high occupation numbers $N$. In Fig.~\ref{fig:HilbertSize}, there are a total of 12869 states, of which 1, 16, 332, 1890, 3931, 3801, 2063,\ldots,1 for $N=0$, 2, 4, 6, 8, 10, 12,\ldots,20 respectively, the maximum being at $N=8$.

Given this exponential haystack of states, are all of them equally important to represent the interacting eigenstates? It turns that the high energy states are less important than the low energy ones. Moreover, the states with high occupation numbers are the least important. Before showing the evidence, let's introduce some terminology. 
Let $|\calE\rangle$ be an interacting eigenstate of the truncated Hamiltonian, which has an expansion
\beq
|\calE\rangle = \sum c_n |n\rangle,
\eeq
where $|n\rangle$ runs over the basis of $\calH_l$ described in appendix \ref{sec:basis}. We will call $w_n=|c_n|^2$ the \emph{weight} of the given basis state inside $|\calE\rangle$. We assume $|\calE\rangle$ is unit normalized so the weights sum to one. The most important basis states are those which carry most weight. Which are those states?

We will now show a series of plots concerning the weight composition of the interacting vacuum (the lowest eigenstate in the $\bZ_2$ even 
sector).\footnote{Very similar conclusions are reached looking at any low eigenstate, $\bZ_2$ even or odd.}
 We will choose three representative values of the coupling $g=1,2,3$. These couplings are all strong, and $g\approx 3$ roughly corresponds to the end of the $\bZ_2$ invariant phase \cite{Lorenzo1}.
 
\begin{figure}[h!tbp]
\begin{center}
\includegraphics[width=0.49\textwidth]{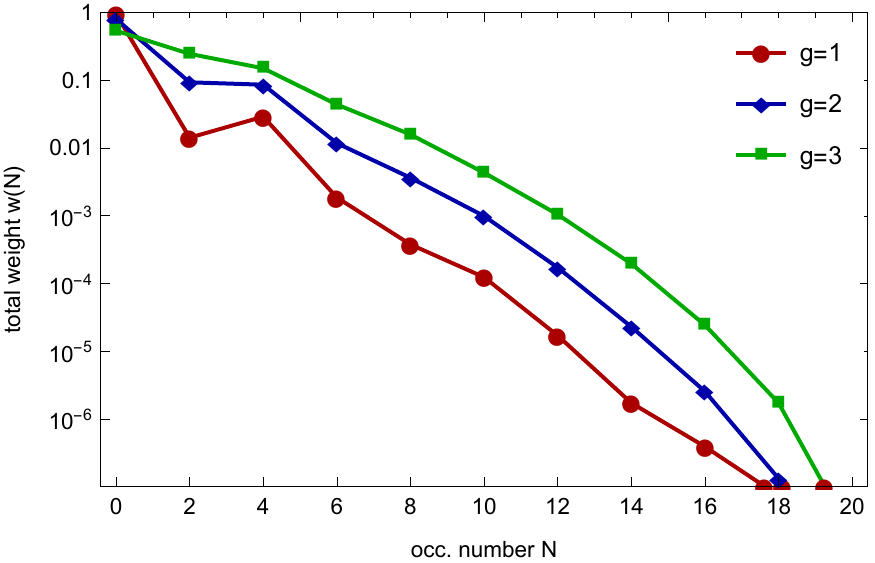}
\includegraphics[width=0.475\textwidth]{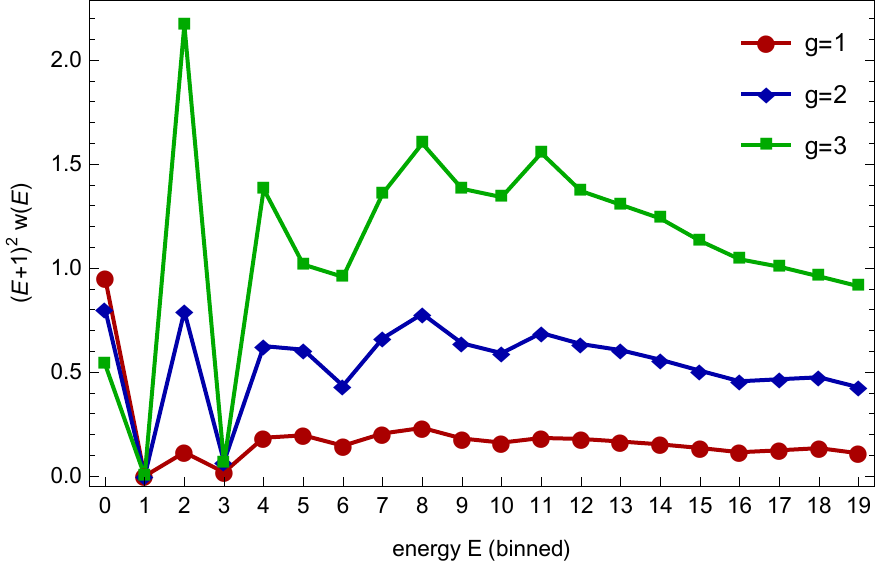}
\caption{Weights as a function of the occupation number (left) and of the state energy (right).} 
\label{fig:OccDist}
\end{center}
\end{figure} 

 The energy $E$ and the total occupation number $N$ are two principal parameters of a basis state. How do they correlate with the weight? Starting with the occupation number, let $w(N)$ be the total weight of all states of occupation number $N$. As is clear from Fig.~\ref{fig:OccDist} (left), $w(N)$ decreases exponentially with $N$. This tendency is especially pronounced at $g=1,2$, but it is noticeable at $g=3$ as well. The free vacuum ($N=0$) dominates the interacting ground state for all three couplings (for the reference, its weight $w_0=0.96, 0.80, 0.54$ for $g=1,2,3$ respectively).\footnote{Since this plot is done at finite $E_T$, the values of $w(N)$ for $N$ close to $E_T$ have some cutoff dependence. However, we believe that the exponential decrease of $w(N)$ is robust, as it can be observed already at $N\lesssim E_T/2$, where the cutoff dependence is minimal.}

We next study the distribution of weights in energy. Let $w(E)$, $E=0,1,2,\ldots$ be the total weight of states whose $H_0$ energy belongs to the interval $[E,E+1)$. It turns out that this distribution also decreases, although not exponentially, but rather like a powerlaw $\sim E^{-2}$ for large $E$. This is clear from Fig.~\ref{fig:OccDist} (right), where we plot $w(E)$ multiplied by $(E+1)^2$. 
 
\begin{figure}[h!tbp]
\begin{center}
\includegraphics[width=0.5\textwidth]{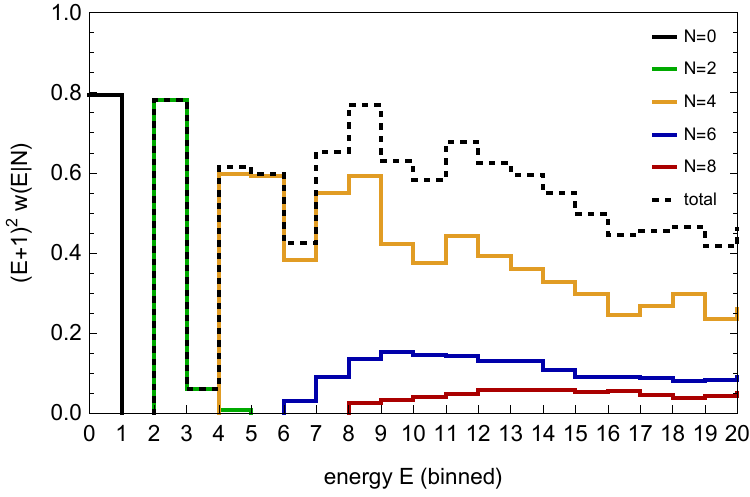}
\caption{This plot refers to the interacting 
ground state for $g=2$. It shows the histogram of weights in an interval of energy, for each occupation number separately (up to $N=8$). The dashed line (same as the $g=2$ line in Fig.~\ref{fig:OccDist} (right)) shows the total weight per the same energy interval.}
\label{fig:EnOccDist}
\end{center}
\end{figure}

Next let us combine Figs.~\ref{fig:OccDist} and see how the weight is distributed both in energy and in the occupation number. Let $w(E|N)$ be like $w(E)$ from the previous plot, but limited to states of fixed total occupation number $N=0,2,4,\ldots$.  This set of distributions is shown in Fig.~\ref{fig:EnOccDist}, where we take $g=2$, the other values of the coupling being similar. Like in Fig.~\ref{fig:OccDist} (right), we multiply by $(E+1)^2$. This plot reveals that for every $N$ the function $w(E|N)$ follows the same powerlaw $\sim E^{-2}$ (the only exception is $N=2$, where the decrease with $E$ seems faster). The total weight per $N$ decreases rapidly with $N$, consistently with Fig.~\ref{fig:OccDist} (left).

\begin{figure}[tbp]
\begin{center}
\includegraphics[width=0.49\textwidth]{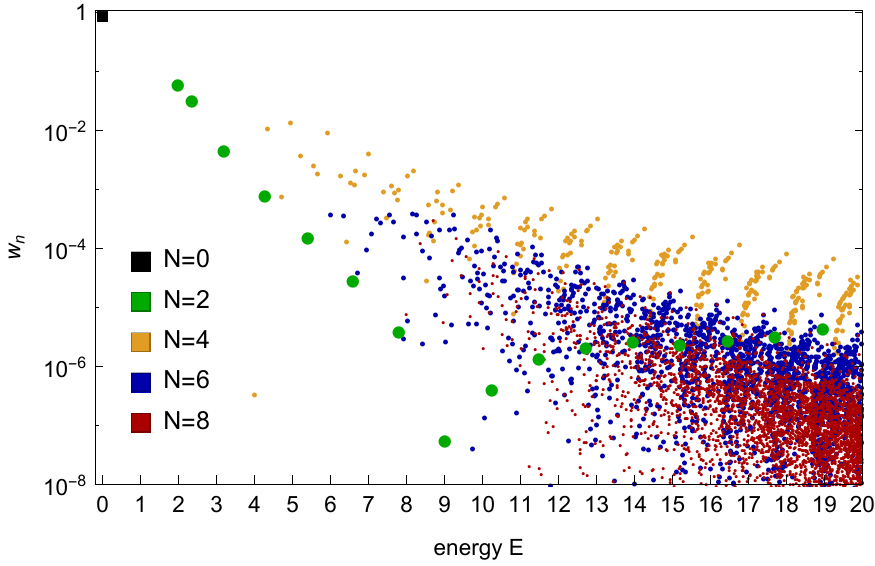}
\includegraphics[width=0.50\textwidth]{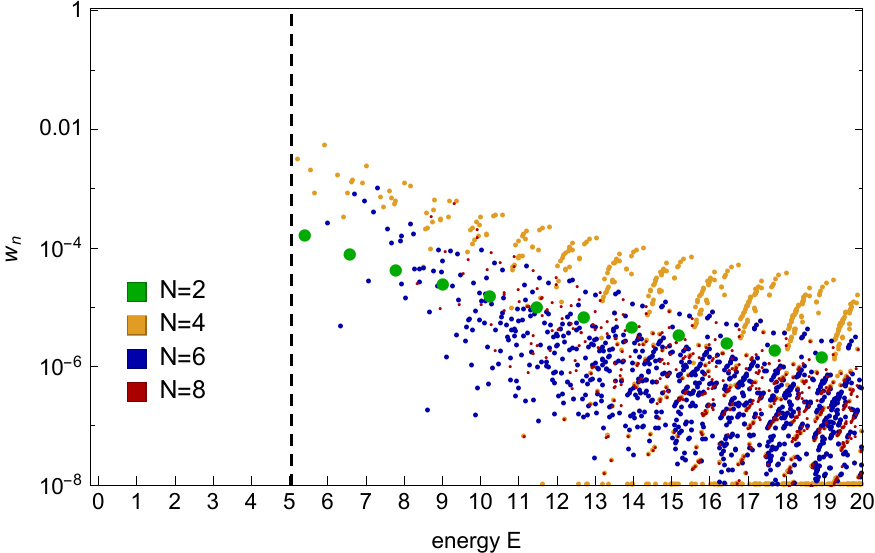}
\caption{\emph{Left:} Weights inside the interacting ground state for $g=2$ from numerical diagonalization. 
\emph{Right:} Weights for states of energies $E>5$ obtained by formula \reef{eq:exer1}. } 
\label{fig:Scatter}
\end{center}
\end{figure}

In the above histograms we grouped states by energy or by occupation number or both. It's important to realize that there is further significant variation of individual weights within the histogram bins. This is clear from Fig.~\ref{fig:Scatter} (left) which shows each state separately for the interacting ground state at $g=2$. For example, weights of 4-particle states (golden points) with nearby energies fluctuate by as much as two orders of magnitude. 

It's instructive to try to understand this plot using Eq.~\reef{eq:tail} which expresses the high energy part of the eigenvector in terms of the low-energy components. For the purpose of this exercise ``low" will denote all energies below 5 (say), and ``high" all energies between 5 and 20. In the spirit of our approximate tail formula \reef{eq:tailstate}, we will also approximate $H_{hh}$ by $H_{0\,hh}$ in \reef{eq:tail}. Let then $c_l$ be the part of the raw eigenvector corresponding to states of energies $E\le 5$, and define $c_h$ by the formula:
\beq
\label{eq:exer1}
c_h = \frac{1}{\calE_0-H_{0\,hh}} V_{hl}.c_l\,,
\eeq
where $\calE_0$ is the raw eigenvalue. The resulting $c_h$ is shown in Fig.~\ref{fig:Scatter} (right). Comparing to the left panel, we see that the periodic variation of the 4-particle component is largely reproduced. This variation is explained by the spread of the $V_{hl}$ matrix elements. The order of magnitudes of $N=2$ and $N=6$ weights are also reproduced (although not the change of sign of $c_n$ in the $N=2$ component which is responsible for the dip at $E=9$). The $N=8$ component is captured poorly, which is not surprising given that too few 4-particle states have been included into $c_l$.

{The observed exponential decoupling of high occupation numbers $N$ is asking to be explained.} Is it related to the fact that $N$ changes by at most a finite amount (four) in each $\phi^4$ interaction? At the moment there is no proof.\footnote{\label{anh1}Compare to the anharmonic oscillator $\hat p^2+\hat q^2+ \lambda \hat q^4$ in quantum mechanics. When it is treated via the Hamiltonian Truncation (Rayleigh-Ritz) in the harmonic oscillator basis, as reviewed in \cite{Hogervorst:2014rta,Pedro}, high occupation numbers ($=$ high energies, as we are in $0+1$ dimensions) are also observed to decouple exponentially. In this simpler problem, this phenomenon can be understood analytically via the analyticity properties of the exact wavefunction in the coordinate representation, or directly in the occupation number representation \cite{Rychkov-unpublished}. See also note \ref{anh2}.} {The exponential decoupling is also asking to be exploited.} Can we take different energy cutoffs in each occupation number sector? Looking at Fig.~\ref{fig:EnOccDist}, it would seem natural to increase the cutoff in the 4-particle sector and reduce the cutoff for $N\ge 6$. One possible rule is that the near-cutoff states in each sector should contribute comparably. There is no guarantee that that this will work, given that some Hamiltonian matrix elements grow with $N$. Still, this is something that needs to be explored.
 
 In this work, as in \cite{Lorenzo1,Lorenzo2,Elias-Miro:2015bqk}, we took a common energy cutoff for all sectors. 
This was convenient for implementing the renormalization corrections. The price to pay is that there's a huge number of states in the Hilbert space - those with high occupation numbers - which have very little weight in the interacting eigenstates. Notice however that we cannot neglect them altogether because their \emph{integrated} weight is not negligible.

\begin{figure}[tbp]
\begin{center}
\includegraphics[width=0.5\textwidth]{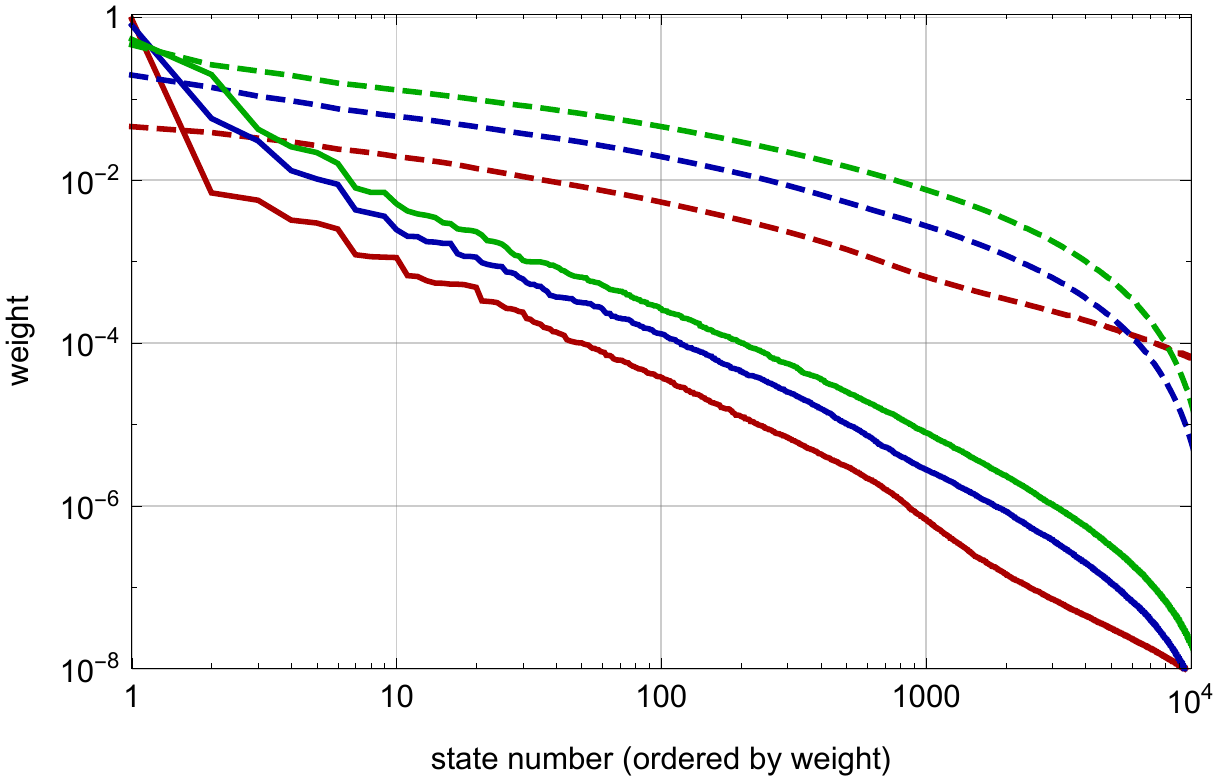}
\caption{Solid lines: all weights inside the interacting vacuum, ordered from large to small. Dashed lines: the total weight of all states past the $n$-th largest.} 
\label{fig:FirstImp}
\end{center}
\end{figure}

Our final plot is relevant for thinking about the idea of optimizing the choice of tail states, mentioned in note \ref{note:tails}. Suppose that we pick a weight threshold $\eps$. How many states are there whose weight is $<\eps$, and how large is the cumulative weight of all remaining states? The answers can be read off Fig.~\ref{fig:FirstImp}. The solid lines plot the sequence $w_n$ ordered from large to small weights. The dashed lines represent the total weight of all states in this ordered list subsequent to the $n$-th.

\subsection{On the orthogonality catastrophe} 

 The orthogonality catastrophe\footnote{Early examples were considered by van Hove \cite{VanHove} and Anderson \cite{Anderson}. This discussion is also related to Haag's theorem \cite{Haag}. In this context, for a formal proof of unitary non-equivalence of two free massive scalars fields with masses $m_1\ne m_2$ in infinite volume, see Theorem X.46 in \cite{reed2}.} 
is the notion that infinite volume interacting eigenstates have zero overlap with the non-interacting ones. 
Since the HT works in a finite but large volume, one may have thought that we will see exponentially small overlaps, while we have seen in the above plots that overlaps remain $O(1)$ even for $Lm=10 - 20$. We would like to discuss how this apparent contradiction gets resolved.

Consider the overlap between the interacting vacuum $|\Omega\rangle$ and the perturbative vacuum $|0\rangle$ in a finite but large volume $L$. In general we expect that in $1+1$ dimensions it will go to zero as 
\beq
\label{eq:over}
|\bra{\Omega}0\rangle|^2 \sim e^{-\alpha L m /(2\pi)},
\eeq 
where $\alpha=\alpha(g/m^2)$ is expected to be order 1 for moderate couplings. The $1/2\pi$ has the usual phase space origin, since the suppression originates from the accumulation of normalization factors of different momentum modes. A toy example is the free massive scalar perturbed by the $\phi^2$ interaction, which amounts to a change in mass. This example can be solved in finite volume via a Bogoliubov transformation. The interacting ground state is a kind of a coherent state. The overlap with the free vacuum can be computed exactly, and $\alpha=O(1)$ confirmed.\footnote{
In the notation of \cite{Lorenzo1}, section 3.4, we have
\beq
\label{eq:g2}
|\Omega\rangle = \prod \frac{1}{\sqrt{\cosh\eta_k}}\exp(-\half \tanh\eta_k\,a_k^\dagger a_{-k}^\dagger) |0\rangle,\quad \frac{\alpha}{2\pi} = 
 \int \frac{dk}{2\pi} \log \half \bigl(1+\frac{1+x/(k^2+1)}{\sqrt{1+2x/(k^2+1)}}\bigr) ,\quad x=g_2/m^2.
\eeq
For $x=0.1,1,10$ we get $\alpha(x)=0.003,0.15,1.8$.
}

In HT we are not interested in taking the mathematically strict infinite volume limit---it suffices to have a volume large enough so that we can extract infinite volume limits of physical quantities, like the particle spectrum.\footnote{For another recent discussion of the infinite volume limit in HT see \cite{Konechny:2016eek}.} By L\"uscher's theorems \cite{Luscher:1985dn}, corrections to stable particle masses in $1+1$ dimensional field theories on a finite circle of length $L$ go as 
\beq
\label{eq:Luscher}
e^{-\beta L m_{\rm ph}}\,,
\eeq
where $m_{\rm ph}$ is the physical particle mass one is trying to extract, and $\beta=\sqrt{3}/2$ or 1 depending on whether the particle appears as a pole in its own $2\to2$ scattering amplitude or not.\footnote{Note that for a massive QFT in $d+1$ dimensions compactified on a flat torus, Eq.~\reef{eq:Luscher} remains valid as written while in Eq.~\reef{eq:over} one has to change $Lm/(2\pi)\to(Lm/(2\pi))^d$ in the exponent. In particular, the sweet window \reef{swindow} is expected to survive.} 

Suppose now we stay away from the critical point so that $m_{\rm ph}$ is order $m$ (this is satisfied for $g\lesssim 2$ for the $(\phi^4)_2$ theory). Comparing \reef{eq:over} with \reef{eq:Luscher} we see that there is a ``sweet window" 
\beq
1\ll Lm  \ll 2\pi/\alpha\,, \label{swindow}
\eeq
where the spectrum is already accurate, but the interacting eigenstates are still dominated by the low-energy non-interacting states.\footnote{For $Lm\gg 2\pi/\alpha$, we expect that the maximum of the distribution of weights of interacting eigenstates will shift to nonzero occupation numbers. It would be interesting to explore this phenomenon in more detail.} This is the range where the HT is expected to work best, and the $g=1$ and $g=2$ plots from this appendix fall precisely into this range.
So we see that the above mentioned apparent contradiction is explained by the 
extra $\alpha/(2\pi)$ in the overlap exponent. 
We expect $\alpha=O(g/m^2)$ for small $g$, so that the window in 
\reef{swindow} widens, while for moderately large couplings we expect $\alpha = O(1)$.

This discussion brings to mind the following question (more theoretical than practical). Suppose that we computed volume $L$ eigenstates, with $L$ in the sweet window. Is it then possible to ``exponentiate" them and construct approximate eigenstates in any volume $L'\gg L$, which would then exhibit the orthogonality catastrophe? We do not know.

\section{Problems with the naive truncation}
\label{FE}

In this appendix we elaborate on the difficulties found when trying to approximate accurately the operator $\Delta H$ by truncating the series expansion \reef{eq:formal}, which we copy here:
\beq
\label{copy}
%\label{eq:formal}
\Delta H(\calE_*)=\sum_{n=2}^\infty \Delta H_n(\calE_*)\,,
\qquad
[\Delta H_n(\calE_*)]_{rs}=\sum V_{rj_{n-1}} \frac1{\calE_*-E_{j_{n-1}}}\ldots V_{j_2 j_1}\frac 1{\calE_*-E_{j_{1}}}V_{j_{1} s}\,,
\eeq
where the sum is taken over all states $j_i$ above the cutoff $E_T$.

Consider this series in the $(\phi^4)_2$ theory. The naive dimensional analysis suggests that each next term in the series is suppressed by $O(g/E_T^2)$. However, this expectation turns out incorrect. There are some intermediate states which violate this power-counting. Because of these states, matrix elements $[\Delta H_n(\calE_*)]_{rs}$ exhibit anomalous growth with $n$. This growth first becomes visible for states $r,s$ just below the cutoff $E_T$, but for sufficiently large $n$ it propagates to all external states. As a result the expansion does not converge; it is only asymptotic.

These effects were first discussed in \cite{Elias-Miro:2015bqk}, and we will review them here. The culprits are intermediate states with large occupation numbers $N$. An oscillator acting on such a state gives an extra factor of $\sim \sqrt{N}$, and the accumulation of such factors skews the asymptotics. We will demonstrate the phenomenon using the states $|N\rangle$ consisting of $N\gg 1$ particles at rest.\footnote{Similar phenomena will happen for other intermediate states with large occupation numbers, e.g.~containing $N/2$ particle pairs of momenta $k,-k$.}
In this section we use $\sim$ to denote order of magnitude estimates.

As a first example, consider equal initial and final states $r=s=|N\rangle$. We choose $N= \lfloor E_T/m\rfloor$ so that this state is at or just below the cutoff. Then
\beq
(H_0)_{ss}\sim Nm \sim E_T, \quad V_{ss}\sim g N^2/(L m^2) \sim f_N E_T, \qquad f_N = \frac{gN}{Lm^3}\,.
\eeq
We see in particular that for any $g$ there exists a large enough $N$ such that the perturbation $V$ is not suppressed with respect to $H_0$ in this matrix element. As we will see now, $\Delta H_2$ will pick up a further factor of $f_N$. Indeed, the state $|N\rangle$ will be connected by $V$ to the states $|N+2\rangle$ and $|N+4\rangle$ which lie above $E_T$. The connecting matrix elements are of the same order as $V_{ss}$. Taking into account the contribution of just these states to $\Delta H_2$, we get:
\beq
\label{eq:lowerb}
|(\Delta H_2)_{ss}|\gtrsim V_{ss}^2/(Nm) \sim f_N^2 E_T.  
\eeq
Notice that all terms entering the expression for $[\Delta H_n]_{rs}$ have the same sign, namely $(-1)^{n-1}$, as long as $\calE_*<E_T$ as we assume. This is because the matrix elements $V_{ij}$ are positive by inspection, and all denominators are negative. 
So if we focus on just some intermediate states, we obtain a lower bound on the absolute value, as in \reef{eq:lowerb}.
Going to higher orders, we will keep getting the same relative factor:
\beq
\label{fNgrowth}
|(\Delta H_{n})_{ss}|\gtrsim f_N^{n+1} E_T\,,
\eeq
totally unlike the naively expected suppression by powers of $g/E_T^2$. For sufficiently large $N$ (i.e.~for sufficiently large $E_T$) we will have $f_N>1$ and the series for this matrix element will then diverge. 

The above example can be generalized to show that the situation is in fact even worse, namely that the series diverges for \emph{any} $E_T$ and for \emph{any} nonzero matrix element. For this we argue as follows. We pick $s,r$ in the same $\bZ_2$ sector, for definiteness even. Pick an even $N$ so large that the state $|N\rangle$ is above $E_T$ and that $f_N>1$. It's easy to see that any even state can be connected to the state $|N\rangle$ by a finite sequence of intermediate Fock states $|j\rangle$ which are above $E_T$ and are obtained by repetitive actions of $V$, i.e.~so that the matrix elements $V_{j_{i+1} j_i}$ are nonzero.\footnote{Here's one way to do this. Recall that we assume that $s,r$ have zero momentum. There are four stages:  (1) Act on $s$ with $V$ once just to get above $E_T$; (2) Pick one particle, say of momentum $p$, and act on it with $(a^\dagger_0)^2 a_p^\dagger a_p$ monomial inside $V$ repeatedly, increasing the zero momentum occupation number up to $N$; (3) Eliminate the nonzero momentum particles by acting repeatedly with $a^\dagger_0 a^\dagger_{p_2+p_1} a_{p_1} a_{p_2}$, picking particle pairs with $|p_2|\ge |p_1|$ and $p_2$, $p_1$ of opposite sign. (4) Annihilate unnecessary zero momentum particles.} If $n_s$ and $n_r$ are the number of steps necessary to connect $s,r$ to $|N\rangle$, then starting from $n=n_s+n_r$ each following $[\Delta H_{n}]_{rs}$ will pick up at least a factor of $f_N$ by the same argument as the one leading to \reef{fNgrowth}. Since $f_N>1$ the series will diverge.

The above effects show that the strategy of systematically improving the accuracy of the spectra by truncating \reef{copy} at increasingly higher orders $n_{\rm max}$ is problematic.
So what can we do about all this? It's important that we are mostly interested in the low-energy eigenstates, and as discussed in appendix \ref{sec:struct}, those have large overlap mostly with the low-energy noninteracting states. In particular, the states with large occupation numbers, like the state $|N\rangle$ close to the cutoff, have contributions which are exponentially suppressed (see Fig.~\ref{fig:OccDist}). One could hope that the problem of the overall divergence of the $\Delta H_n$ series is irrelevant if one is mostly interested in the low-energy entries of the Hamiltonian matrix and if one truncates the series at low $n$. One could also hope that even though $\Delta H_2$ is not small for some states close to the cutoff, this is not important because those states contribute very little to the interacting eigenstates. In other words, 
\emph{Hope: the series \reef{copy},
truncated at low $n$, approximates the low-energy part of the matrix $\Delta H$ well, and the part close to the cutoff which is not well-approximated is unimportant.}

However, as numerical experimentation shows, this hope does not seem to materialize, at least for the values of $E_T$ which are computationally feasible. For example, if one truncates the expansion at $n=2$,
computes $\Delta H_2$ exactly, and then uses it to diagonalize $H+\Delta H_2$ in \reef{eq:eigequiv}, then one finds the following.
First of all, one finds spurious eigenvectors which live close to the cutoff. Since $\Delta H_2$ is negative and large near the cutoff, these spurious eigenvectors have eigenvalues smaller than the physical eigenvalues. Even if one eliminates these and focuses on the eigenvectors which can be interpreted as corrections to the raw truncated eigenvectors, one finds that the corrections are erratic and not always small. The conclusion is that the matrix $\Delta H_2$ near the cutoff is really too large compared to the true $\Delta H$, and this messes up the physical spectrum.\footnote{\label{anh2}By the way, the described effects appear in some form even for the anharmonic oscillator in quantum mechanics (note \ref{anh1}). For that theory the raw HT converges exponentially, but if one tries to improve convergence using renormalization one runs into the problem that the $\Delta H$ series diverges and the renormalized result is worse than the truncated one.} This problem did not influence the results of \cite{Lorenzo1,Lorenzo2} because in those papers the local approximation was used for $\Delta H_2$, suppressing the anomalously large matrix elements near the cutoff. The problem was instead realized by the authors of \cite{Elias-Miro:2015bqk}, who were the first to compute $\Delta H_2$ exactly. {The problem was dealt with in \cite{Elias-Miro:2015bqk} by introducing an auxiliary cutoff $E_W\lesssim E_T/2$ and setting $\Delta H_2$ to zero above this cutoff.}
This temporary solution did not allow to fully take advantage of the exactly known $\Delta H_2$, nor
to include the $\Delta H_3$ corrections.

\subsection{Taming the divergence?}
\label{sec:alternative}

We would like to describe here an idea which might tame the divergence of the perturbative series \reef{copy}.
The idea was not used in this paper, but in the future it may be used either as an alternative to NLO-HT from section \ref{sec:ren-tails} or in combination with that method.

The key observation is that the problematic growth of \reef{eq:lowerb} and \reef{fNgrowth} for large occupation numbers can be cured
if one performs an expansion not around $H_0$ but around $\bar{H}_0\equiv H_0+ \text{diag}\,V$, with $ \text{diag}\,V$ the diagonal part of the potential $V$.
So, consider splitting the Hamiltonian
as
\be
H= \bar{H}_0+ \bar{V} \, , \label{hdiag}
\ee
where we introduced the notation $\bar V=V-\text{diag}\,V$. This is a reasonable split because $\bar{H}_0$ is still an exactly solvable Hamiltonian, diagonal in the same Fock space in which $H_0$ is diagonal. On the other hand, by moving the diagonal part of $V$ into $\bar{H}_0$, one can hope that the series for the correction term will be better behaved.\footnote{A straightforward generalization is to consider instead $\bar H_0=H_0+\lambda (\text{diag}\,V)$ and  $\bar V=V-\lambda (\text{diag}\,V)$ with $\lambda\ne 1$. }

The derivation \reef{p1}-\reef{eq:DeltaH} goes through with the corresponding substitutions, so that one obtains the formal expansion \reef{eq:formal} for the correction term with the replacements
\beq
\frac{1}{\cE - H_{0\,hh}}\rightarrow\frac{1}{\cE - \bar H_{0\, hh}} \,,\qquad V \rightarrow \bar V \label{subs2} \, .
\eeq
These replacements produce higher powers of the occupation numbers in the denominators in such a way that the r.h.s.~of \reef{eq:lowerb} gets replaced by 
 \be
\sim \frac{V_{ss}^2}{Nm + V_{ss}} \,,
  \ee
which is at most order $V_{ss}$ no matter how high $N$ is.

Similarly, in the mechanism for the divergence of any matrix element we will no longer encounter arbitrary large factors $f_N$. At most we get $O(1)$ factors starting from $n=n_s+n_r$. Notice that those factors will not come from the diagonal matrix elements $\langle N|\bar V|N\rangle$, since those are zero, but one can get similar factors oscillating between $|N\rangle$ and $|N+2\rangle$, say. Actually, we believe the series is still divergent (as our numerical experiments and the study of the anharmonic oscillator example show), but it diverges much more slowly and one could think that the above-stated \emph{Hope} perhaps has a chance to be true in this modified setup.

Concerning the technical realization of this possible solution, notice that the Hamiltonian $\bar H_0$, although diagonal in the free Fock space, does not allow a natural formulation in terms of fields. In particular, the increase in the energy of a state acted upon by the oscillator depends on the initial energy and not only on the oscillator frequency. Still, diagrammatic rules from appendix \ref{DT} apply with appropriate changes. Namely, those vertices with two lines to the left and two to the right which correspond to $\text{diag}\,V$ are forbidden, and the energy of the states between any two vertices gets replaced by the $\bar H_0$ eigenvalue.
  
Initial numerical tests of the described procedure looked promising (in particular we had nice results truncating to $n_{\max}=2$ and evaluating the correction term $\Delta \bar H_2$ exactly). A more complete exploration is left for the future.
 
 \section{Relations to other expansions}
 \label{sec:other-expansions}
 
 \subsection{Brillouin-Wigner series}
 
 One particular case where the equation for the effective Hamiltonian \reef{eq:eigequiv} is used in quantum mechanics is when the low Hilbert space
 $\calH_l$ consists of a single element, of non-interacting energy $E_1$, say. In this case there is nothing to diagonalize, and Eq.~\reef{eq:eigequiv} directly expresses the interacting eigenvalue as a solution of the non-linear equation:
 \beq
 \calE = E_1+V_{11}+V_{1h}\frac{1}{\calE - H_{0\, hh} - V_{hh}}V_{h1} \,.
 \eeq
 We can then expand the denominator in $V_{hh}$ and obtain the analogue of Eq.~\reef{eq:formal}:
 \beq
 \label{eq:BW}
\calE = E_1+V_{11}+\sum_{n=2}^\infty T_n(\calE),\qquad T_n =  V_{1h}\frac{1}{\calE - H_{0\, hh}} \left(V_{hh}\frac{1}{\calE - H_{0\, hh}}\right)^{n-2}V_{h1} \,.
\eeq
 This perturbative series is called the Brillouin-Wigner (BW) series \cite{Brillouin,Wigner} and represents a way to organize quantum-mechanical perturbation theory which is somewhat different from the usual Rayleigh-Schr\"odinger (RS) series. Of course, if we further expand $\calE$ in the denominator and  solve the equation order-by-order in $V$, we  get back to the RS series. But if we truncate the BW series at a certain finite order and then solve the resulting equation for $\calE$ exactly, we will get an approximation to the true eigenvalue which is different from the same-order RS approximation.
 
As proved by Wigner \cite{Wigner}, the BW approximations of odd order allow the variational interpretation. Namely, let $\calE=\calE_{2N+1}$, $N\ge 1$, be an odd-order BW approximation, i.e. the smallest solution of the truncated equation
\beq
 \label{eq:BWtrunc}
\calE = E_1+V_{11}+\sum_{n=2}^{2N+1} T_n(\calE)\,.
\eeq
Then there exists a wavefunction $\psi=\psi_{2N+1}$ such that 
\beq
\label{eq:var}
 \langle \psi|H_0+V|\psi \rangle = \calE_{2N+1} \langle \psi|\psi \rangle\,.
\eeq
This wavefunction can be given explicitly:
\beq
\label{eq:psi}
|\psi \rangle = |1\rangle +  \sum_{n=0}^{N-1} \left(\frac{1}{\calE - H_{0\, hh}}V_{hh}\right)^{n} \frac{1}{\calE - H_{0\, hh}}V_{h1}|1\rangle\,.
\eeq
The proof consists in plugging \reef{eq:psi} into \reef{eq:var}. 
Various cancellations and simplifications occur as a consequence of \reef{eq:BWtrunc} and the identity follows.
If the eigenvalue in question is the ground state, the existence of the variational interpretation \reef{eq:var} implies that the BW approximations are always overestimates. Notice that there is no claim that the accuracy of approximation increases with $N$, as unlike in the RR method the trial Hilbert space is not enlarged.

It's instructive to compare the above discussion with our section \ref{sec:ren-tails}. To allow for the comparison, we specialize section \ref{sec:ren-tails} to the case when $\calH_l$ consists of a single state, to which we add a tail. The effective Hamiltonian correction $\Delta \widetilde H$, which is simply the eigenvalue correction in the single state case, is then given by (see Eq.~\reef{eq:mainresult}, where $\Delta H_2$, $\Delta H_3$ are numbers in the case at hand)
\begin{gather}
\Delta \widetilde H = \frac{\Delta H_2}{1-\Delta H_3/\Delta H_2}\,,\\
\Delta H_2 = V_{1h}\frac 1{\calE - H_{0\, hh}}V_{h1},
\quad \Delta H_3 = V_{1h}\frac 1{\calE - H_{0\, hh}}V_{hh}\frac 1{\calE - H_{0\, hh}}V_{h1}\,.
\end{gather}
This can be compared to the BW correction for $N=1$, which takes the form:
\beq
\Delta H_{\rm BW} = \Delta H_2+\Delta H_3\,\quad(N=1)\,. \label{bweff}
\eeq
Both our correction and the BW correction have a variational interpretation. For the BW it's \reef{eq:psi} with $N=1$. For us it's the same equation except that the normalization of the tail, which is the second term in \reef{eq:psi}, is not kept fixed to 1 but is a free parameter which is determined dynamically (see Eq.~\reef{eq:psi+tail}, where $c_t$ and $c_l$ are independent variables). This means that our procedure is bound to give a better approximation. In the case of the ground state, the variational interpretation implies that our correction has to be always smaller than BW. This can also be seen formally from the above equations: for the ground state $\Delta H_2<0$ and so $\Delta \widetilde H\le \Delta H_{\rm BW}$ independently of the sign of $\Delta H_3$.

 \subsection{Schrieffer-Wolff transformation}
   
The renormalization correction  $\Delta H_2$ in \reef{firstDeltaH2} is closely related to another kind of renormalized effective Hamiltonian used in  condensed matter physics. Let us describe briefly the idea behind it. 

Consider a Hamiltonian having the following block structure
\be
H=\left(\begin{array}{cc} H_L & V^\dagger \\  V & H_H\end{array}\right) \, ,
\ee
where the interaction $V$ that mixes the low and high energy Hilbert spaces spanned by the eigenvalues $E_i$ of the free Hamiltonian. $H_L$ and $H_H$ act on the low and high Hilbert spaces, respectively. $V$ is assumed small, in the sense specified below, and the method will involve an expansion in $V$.
 
We want to derive an effective Hamiltonian in the low energy Hilbert subspace.
The idea then is to perform a canonical transformation to $H$  to bring it into block diagonal form
\be
H \rightarrow U H U^\dagger = \left(\begin{array}{cc} H_{\rm eff} & 0 \\ 0 & H_H^\prime\end{array}\right) \,.   \label{blockD}
\ee
Since \reef{blockD} is block diagonal, $H_{\rm eff}$ is the renormalized effective Hamiltonian that describes the low energy physics taking into account the mixing with the states in the high energy Hilbert space. 
A practical way to find the unitary transformation matrix $U$ is to plug in the ansatz
\be
U=e^{ S} \,, 
\ee
with $S$ antihermitean, in \reef{blockD} and solve perturbatively for $S= S^{(1)}+ S^{(2)}+\dots$, where $S^{(i)}=O(V^i)$, by requiring $U^\dagger H U$ to be block-diagonal \cite{SWtrans}.
At leading order $S\approx S^{(1)}$ and \reef{blockD} is solved by
\be
S^{(1)}= \left(\begin{array}{cc} 0 & -s^\dagger \\  s & 0\end{array}\right) \quad \text{with} \quad s_{ki}= \frac{V_{ki}}{E_k-E_i}\,.
\ee
Projecting $e^{S^{(1)}}H e^{-S^{(1)}}$ in the low Hilbert space gives
\be
H_{\rm eff}= H_L+\Delta H_2^{SW}\,,
\ee
with the Schrieffer-Wolff (SW) correction given by
\be(\Delta H_2^{SW})_{ij}= \frac{1}{2}\sum_{k}\left\{ V_{ik}\frac{1}{E_{i}-E_k}V_{kj}+V_{ik}\frac{1}{E_{j}-E_k}V_{kj}   \right\} \label{transSW} \, ,
\ee
where the sum over $k$ is over the high energy Hilbert space. This has to be compared to $\Delta H_2$ in \reef{firstDeltaH2}, which we recall here:
\be
(\Delta H_2)_{ij} = \sum_k V_{ik}\frac{1}{\cE-E_k}V_{kj} \, . \label{lDeltaH2}
\ee
The key difference is that \reef{lDeltaH2} corresponds to the two terms in $\Delta H_2^{SW}$ with $\cE$ replaced by the free energies $E_i$ and $E_j$. In fact the Hamiltonian $H_{\rm eff}$ constructed via the SW procedure is $\calE$-independent, unlike \reef{eq:DeltaH} which was the starting point of our discussion.

The perturbative solution to the canonical transformation \reef{blockD}  was worked out by Schrieffer and Wolff in \cite{SWtrans}. There it was used to relate the Anderson impurity model to the Kondo model. The Anderson impurity model
describes the interaction of conducting electrons in a metal with localized atoms  in it (impurities) that can lead to localized magnetic moments. The highest atomic states of the Anderson model can be integrated out, following for instance the procedure just reviewed. This leads to an effective Hamiltonian that couples the spin density of the conducting electrons with a localized spin, namely the Kondo effective model. 
In this physical system the use of \reef{transSW} and the truncation of the series is well justified because the energy difference in the denominators is large, namely  the  energy gap of the atomic transition into excited states. In other words the dimensionless expansion parameter $V_{ki}/(E_k-E_i)\ll 1$. This has to be contrasted with QFT applications that we have in mind in this paper. In the QFT context, the spectrum is dense at the cutoff and there is no parametric separation between the low and high energy Hilbert spaces. If we introduce a cutoff $E_T$ and take states $E_i$ and $E_k$ just below and just above the cutoff, the ratio $V_{ki}/(E_k-E_i)$ can be arbitrarily large. For this reason the SW procedure does not seem adapted for our problem.

\section{Diagram technique}
\label{DT}

This appendix reviews the diagram technique \cite{Elias-Miro:2015bqk} for a systematic expansion for the matrices $\Delta H_n$ in Eqs.~\reef{eq:formal}, \reef{copy}. Although only $n=2,3$ is needed for this work, we will consider general $n$.

\begin{figure}[h]
\begin{center}
\includegraphics[scale=1]{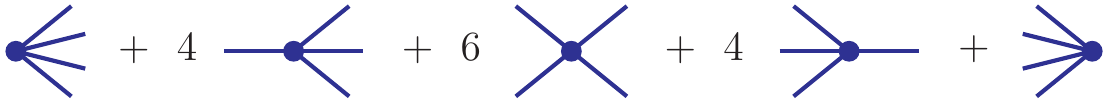}
\caption{The vertices representing the quartic interaction; see the text. }
\label{fig-V}
\end{center}
\end{figure}

The diagram technique follows from Wick's theorem. We represent each $V$ insertion by a vertex with 4 lines exiting. Lines exiting towards right (resp. left) will represent $a_k$ (resp.~$a_k^\dagger$).\footnote{Notice that this assignment is the opposite from the one in Ref.~\cite{Elias-Miro:2015bqk}.} Momentum and time flows from right to left and there is momentum conservation in each vertex. There is also a factor of $1/\sqrt{2L\omega_k}$ for each line. In this notation $V$ is shown in Fig.~\ref{fig-V}.

\begin{figure}[h!tbp]
\begin{center}
\includegraphics[scale=0.9]{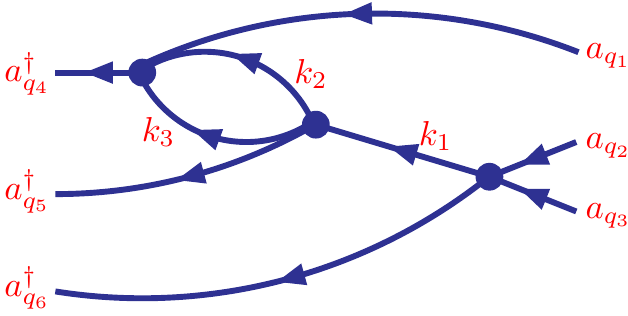}
\caption{A diagram producing operator \reef{eq:ex1}.} 
\label{fig-exVV}
\end{center}
\end{figure}

To construct a diagram we put $n$ vertices time-ordered from right to left in the same order as in \reef{copy}. Some lines exiting from a vertex can be contracted with lines entering into a later vertex. These are produced by oscillator contractions when using Wick's theorem. The uncontracted lines are extended to the ends of the diagram; they correspond to the remaining creation and annihilation operators. This is better illustrated by examples rather than formalized. E.g.~the diagram in Fig.~\ref{fig-exVV} produces the operator
\beq
a^\dagger_{q_4} a^\dagger_{q_5}a^\dagger_{q_6} a_{q_1} a_{q_2} a_{q_3}
\label{eq:ex1}
\eeq
with momenta subject to 
\beq
q_2+q_3=k_1+q_6\,,\quad k_1= k_2+k_3+q_5\,,\quad q_1+k_2+k_3=q_4\,.
\eeq
More precisely, the constraint of momenta conservation is imposed by a delta function times the length $L$ of the cylinder circle. For instance we have the factor $L\delta_{k_3+k_2+q_1-q_4}$ for the leftmost vertex in Fig.~\ref{fig-exVV}. The internal momenta should be summed over. There is also a scalar factor $1/\sqrt{2L\omega_q}$ for each external and $1/(2L\omega_k)$ for each internal line. 

\begin{figure}[h!tbp]
\begin{center}
\includegraphics[scale=0.9]{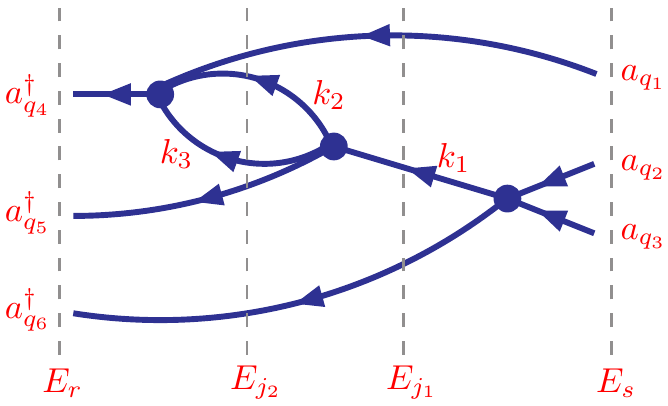}
\caption{The same diagram as in Fig.~\ref{fig-exVV} where we indicated the energies of the external and the intermediate states.  }
\label{fig-exVV1}
\end{center}
\end{figure}

We also have to multiply by factors $1/(\calE_*-E_{j_i})$ produced by the corresponding insertions in \reef{copy}. Here $E_{j_i}$ are energies of the intermediate states, which can be found in terms of the final and initial energies $E_{r}$ and $E_{s}$, taking into account that each vertex changes the flowing energy by the total frequency of all creation minus all annihilation operators.
For example, in Fig.~\ref{fig-exVV} we have two intermediate states denoted by vertical dashed lines in Fig.~\ref{fig-exVV1}, and their energies are related to $E_{r,s}$ by:
\begin{align}
E_{j_1}&=E_s+\delta V_1,\quad \delta V_1= (\omega_{k_1}+\omega_{q_6}) -(\omega_{q_2}+\omega_{q_3})\label{forpoint}\,,\\
E_{j_2}&=E_{j_1}+\delta V_2,\quad \delta V_2 = (\omega_{k_2}+\omega_{k_3}+\omega_{q_5}) -\omega_{k_1}\,,\\
E_{r}&=E_{j_2}+\delta V_3,\quad \delta V_3 = \omega_{q_4}-(\omega_{q_1}+\omega_{k_2}+\omega_{k_3})\,,
\end{align}
where $\delta V_i$ are energy changes in the vertices. 

\begin{figure}[h!tbp]
\begin{center}
        \includegraphics[scale=0.9]{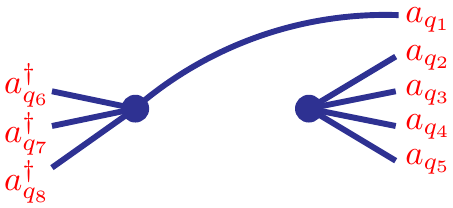}
\caption{This diagram does not contribute, since the intermediate state is always below $E_T$.}
\label{fig-not-contr}
\end{center}
\end{figure}

One way to write a compact solution for these energy conservation constraints, for any diagram, is as follows. Denote by $W_j$ the sum of frequencies of all oscillator lines crossing the dashed line $j$ (which can be an intermediate or external state line). We can move from a external state to an intermediate state in a number of steps, and every time we have to subtract $W_j$ and add $W_{j+1}$. When we add all the steps all increments except the first and the last cancel. Thus the energy of an intermediate state $j_i$ is related to the external state energies by
\begin{gather}
E_{j_i}=E_s-W_s+W_{j_i}\,,\label{eq:interm1}\\ 
E_r=E_{j_i}-W_{j_i}+W_r\,.\label{eq:interm2}
\end{gather}
Taking the difference of these two equations we obtain a more symmetric expression \cite{Elias-Miro:2015bqk}
\beq
E_{j_i}=\half(E_r+E_s)- \half(W_r+W_s)+W_{j_i}\,.  \label{symex}
\eeq
We have to impose the constraints that all of these intermediate state energies are above $E_T$. This will translate into the restrictions on the internal line momenta. 
For some diagrams this constraint cannot be satisfied at all, and such diagrams won't contribute. One example is the diagram in Fig.~\ref{fig-not-contr}, for which the energy of the intermediate state is $E_s-(\omega_{q_2}+\omega_{q_3}+\omega_{q_4}+\omega_{q_5})\le E_T$ since $E_s\le E_T$.
%\be
% \begin{minipage}[h]{0.3\linewidth}
%        \vspace{0pt}
%   \end{minipage}
%   \label{fig-not-contr}
%\ee

Finally, there is a combinatorial factor for each diagram which is computed as usual.

The original derivation of the diagram technique \cite{Elias-Miro:2015bqk} was different. It used an auxiliary operator $\Delta \widehat H_n$ defined as in \reef{copy} but summing over all $j_i$ (not just those with $E_{j_i}>E_T$). This $\Delta \widehat H_n$ is expressed as an iterated integral of a time-ordered $n$-point correlation function of the $\NO{\phi^4}$ interaction: 
\be
\Delta \widehat H_n(\cE_*)_{rs} =(-1)^{n-1} \int_0^{\infty } d\tau_{1} \ldots d\tau_{n-1} e^{(\cE_*-E_r)(\tau_1+\cdots +\tau_{n-1})}\,  [V(T_{n-1})\cdots V(T_1) V(0)]_{rs} \, ,\label{deltaHhat}
\ee
where $T_k=\sum_{i=1}^{k}\tau_i$.\footnote{This is the same $\Delta \widehat H_n$ as in \cite{Elias-Miro:2015bqk} but the definition has been Wick-rotated. Here we work in Euclidean time as in \cite{Lorenzo1}. The time-dependence of operators is in the interaction representation: $V(T)=e^{H_0 T} V(0) e^{-H_0 T}$. The integrals in \reef{deltaHhat} converge for $\calE_*<0$, and for other $\calE$ analytic continuation is understood.} 
Wick's theorem is then used at the level of fields, giving rise to diagrams. Only then one passes from $\Delta \widehat H_n$ to $\Delta H_n$, imposing the restriction that all intermediate states be above $E_T$. This is neatly achieved by considering the analytic dependence of any diagram on $\calE$, viewed as a fiducial variable. The needed terms are those for which the poles in $\calE$ are above $E_T$.

The derivation given here, based on Wick's theorem for oscillators, is more direct than the one in \cite{Elias-Miro:2015bqk}. Both derivations have virtues. The original derivation of \cite{Elias-Miro:2015bqk} has a useful spinoff by allowing to focus directly on the diagrams giving rise to the local approximation, as discussed in appendix~\ref{viaf}. Also, as emphasized in \cite{Elias-Miro:2015bqk}, the poles of $\Delta \widehat H_n$ at $\calE_*<E_T$, although not needed for renormalization, can be used to set up an efficient test for the code. On the other hand, the derivation given here is useful if one wants to play with splitting $H$ into the diagonal and off-diagonal part differently from \reef{eq:Hdef}, see appendix \ref{sec:alternative}.

\subsection{Bound on the intermediate energies for $\Delta H_2$}
\label{Ej}

In this section we will prove an auxiliary result which will be needed in appendix \ref{dh2}. Consider the diagrams contributing to $\Delta H_2$. Some of these have loops, others are tree-level or disconnected. We claim that: \emph{there is an upper bound $2E_T+m$ on the intermediate state energy $E_j$ for tree-level and disconnected diagrams contributing to $\Delta H_2$.} The proof is based on the formula \reef{symex}:
\beq
\label{eq:base}
E_j=E_{rs}-(W_r+W_s)/2+W_{j}\,,
\eeq
where we use the notation $E_{rs}=\half (E_r+E_s)$.

Consider first the disconnected diagrams. For such diagrams we have $W_j\le W_r+W_s$ (with equality if all lines from the left vertex go right, and all lines from the right vertex go left). So $E_j\le E_{rs}+(W_r+W_s)/2$. To have a nonzero matrix element, we must have $W_r\le E_r$, $W_s\le E_s$, since all particles acted upon by the oscillators must be present in the initial and final state. So the nonzero matrix elements have $E_j\le 2E_{rs}\le 2E_T$, which is even stronger than the claimed bound.

The proof for the tree level diagrams is slightly more difficult as one has to keep track of the line connecting the vertices. It will be convenient to condense diagrams into ``thick line" diagrams, carrying the essential information. For this we replace all lines entering or exiting the vertex from the same direction by a ``thick line" carrying the momentum and energy equal to sum of united line momenta and energies. For a thick line carrying momentum $Q$ we will denote the energy carried by it $E(Q)$. Although this energy depends not only on $Q$ but also on how the momenta are distributed, this information will not be needed in the proof and is omitted.

The most general thick line diagram corresponding to a tree-level $\Delta H_2$ diagram is:
\beq
        \includegraphics[width=0.35\textwidth]{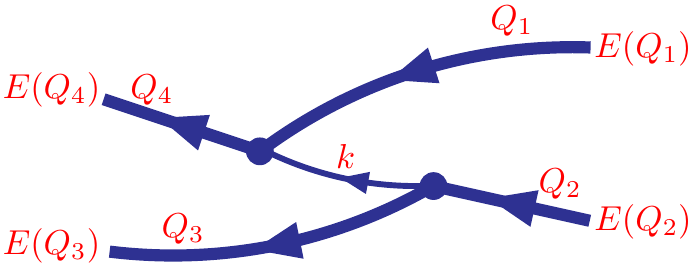} 
        \label{eq:thick}
\eeq
For example the diagram
\beq
        \includegraphics[width=0.3\textwidth]{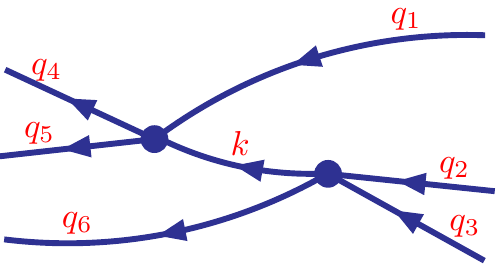} 
\eeq
 will correspond to the thick line diagram \reef{eq:thick} with
\begin{gather}
Q_1=q_1,\quad Q_2=q_2+q_3,\quad Q_3=q_6,\quad Q_4=q_4+q_5,\\ 
E(Q_1)=\omega_{q_1},\quad E(Q_2)=\omega_{q_2}+\omega_{q_3},\quad E(Q_3)=\omega_{q_6},\quad E(Q_4)=\omega_{q_4}+\omega_{q_5}.
\end{gather}
Some of the thick lines may be missing. E.g.~for the diagram 
\beq
        \includegraphics[width=0.3\textwidth]{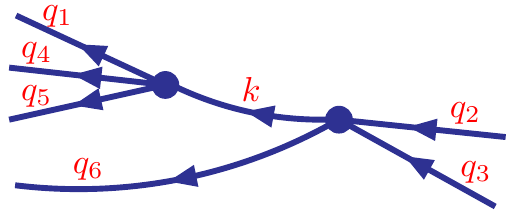} 
\eeq
the thick line diagram would be missing the $Q_1$ thick line, since there are no thin lines coming into the vertex from that direction. In order not to treat such cases separately, we represent them by the same thick line diagram \reef{eq:thick} with the understanding that the ``missing" lines have associated momentum $Q=0$ and energy $E(Q)=0$.
 
Now to the proof. We have
\begin{gather}
W_s= E(Q_1)+E(Q_2),\quad W_r=E(Q_3)+E(Q_4),\quad
W_j=\omega_k+E(Q_1)+E(Q_3) \,.
\end{gather}
We also have 
\beq
\label{eq:Wsr}
 E_s = W_s+ E(-Q_1-Q_2)\,,\quad E_r = W_r + E(-Q_3-Q_4)\,.
\eeq
In the first equation, $E(-Q_1-Q_2)$ stands for the total energy of the constituents of the $s$ state apart from those which are acted upon by the oscillators in the diagram. Their momentum is $-Q_1-Q_2$ since the total state momentum is zero. The second equation is analogous. 

Using the above equations in \reef{eq:base} and eliminating $E(Q_1)$ and $E(Q_3)$, we obtain:
\begin{gather}
E_j=2 E_{rs}-\delta,\\
\delta =  \{\half[E(-Q_1-Q_2)+E(-Q_3-Q_4)]+E(Q_2)+E(Q_4)\}-\omega_k\,. 
\end{gather}
We claim that 
\beq 
\label{eq:lower}
E(-Q_1-Q_2)+E(Q_2)+E(Q_4)\ge \omega_k-m\,.
\eeq
Indeed, in the l.h.s.~we have a sum of energies of a group of particles whose momenta sum to $k$.
Using convexity properties of the function $\omega_k$ it's not hard to prove that
\beq
\sum \omega(k_i) \ge \omega\left(\sum k_i\right)\,,
\eeq
which implies \reef{eq:lower}, in its stronger version without $-m$ in the r.h.s. This $-m$ is needed in the special case when the l.h.s.~of \reef{eq:lower} is actually empty because all three groups of particles are empty (in particular if $Q_2$ and $Q_4$ are ``missing lines''). If this happens then $k=0$ and adding $-m$ we restore the inequality.

Eq.~\reef{eq:lower} and its analogue for $E(-Q_3-Q_4)$ imply that $\delta\ge -m$, and so as claimed
\beq
E_j\le 2 E_{rs}+m \le 2E_T+m\,.
\eeq

\section{Local approximation}
\label{la}

The diagram technique from appendix \ref{DT} leads to exact expressions for the matrix elements of $\Delta H_n$, but evaluating these exact expressions can be demanding. There are many diagrams, and diagrams with loops involve sums over intermediate momenta, with the cutoff that the intermediate energies be above $E_T$. Each diagram corresponds to a product of a certain number of creation and annihilation operators, but the coefficients have a complicated dependence on their frequencies. As a result, $\Delta H_n$ cannot be exactly expressed as an integral of an operator local in the field $\phi$; as we say, it's a \emph{non-local} operator.

However, if we are interested in matrix elements between low-energy states, then one can hope that $\Delta H_n$ may be \emph{approximated} by a local operator. In fact, the calculation of the diagrams can be greatly simplified when the energy exchanged between the different vertices of the diagrams is much larger than the frequencies of the external particles. In this limit, according to  the usual effective field theory intuition, we may expect that the processes described by the non-local diagrams can be approximated by collapsing the loops over ultrahigh momenta into point-like interactions, i.e.~local operators. It's definitely true for $\Delta H_2$ \cite{Lorenzo1,Elias-Miro:2015bqk}, but as we will see there are subtleties for $\Delta H_3$. It's instructive to proceed carefully and see how the local operators arise as a good approximation starting from the diagrams. We will focus on $n=2,3$ as needed in this work.

\subsection{Local approximation for $\Delta H_2$ via diagrams}
\label{dh2}

The diagrams for $\Delta H_2$ have two vertices. For the quartic interaction case considered here, depending on the number of contractions, the resulting terms have 0,2,4,6, or 8 oscillators; see appendix C.1 of \cite{Elias-Miro:2015bqk} for the full list.

To see how the local approximation arises, we start by considering the diagram with 2 external legs, hence 2 oscillators.
There are four such diagrams:
\be
 \begin{minipage}[h]{0.7\linewidth}
        \vspace{0pt}
       \includegraphics[width=\linewidth]{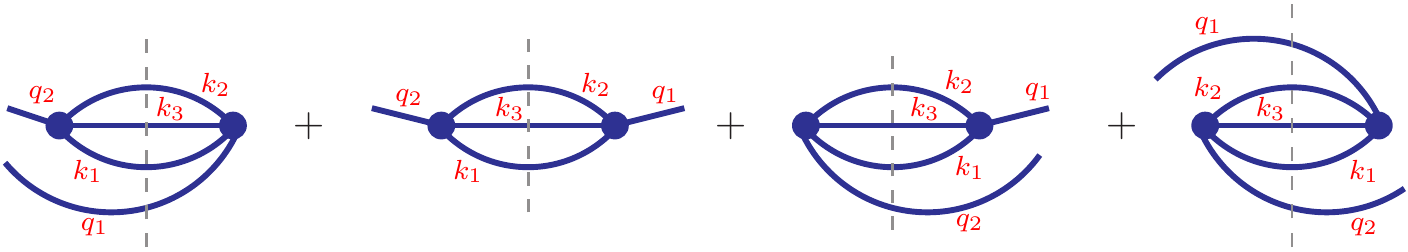}   
   \end{minipage}
   \label{exdiag}
\ee
Let's start with the first of these. By the rules of appendix \ref{DT}, it corresponds to the operator
\be
   96 g^2 \sum_{k_i, q_i} \frac{ L^2\delta_{q_1+k_1+k_2+k_3}\, \delta_{q_2-k_1-k_2-k_3} }{2L \om_{k_1} \, 2L\om_{k_2}\, 2L \om_{k_3}}\frac{\theta (E_j-E_T)}{\cE_*-E_j}\frac{a_{q_1}^\dagger a_{q_2}^\dagger}{\sqrt{2L\om_{q_1} 2L\om_{q_2}}} \, .
   \label{exex}
\ee
Here $E_j$ is the energy of the intermediate state, subject to $E_j>E_T$. We have $E_j=E_s +\om_{q_1}+\om_{k_1}+\om_{k_2}+\om_{k_3}$ by Eq.~\reef{eq:interm1}. As in appendix \ref{DT}, $E_r$ and $E_s$ denote energies of the external states in the considered matrix element, $q$'s are the external and $k$'s the internal momenta. Momentum always flows from right to left. The combinatorial factor for this diagram is $96=\binom{4}{3}^2 3!$.

As explained in section \ref{PI}, we will introduce another energy scale $E_L> E_T$ (we would like it to be much larger than $E_T$ but in practice we can afford $E_L=(\text{2 - 3}) E_T$). We will split $\Delta H_2$ as in Eq.~\reef{eq:splitDH2} depending on whether the intermediate energy is below or above $E_L$. The part of diagram \reef{exex} with $E_j\le E_L$ will be included in $\Delta H_2^<$ and will be evaluated exactly. Here we will be concerned with the part with $E_j> E_L$, included in $\Delta H_2^>$. In this case we will approximate \reef{exex} by dropping the $q_1$ dependence in the momentum conserving $\delta$-functions, and also by neglecting $E_s$ and $\omega_{q_1}$ in $E_j$ with respect to 
$\om_{k_1}+\om_{k_2}+\om_{k_3}$. In this way we conclude that the $E_j>E_L$ part of the diagram is approximated by\be
C \sum_{ q_1, q_2} L\delta_{q_1+q_2} \frac{a_{q_1}^\dagger}{\sqrt{2L\om_{q_1}}}\frac{a_{q_2}^\dagger}{\sqrt{2L\om_{q_2}}} \,   = C \int _0^Ldx[\phi^+(x)]^2 \label{lc1}\,,
\ee
where $\phi^+(x)=\sum_q  a_q^\dagger/\sqrt{2L\om_q} e^{-iqx}$ is the positive-frequency part of $\phi(x)$, and $C$ is just a constant without $q$-dependence:
\be
C = 96\, g^2  \sum_{k_i}  \frac{L \delta_{k_1+k_2+k_3}}{2L \om_{k_1}\, 2L \om_{k_2} \, 2L\om_{k_3}} \frac{\theta (\om_{k_1}+\om_{k_2}+\om_{k_3}-E_L)}{\cE_*-\om_{k_1}-\om_{k_2}-\om_{k_3}} \label{lc1-C}\,.
\ee

So we see that the first diagram in \reef{exdiag} produced a piece of $\phi^2$, and it's not hard to guess that the remaining pieces will come from the remaining three.
Their exact expressions differ from \reef{exex} in how $q$'s and $\omega_q$'s enter into the momentum conserving $\delta$-functions and into $E_j$. However, when we consider the $\Delta H_2^>$ parts and carry out the approximation described above, these differences disappear. So each of these diagrams is approximated by another piece of $\phi^2$ ($\phi^+\phi^-$ for the second and fourth, $[\phi^-]^2$ for the third), times the same constant $C$ as above. The pieces combine neatly when we sum the diagrams, producing
\be
C\int_0^L dx\, \NO{\phi^2}\label{phi2nl}\,.
\ee
Hence the ultrahigh energy part of these four diagrams renormalizes the local operator $\NO{\phi^2}$.
 
When the above procedure is carried out systematically for other classes of diagrams, it gives rise to the approximate expression \reef{eq:loc>}. To be precise, the ultrahigh energy part of diagrams with $p$ contractions, $p=2,3,4$, renormalizes $V_{8-2p}$. The coefficients are given by:
\be
\kappa_{8-2p}(E_L) =  s_{p} g^2  \sum_{k_i}  \frac{ L \delta_{\Sigma k_i}}{\prod_{i=1}^{p}2 L \om_{k_i} } \frac{\theta (\sum \om_{k_i}-E_L)}{\cE_*-\sum \om_{k_i}}\,,   \label{coefsg2}
\ee
with $s_p=\binom{4}{p}^2p!$ the combinatorial factor. This can be rewritten in terms of the relativistic phase space in finite volume:
\begin{gather}
\kappa_{8-2p}(E_L) =  s_{p} g^2 \int_{E_L}^\infty \frac{dE}{2\pi} \frac{\Phi_{p}(E)}{\cE_*-E}\,,  \label{ps1}\\
\Phi_{p}(E) =      \sum_{k_i}  \frac{ L \delta_{\Sigma k_i}}{\prod_{i=1}^{p}2 L \om_{k_i} } 2\pi \delta\left(\normalsize\sum \om_{k_i} -E\right)  \, .        \label{ps2}
\end{gather}
Hence 
\be
\mu_{8-2p}(E)=\frac{g^2 s_{p}}{2\pi} \Phi_{p}(E)  \,  . 
\ee
in \reef{eq:kappamu}. In finite volume, the spectrum is discrete and phase spaces $\Phi_{p}(E)$ are sums of $\delta$-functions. We will be mostly interested in the $L\to\infty$ limit, $Lm\gg 1$. For the purposes of evaluating the integral \reef{ps1} we can then replace $\Phi_{p}(E)$ by its infinite-volume limit:
\be
\Phi_{p}(E) \rightarrow   \int \left( \prod _{k=1}^p \frac{dk_i}{4 \pi \om_{k_i} } \right)(2\pi)^2 \delta\left(\normalsize\sum k_i \right) \delta\left(\normalsize\sum \om_{k_i} -E\right)      \, .        \label{pss4}
\ee
Eqs.~\reef{mus} arise from the leading terms of \reef{pss4} in the $m/E$ expansion. These expressions can be obtained by the Laplace transform method \cite{Lorenzo1}. For $p=2,3$ one can also expand the known exact expressions for the infinite-volume phase space \cite{Elias-Miro:2015bqk}.

It remains to discuss the diagrams with one and no contractions (six and eight external legs). These diagrams cannot be approximated by local operators, because the energy exchanged between the vertices is of always of the same order as the frequencies of the external particles. Consider for instance  
\be
 \begin{minipage}[h]{0.18\linewidth}
        \vspace{0pt}
        \includegraphics[width=\linewidth]{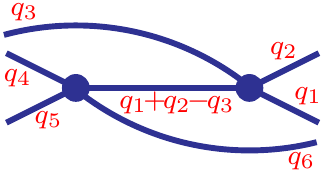}
   \end{minipage}. \label{pureNL}
\ee
The intermediate state energy is
\beq
E_j=E_s -\om_{q_1}-\om_{q_2}+\om_{q_3}+\om_{q_1+q_2-q_3}\,.
\eeq
Since there are no free loop momenta, the intermediate state energy can never become parametrically large compared to the external energies. 
So there is no way to approximate this diagram by local operators; it has to be computed exactly.

The same is true for the rest of the diagrams with six or eight external legs: the intermediate energies is never much larger than $E_T$. 
As shown in appendix \ref{Ej}, the maximal possible energy is $2E_T$ for the disconnected diagrams and $2 E_T+m$ for the tree-level diagrams.

Our strategy will therefore be as follows. In the ``moderately high" energy range $E_T<E_j\le E_L$ our procedure of computing $\Delta H_2^<$ exactly (by multiplying matrices) will amount to taking into account all diagrams, including the `non-local' ones like \reef{pureNL}, without making any approximations. In the ``ultrahigh" range $E_j> E_L$ we will take into account the diagrams with 2,3,4 contractions in the local approximation \reef{eq:loc>}. For this to be a reasonably good approximation we will take $E_L \gtrsim \text{(2 - 3)}E_T$. The `non-local' diagrams like \reef{pureNL} can be ignored when considering the ``ultrahigh" range, in view of the discussed upper bound on their intermediate state energy.

\subsection{Local approximation for $\Delta H_2$ via correlation functions}
\label{viaf}

In the previous section we explained very concretely how the local approximation arises via the diagrams. We will now review an alternative derivation which produces all the relevant terms quickly without having to sift through the diagrams. This will be especially helpful when we move to $\Delta H_3$ where the number of diagrams is even larger. It also has other uses, e.g.~if one wants to compute or
understand the sub-leading corrections in the $E_T/E_L$ expansion. 

Consider the operator 
\be
\Delta \widehat H_2(\cE_*)_{rs} = \sum_{k}V_{rk}\frac{1}{\cE_*-E_k}V_{ks} =  - \int_0^{\infty} d\tau \, e^{(\cE_*-E_{rs})\tau }\,   [V(\tau/2) V(-\tau/2)]_{rs} \label{d2} \, ,
\ee
which differs from $\Delta H_2$ in that we sum over all intermediate states, not just over those above $E_T$. 
We will first analyze $\Delta \widehat H_2(\cE_*)$. Then, we will obtain $\Delta  H_2^>(\cE_*)$ by picking up the terms in $\Delta \widehat H_2(\cE_*)_{rs}$ which have poles in $\cE_*$ located at $\cE_*>E_L$. This is the trick of \cite{Elias-Miro:2015bqk}. 

Eq.~\reef{d2} is the $n=2$ case  of \reef{deltaHhat}, except that we shifted $V$'s to the symmetric time configuration, which explains the change $E_r\to E_{rs}=(E_r+E_s)/2$ in the exponent.\footnote{Here we follow the notation of \cite{Elias-Miro:2015bqk}, while in \cite{Lorenzo1} $E_{rs}$ denoted a related but a different quantity.}  This will be convenient, as the linear terms in $\tau$ will vanish when doing the local expansion around $\tau = 0$.%, see \eq{loc}.

We compute $\Delta \widehat H_2$ by applying Wick's theorem to express the operator product under the integral sign as a sum of normal-ordered terms:
\be
 - g^2 \int_0^{ \infty } d\tau\,  e^{(\cE_*-E_{rs})\tau} \int_0^L dx\, dz \sum_{m=0}^{4} s_{4-m} [G_L(x,\tau)]^{4-m} \, \NO{\phi^m(x+z,\tau/2)\phi^m(z,-\tau/2)} \, , \label{Dwick}
\ee
where $G_L(x,\tau)$ is the Euclidean propagator in finite volume, for positive times given by
\begin{align}
G_L(x,\tau)= \sum_{k} \frac{1}{2L\om_k}e^{-\om_k \tau}e^{i k x}\qquad(\tau\ge 0)\,. \label{prop}
\end{align}
Recall that \eq{Dwick} can be used as a starting point to produce the diagrammatic expansion of \cite{Elias-Miro:2015bqk}, as we explained in appendix~\ref{DT}. Each diagram has a series of poles in $\calE_*$, which are the intermediate energies. Restricting the diagrams so that all poles be above $E_T$ gives $\Delta H_2$.
Here we would like to emphasize a different fact, namely that \eq{Dwick} can be also used as a starting point to produce the local approximation, bypassing the diagrams. 

The local approximation takes into account the contributions of high-energy intermediate states. Since high energy corresponds to short times, it should be possible to pick up these contributions  by studying correlation functions in the $\tau\to0$ limit, using the operator product expansion \cite{Hogervorst:2014rta,Lorenzo1}. 
So we Taylor-expand the operator insertions of \eq{Dwick} around $x,\tau=0$. 
Keeping only the leading term (the subleading $O( x^2, \tau^2)$ terms can be used to study the $m/E$ expansion) we get
\be
  \sum_{m} \hat \kappa_{2m} \int_0^L dz\, \NO{\phi^{2m}(z,0)}\,,  \label{loc} 
\ee
where the coefficients are given by 
\be
 \hat \kappa_{8-2p} = - s_{p}\, g^2 \int_0^{\infty} d\tau \,  e^{(\cE_*-E_{rs} )\tau} \int_0^L  dx \, [G_L(x,\tau)]^{p} \label{cloc} \, . 
\ee
Plugging in \reef{prop} and performing the integral, we obtain 
\be
\hat \kappa_{8-2p} =  s_{p} g^2  \sum_{k_i}  \frac{ L \delta_{\Sigma k_i}}{\prod_{i=1}^{p}2 L \om_{k_i} } \frac{1}{\cE_*-E_{rs}-\sum \om_{k_i}}\,.   \label{coefsg2-1}
\ee
The local approximation coefficients \reef{coefsg2} are obtained from this by two simple and natural operations.
First, we drop $E_{rs}$ in the denominator, since the external energies were totally neglected in 
  \reef{coefsg2}. Second, we should add a $\theta$-function restricting summation to intermediate states above $E_L$. This is the operation which passes from $\Delta \widehat H_2$ to $\Delta H_2^>$.
  
  Notice that if we apply these operations to $\hat \kappa_{8-2p}$ with $p=0,1$ we get zero. This is not surprising, since we already know that the diagrams with 0 or 1 contractions do not allow local approximation. So $\hat \kappa_{6}$ and $\hat \kappa_{8}$ are unphysical and should be simply dropped.

To summarize, the correlation function method for deriving the local approximation for $\Delta H_2$ proceeds as follows. Write down $\Delta \widehat H_2$, and use the OPE under the integral sign. This gives an expansion in local operators with coefficients given by integrals of products of Green's functions. Do the integrals, drop the external energies, and insert $\theta$-functions to enforce the intermediate energy thresholds. Use a bit of diagrammatic intuition to eliminate terms which come from diagrams without such high energy intermediate states (i.e.~diagrams with 0 or 1 contractions).

\subsection{Local approximation for $\Delta H_3$: general strategy} 

\label{sdh3}

According to \eq{d3org} we organize the calculation of $\Delta H_3$  by splitting it into the ${<<}$, $<>$ and ${>>}$ parts. The $<<$ part will be evaluated exactly by multiplying matrices. In the language of diagrams, this means that contributions of all diagrams, including tree-level and disconnected ones, is taken into account. On the other hand, a local approximation will be used when evaluating $\Delta H_3^{<>}$ and $\Delta H_3^{>>}$. The corresponding cutoffs should be chosen high enough so that the local approximation is accurate. 

The calculation of $\Delta H_3^{<>}$ follows the logic explained after Eq.~\reef{mix2}. It involves the matrix $\Delta H_2^>$, which will be approximated by local operators, as reviewed in the preceding section. Recall though that the coefficients are evaluated at $E_L^{\prime\prime}$ which will be fixed at $E_L^{\prime\prime}/E_L^\prime\sim 1.5$.
 We hasten to add that the introduced scales $E_L^\prime$ and $E_L^{\prime\prime}$ are arbitrary. The final exact result should not depend on them. In practice the use of the local approximation introduces some dependence, but we check that it is quite negligible (appendix \ref{convEL}).
 
We next discuss the calculation of $\Delta H_3^{>>}$. Recall that both intermediate states 
 in $\Delta H_3^{>>}$ have the $H_0$-energies restricted to $E_{j}>E_L^\prime$. As we will now explain, for $E_L'\gg E_T$, $\Delta H_3^{>>}$ is well approximated by the local and bilocal operators in \reef{localexp2}.
In practice it will be sufficient to take $E_L^{\prime}/E_T\gtrsim 2-3$. 
The appearance of bilocal operators is one of several new issues encountered for $\Delta H_3^{>>}$ compared to the $\Delta H_2^{>}$ case.
 
The derivation can use any of the two methods explained in sections~\ref{dh2} or \ref{viaf} for $\Delta H_2^{>}$. The first method starts from the exact diagrams, neglects the energies of the external states, and collects all the pieces that combine into the local operators. Here we will follow the second, equivalent, method which start from the correlation functions and uses the OPE. We consider the operator 
\be
\Delta \widehat H_3(\cE_*)_{rs} = \int_0^\infty d\tau_1\, d\tau_2\, e^{(\cE_*-E_r  )(\tau_1+\tau_2)}\, [ V(T_2)V(T_1)V(0)]_{rs}\,, \, \label{d3r}
\ee
where $T_k=\sum_{i=1}^k\tau_i$. 
Applying Wick's theorem to the operator product $V(T_2)V(T_1)V(0)$ we obtain
 \be
g^3 \int_{0}^{L}dx_2\,dx_1\,dx_0  \sum_{m,n,p=0}^{4} s_{mnp}\, G^{m}_{10}G^{n}_{21}G^{p}_{20} \,   \NO{ \phi^{4-p-n}_{x_2,T_2}\phi^{4-n-m}_{x_1,T_1}\phi^{4-p-m}_{x_0,0}} \ , \label{nonl3}
 \ee
 where $\phi_{x,t}=\phi(x,t)$, $G_{ij}$ is the Green's function \reef{prop} joining points $i$ and $j$, and the symmetry factor is
 \be
 s_{mnp}=\frac{(4!)^3}{(4-m-n)!(4-m-p)!(4-n-p)!\, m!n!p!}  \, . 
\ee 

The next step would be to perform the OPE as in the step from \reef{Dwick} to \reef{loc}. This sets all points at the same time, and would seem to produce a local operator. However, one has to be careful. The leading term will indeed have all three points at the same \emph{time}, but depending on the Wick contraction pattern, not all operators may end up at the same \emph{spatial} point. First consider the fully connected contraction patterns, such as e.g.~$m=2$, $n=2$, $p=1$:
\be
  \begin{minipage}[h]{0.16\linewidth}
        \vspace{0pt}
        \includegraphics[width=\linewidth]{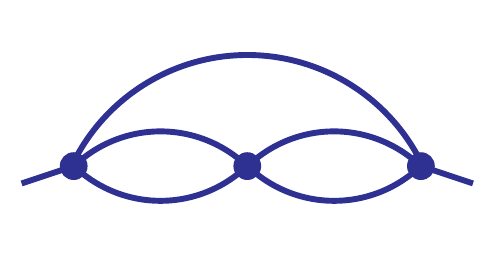}
   \end{minipage} 
   \label{dsh3-ex1}
   \ee
These indeed force all three operators to live near the same $x$ and $t$, giving rise to a local operator ($\NO{\phi^2}$ in this example). But what about not fully connected patterns? Most of these don't contribute to $\Delta H_3^{>>}$, because the intermediate energy constraints are not satisfied. However, those which do contribute can give rise to {\it bilocal} operators. 
There are two patterns for which this happens. The first one is $m=n=0$, $p=3$:
\be
  \begin{minipage}[h]{0.16\linewidth}
        \vspace{0pt}
        \includegraphics[width=\linewidth]{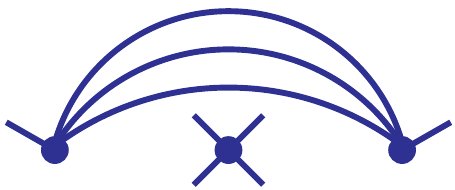}  
   \end{minipage} 
   \ee
This clearly does not represent a local operator. Indeed, the six momenta are split into two groups, $4+2$, which sum to zero independently. A moment's thought shows that the corresponding operator is $\NO{V_{2}V_{4}}$. The second case is $m=n=0$, $p=2$:
\be
\begin{minipage}[h]{0.14\linewidth}
        \vspace{0pt}
        \includegraphics[width=\linewidth]{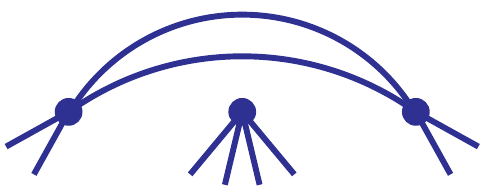},
   \end{minipage} 
        \ee
which gives rise to the operator $\NO{V_{4}V_{4}}$. Another not fully connected pattern which contributes to $\Delta H_3^{>>}$ is $m=n=0$, $p=4$: 
\be
\begin{minipage}[h]{0.14\linewidth}
        \vspace{0pt}
        \includegraphics[width=\linewidth]{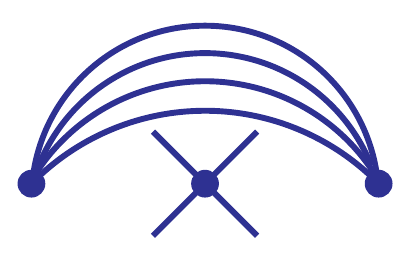}.
   \end{minipage} 
        \ee
However, this one does give rise to a local operator $V_4$ (formally because $\unit.V_{4} = V_4$).

In this way we arrive at the local approximation shown in Eq.~\reef{localexp2}, with the coefficients related to the diagrams representing the various Wick contraction patterns. The $\lambda$-coefficients depend on $E_L'$, since we enforce the constraint that both intermediate state energies be above $ E_L^\prime$. Further details will be provided in appendix \ref{detailsDH3}.

\section{Local approximation for $\Delta H_3$: gory details}
\label{detailsDH3}

In this appendix we analyze in detail the local approximation \reef{localexp2} for $\Delta H_3^{>>}$. The correlation function and OPE method presented in section \ref{sdh3} gives rise to terms \reef{nonl3}, corresponding to the various {\it Wick contraction patterns}. It's not difficult to reconstruct from which {\it diagrams} these terms would come if we started from the diagrammatic expansion rather than from the correlation functions. For example, pattern \reef{dsh3-ex1} would correspond to the four diagrams where the external lines could extend left or right, like in \reef{exdiag}. The exact expressions for the diagrams would be sensitive to this information, but in the local approximation we just get an overall coefficient, common for the four diagrams and represented by the Wick contraction pattern. 

It is understood that both intermediate state energies must be above an auxiliary cutoff $E_L$,\footnote{In this appendix we rename $E_L'$ to $E_L$ for brevity.} enforced by inserting the corresponding $\theta$-functions. Because of these constraints, some contraction patterns do not actually contribute to $\Delta H_3^{>>}$. Here are two examples:
%%%%%
\be
          \begin{minipage}[h]{0.15\linewidth}
        \vspace{0pt}
        \includegraphics[width=\linewidth]{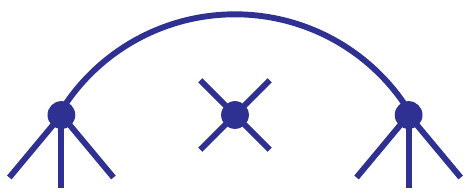}  
   \end{minipage} \,\quad\text{and}\quad
\begin{minipage}[h]{0.15\linewidth}
        \vspace{0pt}
        \includegraphics[width=\linewidth]{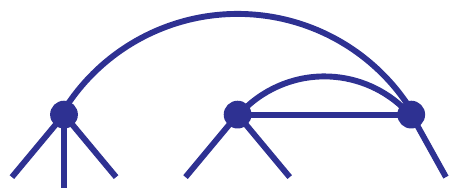} 
   \end{minipage} \,. 
     \ee
For the first case both intermediate energies are $O(E_T)$ so this diagram contributes to $\Delta H_3^{<<}$. The second diagram contributes 
to $\Delta H_3^{<<}$ and $\Delta H_3^{<>}$. None of these diagrams contribute to $\Delta H_3^{>>}$ provided that $E_L$ is sufficiently large. Below we will not take such diagrams into account.

We will now list systematically all {\it contraction patterns} which contribute to the local approximation \reef{localexp2}, and give for each one its contribution to the corresponding coefficient. The rules for evaluating this coefficient are the same as for the {\it diagrams}, except that we neglect the external oscillator momenta and energies, as well as the energies of the external states. We will introduce a few rules to save space in the writing: 
\begin{itemize}

\item The external oscillators with their $1/\sqrt{2 L\omega_k}$ factors are not written, as they are included into $V_N$.
This also concerns one momentum conserving delta-function $L \delta_{\sum k_i}$, or two of those if we are dealing with a bilocal operator.

\item Since dimensions $[V_N]=E^{-1}$, we must have $[\lambda_N]=E^2$, and for bilocals $[\lambda_{N|M}]=E^3$. Below we do not show the factor $g^3$, and the given expressions will have dimensions $[\lambda_N/g^3]=E^{-4}$, $[\lambda_{N|M}/g^3]=E^{-3}$.

\item In this section $q_i$, $p_i$ and $k_i$ will denote the momenta connecting the left and central vertices, the central and right, and the left and right, respectively. Momenta always flow from right to left as in appendix \ref{DT}.

\item The $\theta$-functions imposing intermediate state energies above $E_L$ are understood but not written. They are always uniquely reconstructible, as one intermediate energy involves the sum of $\omega_q$'s and $\omega_k$'s, and the other the sum of $\omega_p$'s and $\omega_k$'s.

\item
Each diagram involves a sum over all finite volume momenta $(2\pi/L)\bZ$, subject to the shown $\delta$-functions. 
We will define the following summation symbol that includes the relativistic normalization and has a finite infinite volume limit. If there are $n$ momenta $P_i$ (be that $q$'s, $p$'s or $k$'s) to sum over, we will write: 
\be
\sum_n \equiv \sum_{P_1\ldots P_n} \frac{1}{\prod_{i=1}^n 2L\om(P_i)}\,.
\ee
Notice that $[\sum_n]=E^0$.
\end{itemize}

\subsection{Coefficients}
\label{dh3coefs}
$\lambda_0$ receives contribution from just one pattern (click on $\lambda$'s to go to the asymptotic analysis of the corresponding diagram in section \ref{intanal}):
\be
\hyperref[d0]{\lambda_0}=\begin{minipage}[h]{0.12\linewidth}
        \vspace{0pt}
        \includegraphics[width=\linewidth]{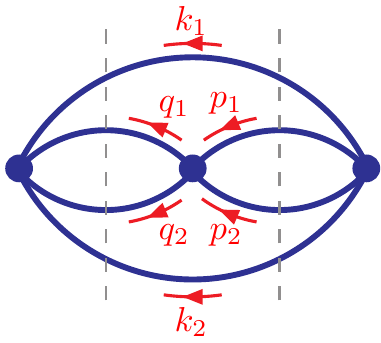} 
   \end{minipage}  = s_{222}    \sum_6   \frac{L \delta_{p_1+p_2+k_1+k_2}}{\cE_*-\om_{p_1}-\om_{p_2}-\om_{k_1}-\om_{k_2}} \frac{L  \delta_{q_1+q_2+k_1+k_2} }{\cE_*-\om_{q_1}-\om_{q_2}-\om_{k_1}-\om_{k_2}}
     \, . \label{l0}
    \ee   
   $\lambda_2$ receives contributions from patterns with 5 contractions. Up to left-right reflection (denoted by h.c.), there are 4 such patterns:
  \begin{align}
&\hyperref[d21]{\lambda_{2.1}}=  \begin{minipage}[h]{0.13\linewidth}
        \vspace{0pt}
        \includegraphics[width=\linewidth]{./figs/Phi4_3Pt/mass1.pdf} 
   \end{minipage} = s_{221}\, \sum_5     \frac{L\delta_{q_1+q_2+k}}{\cE_*-\om_{q_1}-\om_{q_2}-\om_k}    \frac{ L\delta_{p_1+p_2+k}  }{\cE_*-\om_{p_1}-\om_{p_2}-\om_k}\, ,  \label{l21}   \\[.2cm]
   %%%%%%%%%%%%%%%%%%%%%%%%%%%%%%%%%%%
&\hyperref[d22]{\lambda_{2.2}}= \begin{minipage}[h]{0.12\linewidth}
        \vspace{0pt}
        \includegraphics[width=\linewidth]{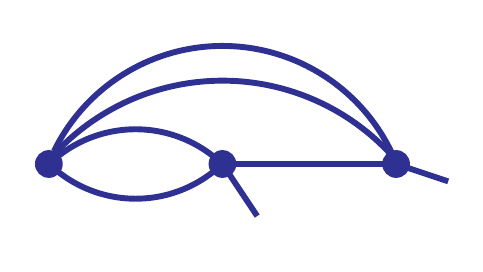} 
   \end{minipage} +\text{h.c.}  =  2s_{212}\,   \sum_5     \frac{L\delta_{q_1+q_2+k_1+k_2}}{\cE_*-\om_{q_1}-\om_{q_2}-\om_{k_1}-\om_{k_2}}    \frac{L \delta_{p+k_1+k_2}}{\cE_*- \om_p-\om_{k_1}-\om_{k_2}} \ \, ,\quad
\label{l22} \\    %    \eea
      %\bea
   %%%%%%%%%%%%%%%%%%%%%%%%%%%%%%%%%%%
&\hyperref[d43]{\lambda_{2.3}}  =   \begin{minipage}[h]{0.13\linewidth}
        \vspace{0pt}
        \includegraphics[width=\linewidth]{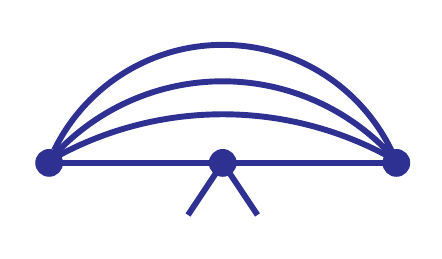} 
   \end{minipage} =  s_{113}\,      \sum_5      \frac{  L\delta_{q+k_1+k_2+k_3}}{\cE_*-\om_q-\om_{k_2}-\om_{k_1}-\om_{k_3}}    \frac{L \delta_{p+k_1+k_2+k_3}}{\cE_*-\om_p-\om_{k_2}-\om_{k_1}-\om_{k_3}} \label{l23}  \, ,
%\nonumber
\\[.2cm]
      %%%%%%%%%%%%%%%%%%%%%%%%%%%%%%%%%%
&\hyperref[d44]{\lambda_{2.4}} =   \begin{minipage}[h]{0.12\linewidth}
        \vspace{0pt}
        \includegraphics[width=\linewidth]{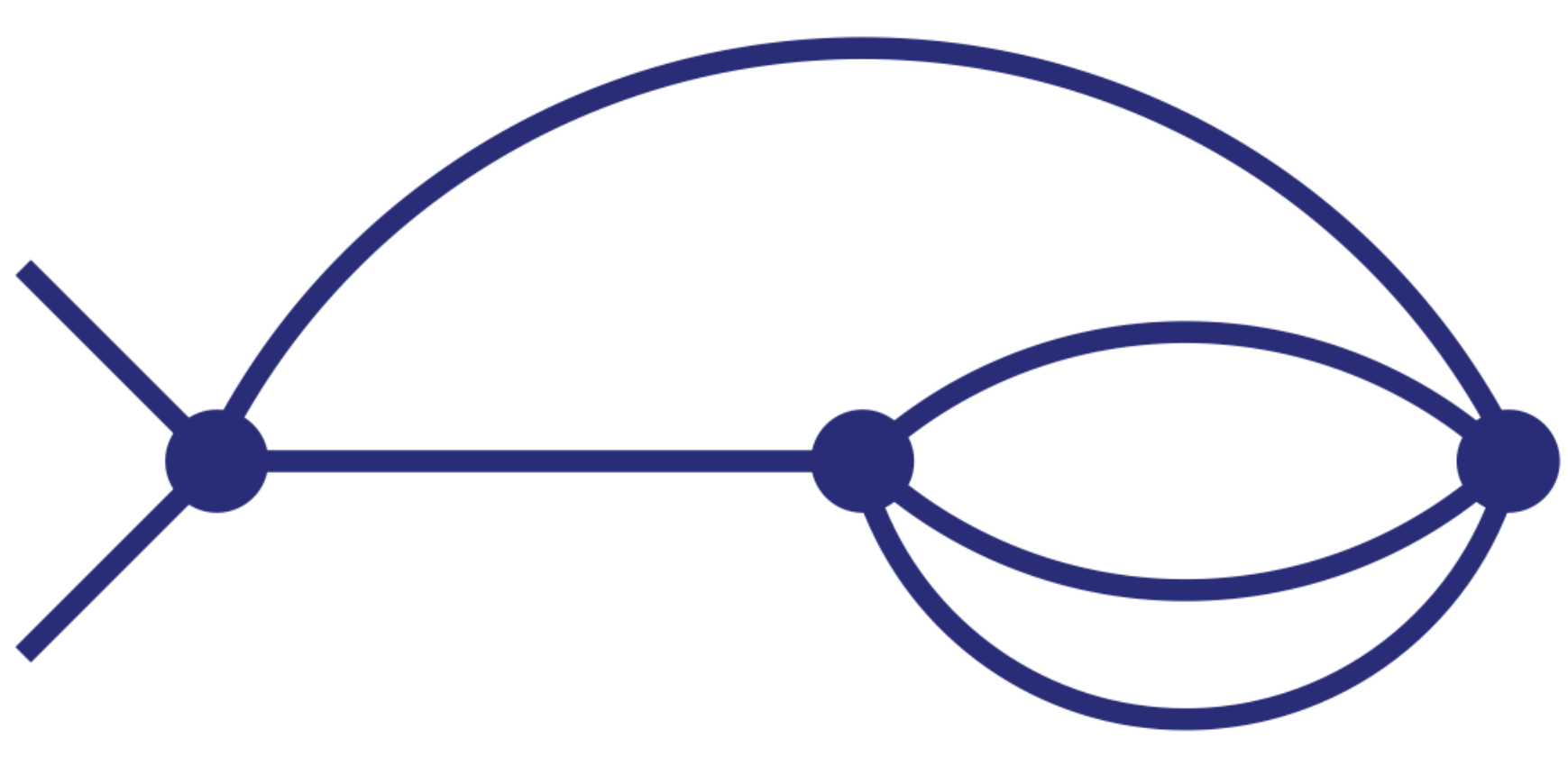} 
   \end{minipage} + \text{h.c.}  =  2 s_{131}\,  \sum_5     \frac{L\delta_{p_1+p_2+p_3+k}}{ \cE_*-\om_{p_1}-\om_{p_2}-\om_{p_3}-\om_k   }    \frac{L \delta_{q+k}}{ \cE_*-\om_q-\om_k}  \, .
   \label{l24}
   \end{align}
   %%%%%%%%%%%%%%%%%%%%%%%%%%%%%%%%%%%%%%%%%%
%%%%%%%%%%%%%%%%%%%%%%%%%%%%%%%%%%%%%%%%%%
%%%%%%%%%%%%%%%%%%%%%%%%%%%%%%%%%%%%%%%%%%
%%%%%%%%%%%%%%%%%%%%%%%%%%%%%%%%%%%%%%%%%%
%%%%%%%%%%%%%%%%%%%%%%%%%%%%%%%%%%%%%%%%%%
$\lambda_4$ receives contributions from 6 patterns with 4 contractions:
   \begin{align}
&\hyperref[d41]{\lambda_{4.1}}=  \begin{minipage}[h]{0.14\linewidth}
        \vspace{0pt}
        \includegraphics[width=\linewidth]{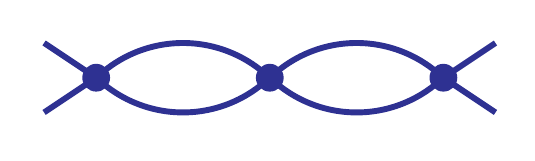} 
   \end{minipage} =   s_{220}\,  \sum_4    \frac{L\delta_{q_1+q_2}}{\cE_*-\om_{q_1}-\om_{q_2}}    \frac{L \delta_{p_1+p_2}}{\cE_*-\om_{p_1}-\om_{p_2}} \, , \label{l41} \\[.2cm]
          %%%%%%%%%%%%%%%%%%%%%%%%%%%%%%%%%%%
&\hyperref[d42]{\lambda_{4.2}} =  \begin{minipage}[h]{0.12\linewidth}
        \vspace{0pt}
        \includegraphics[width=\linewidth]{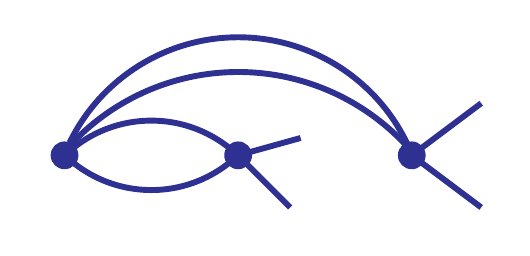} 
   \end{minipage}+ \text{h.c.} =   2 s_{202}\,  \sum_4   \frac{L \delta_{q_1+q_2+k_1+k_2}}{\cE_*-\om_{q_1}-\om_{q_2}-\om_{k_1}-\om_{k_2}}    \frac{L \delta_{k_1+k_2}}{\cE_*-\om_{k_1}-\om_{k_2}} \, , \label{l42}  \\[.2cm]
   %%%%%%%%%%%%%%%%%%%%%%%%%%%%%%%%%%%
 &\hyperref[d43]{\lambda_{4.3}}=  \begin{minipage}[h]{0.14\linewidth}
        \vspace{0pt}
        \includegraphics[width=\linewidth]{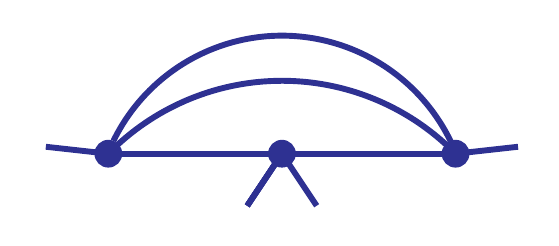} 
   \end{minipage}  =    s_{112}\,  \sum_4    \frac{ L  \delta_{q+k_1+k_2}}{\cE_*-\om_q-\om_{k_2}-\om_{k_1}}    \frac{L \delta_{p+k_1+k_2} }{\cE_*-\om_p-\om_{k_2}-\om_{k_1}} \, ,  \label{l43} \\[.2cm]
   %%%%%%%%%%%%%%%%%%%%%%%%%%%%%%%%%%%
 &\hyperref[d44]{\lambda_{4.4}}=  \begin{minipage}[h]{0.14\linewidth}
        \vspace{0pt}
        \includegraphics[width=\linewidth]{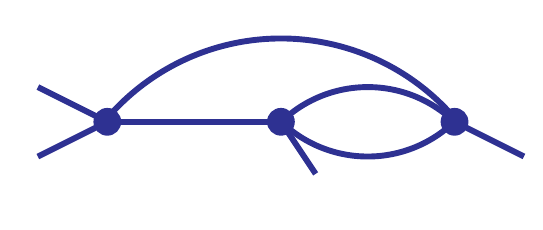} 
   \end{minipage} +\text{h.c.} =  2s_{211}\,   \sum_4   \frac{L \delta_{q+k}}{\cE_*-\om_q-\om_k}    \frac{ L  \delta_{p_1+p_2+k} }{\cE_*-\om_{p_1}-\om_{p_2}-\om_k} \, ,  \label{l44}
\\[.2cm]
   %%%%%%%%%%%%%%%%%
 &\hyperref[d45]{\lambda_{4.5}}=\begin{minipage}[h]{0.12\linewidth}
        \vspace{0pt}
        \includegraphics[width=\linewidth]{./figs/Phi4_3Pt/lambda7.pdf} 
   \end{minipage}  =     s_{004}\,  \sum_4  \frac{   L^2 \delta_{k_1+k_2+k_3+k_4} }{ \left(\cE_*-\om_{k_1}-\om_{k_2}-\om_{k_3}-\om_{k_4}\right)^2} \, , \label{l45}  \\[.2cm]
      %%%%%%%%%%%%%%%%%%%%%%%%%%%%%%%%%%
   %%%%%%%%%%%%%%%%%%%%%%%%%%%%%%%%%%%   
      %%%%%%%%%%%%%%%%%%%%%%%%%%%%%%%%%%%
&\hyperref[d46]{\lambda_{4.6}}= \begin{minipage}[h]{0.12\linewidth}
        \vspace{0pt}
        \includegraphics[width=0.9\linewidth]{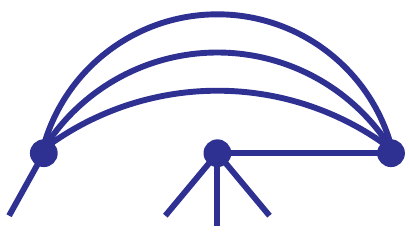} 
   \end{minipage} +\text{h.c.}  =    \frac{ 2 s_{013} }{2m}  \sum_3    \frac{L \delta_{k_1+k_2+k_3}}{\cE_*-\om_{k_1}-\om_{k_2}-\om_{k_3}-m} \frac{ 1  }{\cE_*-\om_{k_1}-\om_{k_2}-\om_{k_3}}  \, . \label{l46}
   \end{align}
%%%%%%%%%%%%%%%%%%%%%%%%%%%%%%%%%%%%%%%%%%
%%%%%%%%%%%%%%%%%%%%%%%%%%%%%%%%%%%%%%%%%%
%%%%%%%%%%%%%%%%%%%%%%%%%%%%%%%%%%%%%%%%%%
%%%%%%%%%%%%%%%%%%%%%%%%%%%%%%%%%%%%%%%%%%
%%%%%%%%%%%%%%%%%%%%%%%%%%%%%%%%%%%%%%%%%%
$\lambda_6$ receives contributions from just two patterns:
\begin{align}
      %%%%%%%%%%%%%%%%%%%%%%%%%%%%%%%%%%%
& \hyperref[d61]{\lambda_{6.1}}=  \begin{minipage}[h]{0.12\linewidth}
        \vspace{0pt}
        \includegraphics[width=\linewidth]{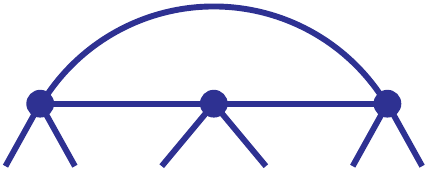} 
   \end{minipage} =   s_{111}\,    \sum_3  \frac{L \delta_{p_1+q_1}}{\cE_*-\om_{q_1}-\om_{k_1}}\frac{L\delta_{p_1+k_2}  }{\cE_*-\om_{p_1}-\om_{k_1}}  \, ,  \label{l61} \\[.2cm]
   %%%%%%%%%%%%%%%%%%%%%%%%%%%%%%%%%
 &\hyperref[d46]{\lambda_{6.2}}  = \begin{minipage}[h]{0.125\linewidth}
        \vspace{0pt}
        \includegraphics[width=\linewidth]{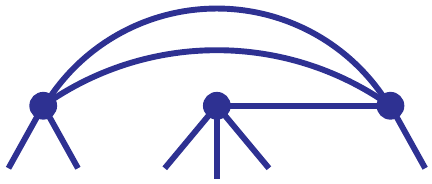} 
   \end{minipage} +\text{h.c.}  =    \frac{ 2 s_{012} }{2m}    \sum_{2}     \frac{L\delta_{k_1+k_2} }{\cE_*-\om_{k_1}-\om_{k_2}-m}    \frac{  1    }{\cE_*-\om_{k_1}-\om_{k_2}}   \, . \label{l62}
          \end{align}  
%%%%%%%%%%%%%%%%%%%%%%%%%%%%%%%%%%%%%%%%%%
%%%%%%%%%%%%%%%%%%%%%%%%%%%%%%%%%%%%%%%%%%
%%%%%%%%%%%%%%%%%%%%%%%%%%%%%%%%%%%%%%%%%%
%%%%%%%%%%%%%%%%%%%%%%%%%%%%%%%%%%%%%%%%%%
%%%%%%%%%%%%%%%%%%%%%%%%%%%%%%%%%%%%%%%%%%
Finally, as explained in section \ref{sdh3}, coefficients of the bilocals are given by the patterns:
     \begin{align}
& \hyperref[d45]{\lambda_{2|4}}= \begin{minipage}[h]{0.13\linewidth}
        \vspace{0pt}
        \includegraphics[width=\linewidth]{./figs/Phi4_3Pt/c2-4.pdf} 
   \end{minipage} =    s_{003}\,     \sum_3    \frac{  L \delta_{k_1+k_2+k_3} }{\left(\cE_*-\om_{k_1}-\om_{k_2}-\om_{k_3}\right)^2}  \, , \label{2|4}\\
 &  \hyperref[d45]{\lambda_{4|4}}= \begin{minipage}[h]{0.14\linewidth}
        \vspace{0pt}
        \includegraphics[width=\linewidth]{./figs/Phi4_3Pt/c4-4.pdf} 
   \end{minipage} =    s_{002} \sum_{2}    \frac{L \delta_{k_1+k_2} }{\left(\cE_*-\om_{k_1}-\om_{k_2}\right)^2 }   \, .  
   \label{4|4}
   \end{align}

A comment is in order concerning the diagrams for $\lambda_{4.6}$ and $\lambda_{6.2}$. They have the middle vertex joined to the rest by a single propagator (the horizontal line). Strictly speaking, this invalidates the local approximation. Indeed, let $p$ be the momentum flowing through this line, which is the sum of momenta entering the middle vertex. The original diagrams will depend on $p$ through the propagator, and also through the energy of the right intermediate state. In \reef{l46} and \reef{l62} this dependence is neglected: $p\to0$, so that $\omega_p \to m$. This is not a problem in the intermediate state, whose energy is dominated by the other energetic particles. But in the propagator this replacement is problematic,  as it changes $1/(2\omega_p)\to 1/(2m)$ and overestimates the matrix elements unless $p=0$.

A moment's thought shows that the diagrams for $\lambda_{4.6}$ and $\lambda_{6.2}$ should be more properly approximated by the following bilocal operators:
\beq
\int_0^L  dx\,dy\, G_L(x-y,0) \NO{\phi^ N(x) \phi^3(y)} \,,
\eeq
with $N=1$ and 3, respectively. However, in this paper we will not try to correct this small error.

For numerical evaluation, expressions \reef{l0}-\reef{4|4} will be further simplified by taking the formal infinite volume limit $L\to\infty$. This will be done by replacing
\beq
\sum_n \to \int \frac{d^nP}{(2\pi)^n},\qquad L\delta_{\sum P_i}\to (2\pi) \delta\left(\sum P_i\right)\,.
\eeq
The validity of this approximation, for the volumes $L$ that we consider in our computations, will be justified below.

So we face the task of evaluating 15 coefficients corresponding to the $L\to\infty$ limits of each diagram.
It would be great if we could find analytic expressions for the spectral densities for both intermediate states, similar to \reef{mus}.
This would allow us to reduce these computations to two-dimensional integrals in the energies of those intermediate states, similar to the one-dimensional integrals in \reef{eq:kappamu}. For 6 diagrams (4.1, 4.5, 4.6, 6.2, $2|4$, $4|4$) the spectral densities trivially reduce to products of spectral densities \reef{mus}. For example, the spectral density for $\lambda_{4.1}$ is $\mu_2(E_1)\mu_2(E_2)$ where $E_1$ and $E_2$ are the two intermediate energies, while for $\lambda_{4.5}$ it's $\mu_4(E_1)\mu_4(E_2)\delta(E_1-E_2)$. For the other diagrams we were not able to find analytic spectral densities. For those diagrams we evaluate the original multidimensional integral, for each needed $E_L$ and $\calE_*$, numerically via Monte Carlo integration  (we use {\tt  vegas-3.2} in {\tt python}).
  
\subsection{$L\to\infty$ limit and the asymptotic estimates}
\label{intanal}

In this section we will carry out a rough asymptotic analysis to determine how various $\lambda$'s scale with $E_L$, $m$, $L$. The accuracy of these asymptotic approximations would be insufficient for practical computations, for which as mentioned we have to resort to numerical integration. Still, this exercise is instructive. It will also help understand the validity and limitations of the described formal $L\to \infty$ limit which replaces sums over momenta by integrals. For brevity of presentation, we will not keep track of $\calE_*$ dependence. I.e.~we assume $\calE_*\ll E_L$ and set $\calE_*\to 0$.

{ \renewcommand{\arraystretch}{1.15} \renewcommand\tabcolsep{6.7pt}
\begin{table}[h]
\begin{center}
\begin{tabular}{L{1.5cm}llclc}  
\toprule
\multirow{2}{*}{} & \multicolumn{2}{c}{{numerical~value} ($\times 10^{-3}$)} && \multirow{2}{*}{asymptotics} &   \multirow{2}{*}{ sensitive to $P_{\rm ext}$} \\
\cmidrule{2-3} 
& $E=20$\phantom{abcabc}  & $E=40$ & \\
\midrule
$ \hyperref[l0]{\lambda_0}$     &   $0.67$ &$ 0.084$ && \footnotesize $ 1/(E^3 m)$ & \\  
\hline
$ \hyperref[l21]{\lambda_{2.1}}$        &   $0.85 $  &  $0.078$   && \footnotesize  $ (\log E/m)^2/E^4$  & \\  
   $ \hyperref[l22]{\lambda_{2.2}}$           &   $3.8$ &  $0.44$   && \footnotesize  $  1/(E^3 m)$ &  $\checkmark$ \\  
$ \hyperref[l23]{\lambda_{2.3}}$            &   $ 1.8$  &   $ 0.25$   &&  \footnotesize  $ (\log E/m)/(E^3 m)$ &  $\checkmark$ \\  
$ \hyperref[l24]{\lambda_{2.4}}$           &   $0.33$ &  $0.032$ && \footnotesize  $ (\log E/m)^2/E^4$ &   \\  
\hline
     %%%%%%%%%%%%%%%%%%%%%%%%%%%%%%%%%%%%%%%%%%%%%%%%%%%%%%
     %%%%%%%%%%%%%%%%%%%%%%%%%%%%%%%%%%%%%%%%%%%%%%%%%%%%%%
$ \hyperref[l41]{\lambda_{4.1}}$             &   $0.034$ &  $0.0021  $ &&  \footnotesize $   1/E^4 $ &  \\  
$ \hyperref[l42]{\lambda_{4.2}}$             &   $1.21$ &  $0.16$ && \footnotesize  \footnotesize $   1/(E^3 m)$ & $\checkmark$ \\  
$ \hyperref[l43]{\lambda_{4.3}}$           &   $ 4.1$ & $0.45$  && \footnotesize $  1/(E^3 m)$  & $\checkmark$ \\  
$ \hyperref[l44]{\lambda_{4.4}}$          &   $ 1.5$ & $ 0.12$   & &\footnotesize $   (\log E/m)/E^4 $  & \\ 
$ \hyperref[l45]{\lambda_{4.5}}$            &   $0.25$ &  $0.046 $ && \footnotesize $L (\log E/m)^2/E^3$  &  \\ 
$ \hyperref[l46]{\lambda_{4.6}}$          &   $ 3.9$ &   $0.60$  && \footnotesize $(\log E/m)/(E^3 m)$   & $\checkmark$ \\  
\hline
     %%%%%%%%%%%%%%%%%%%%%%%%%%%%%%%%%%%%%%%%%%%%%%%%%%%%%%
     %%%%%%%%%%%%%%%%%%%%%%%%%%%%%%%%%%%%%%%%%%%%%%%%%%%%%%
$ \hyperref[l61]{\lambda_{6.1}}$            &   $0.39$ &  $0.026$  &&  \footnotesize$1/E^4$  &  \\  
$ \hyperref[l62]{\lambda_{6.2}}$           &   $ 3.6$  &  $ 0.46$   &&  \footnotesize$ 1/(E^3 m)$  &  $\checkmark$ \\ 
     %%%%%%%%%%%%%%%%%%%%%%%%%%%%%%%%%%%%%%%%%%%%%%%%%%%%%%
     \hline
     %%%%%%%%%%%%%%%%%%%%%%%%%%%%%%%%%%%%%%%%%%%%%%%%%%%%%%
 $\hyperref[2|4]{ \lambda_{2|4}}$           &   $1.0$ &   $0.15$ && \footnotesize $ (\log E/m)/E^3$   &  \\   
     %%%%%%%%%%%%%%%%%%%%%%%%%%%%%%%%%%%%%%%%%%%%%%%%%%%%%%
     %%%%%%%%%%%%%%%%%%%%%%%%%%%%%%%%%%%%%%%%%%%%%%%%%%%%%%
$ \hyperref[4|4]{\lambda_{4|4}}$            &   $0.42$  &   $0.056$ && \footnotesize  $  1/E^3 $  &  \\
\midrule
$ \kappa_0$    &   $-  8.4 $   & $  -3.0   $ & &  \footnotesize$(\log E/m)^2/E^2$  &  \\
$ \kappa_2$  &     $  -  31.8$  &$  -8.5  $    &&\footnotesize$(\log E/m)/E^2$   &  \\
$ \kappa_4$            &   $- 14.3  $ &  $- 3.5 $ &&  \footnotesize$1/E^2$   & \\
\bottomrule
\end{tabular}
\caption{\label{tabasym}Representative values for $\lambda$'s and the asymptotic behavior for $E=E_L\gg m$, $Lm\gg 1$. We only give leading-log asymptotics. The approximate numerical values are given for $g=1$, $m=1$, $L=10$, $\calE_*=0$, in units of $10^{-3}$. For comparison the last three lines report $\kappa$'s from \reef{eq:kappamu} in the same format. See the text for the meaning of the last column. Click on $\lambda$'s to go back to the drawn diagrams in section \ref{dh3coefs}.
 }
\end{center}
\end{table}
}

The results of this analysis are summarized in Table \ref{tabasym}. Below we explain how the entries of this table are obtained. We start from the simple diagrams and proceed to the more complicated ones.  $E_1$ and $E_2$ will  denote the energies of the two intermediate states, counting from the left.
Depending on the context, the symbol $\sim$ in this section means proportionality, asymptotic equality, or leading-log asymptotics.  Click on diagram's number to go back to its drawing in section \ref{dh3coefs}. 

{\bf Diagram \hyperref[l41]{4.1}.} \label{d41}
This is the simplest diagram since the two intermediate states are independent. The spectral densities are just two particle spectral densities, expressed in terms of two particle phase space $\Phi_2(E)$, see \reef{pss4}, which in the limit $L\to\infty$ is given by 
\beq
\Phi_2(E,P)=\frac{\theta(s-4m^2)}{\sqrt{s(s-4m^2)}},\quad s=E^2-P^2\,.
\eeq
We will write $\Phi(E)=\Phi_2(E)$ if the total pair momentum is $P=0$. This is the case for diagram 4.1 since we neglect the external momenta. So omitting the prefactors and setting $\calE_*\to 0$ we get 
\beq
\lambda_{4.1}\sim \int _{E_L}^\infty \frac{dE_1}{E_1} \Phi(E_1) \int _{E_L}^\infty \frac{dE_2}{E_2} \Phi(E_2)\sim \frac{1}{E_L^4}\,,
\label{as41}
\eeq 
where we used that $\Phi(E)\sim 1/E^2$ for $E\gg m$. Notice that the $L\to\infty$ approximation is justified. Indeed, since both intermediate states have large energy and the pair momenta is zero, it follows that both pair components have large momentum, and the spectrum is dense in that region. Thus it's clearly justified to replace sums by integrals. 

{\bf Diagram \hyperref[l42]{4.2}.} \label{d42}
In this case the intermediate state $E_1$ is made of two groups of two particles, one of which is $E_2$. So $E_1>E_2$. The joint spectral density is $\Phi(E_1-E_2)\Phi(E_2)$, and we get 
\beq
\lambda_{4.2}\sim \int_{E_L}^\infty \frac{dE_2}{E_2} \Phi(E_2) \int_{E_2}^\infty\frac{dE_1}{E_1} \Phi(E_1-E_2)\,.
\eeq 
The crucial question is what's the typical value of $E_1-E_2$. Suppose first that $E_1-E_2\sim E_2$. Then we can rescale 
$E_1=E_2(1+x)$ and write
\beq
\lambda_{4.2}\sim \int_{E_L}^\infty \frac{dE_2}{E_2}\int_{0}^\infty \frac{dx}{1+x} \Phi(E_2 x)\Phi(E_2)\sim \int_{E_L}^\infty \frac{dE_2}{E_2} \frac{1}{E_2^4}  \int_{0}^\infty \frac{dx}{x^2(1+x)} \sim \frac{1}{E_L^4}  \int_{0}^\infty \frac{dx}{x^2(1+x)}\,, \label{xint}
\eeq
where we used the asymptotics $\Phi(E)\sim 1/E^2$ for both $\Phi$'s. However, the end result is inconsistent since
the integral over $x$ diverges at $x=0$. 
This means that in fact the leading contribution to $\lambda_{4.2}$ comes from the region where $E_1-E_2$ is very close to the two particle threshold. In this region the approximation $\Phi(E)\sim 1/E^2$ is invalid. Instead we denote $E=E_1-E_2$ and approximate
\beq
\lambda_{4.2}\sim \int_{E_L}^\infty \frac{dE_2}{E_2} \Phi(E_2) \frac{1}{E_2}\int_{0}^\infty dE\,\Phi(E)\sim \frac{1}{E_L^3m}\,.
\label{4.2appr}
\eeq
where we used that $\int_{0}^\infty dE\,\Phi(E) \sim 1/m$. It's important that this latter integral converges at the upper limit, otherwise we would not be able to approximate $E_1\approx E_2$ in the measure $dE_1/E_1$.

The main lesson is that singularities at the boundary of the phase space give rise to $1/(E^3m)$ dependence where naive dimensional analysis ignoring the $m$ scale would predict $1/E^4$. 

One might worry about the validity of the naive $L\to \infty$ approximation (replacing all sums by integrals) for this diagram, since as we have seen the dominant contribution involves a two particle state close to the threshold. There are not so many such states in finite volume, and one might worry about higher sensitivity to finite $L$ effects compared say to diagram $4.1$. 
However, a closer inspection of the exact expression for the relevant integral in finite volume (see \reef{ps2}) shows that finite $L$ effects are exponentially suppressed:
 \be
 \int_0^\infty dE \, \Phi_2(E) = \sum_k  \frac{2\pi}{4\om_k^2 L}=\frac{\pi}{4m}  \coth \frac{Lm}{2} = \frac{\pi}{4m} \left(1+2e^{-Lm}+\dots \right) \,.
 \label{Lsens}
  \ee
This is not accidental. In fact the sum can be expressed as an integral of the propagator \reef{prop}:
  \beq
  \sum_k  \frac{1}{\om_k^2}\propto \int_0^L dx\, [G_L(x,0)]^2\,,
  \eeq
  and the finite and infinite volume propagators differ in position space by exponentially small ``winding" terms. Similar reasoning will apply for the other diagrams, and in the end we will show that the $L\to\infty$ approximation is justified for all of them.
  
There is however another effect related to the importance of low momenta states for this diagram, which is not so innocuous.
This concerns the dependence on the external momenta, marked by $\checkmark$ in the last column of the table. Denote by $P_{\rm ext} = P$ the momentum flowing into the diagram through the middle vertex. Naively if $P=O(m)\ll E_L$ it can be neglected (and it was neglected above). However for this diagram this neglect is not valid, because the small loop is very sensitive to this momentum. If $P\ne0$ we must replace $\Phi(E)$ by $\Phi(E,P)$ in \reef{4.2appr}. Since
\beq
\int_{0}^\infty dE\,\Phi(E,P) = \int_{0}^\infty \frac{ds}{2\sqrt{s+P^2}}\Phi(s) \,, 
\eeq
we see that even $P=O(m)$ leads to an $O(1)$ change (suppression) of this diagram. So strictly speaking it's not allowed to neglect the $P_{\rm ext}$ dependence. 

We will see below several other diagrams exhibiting $P_{\rm ext}$ dependence, by the same mechanism (4.3, 2.2, 2.3), or by a slightly different one (4.6, 6.2). Although it's certainly possible to include this dependence in our numerical calculations, it's a bit tedious, and in this paper we will not take it into account, setting $P_{\rm ext}\to 0$. This can be improved in the future work if needed.

{\bf Diagrams \hyperref[l43]{4.3} and \hyperref[l23]{2.3.}} \label{d43}
Denoting by $p$ momentum flowing through the horizontal line of $\lambda_{4.3}$, this diagram is given by
\beq
\lambda_{4.3}\sim \int \frac{dE_1}{E_1} \int \frac{dE_2}{E_2} \int \frac{dp}{\omega_p^2}\, \delta(E_1-E_2) \Phi(E_1-\omega_p,p)\,.
\eeq
The integral over $p$ converges at $p=O(m)$ so 
\beq
\lambda_{4.3}\sim \int_{E_L}^\infty \frac{dE_1}{E_1^2} \Phi(E_1) \int \frac{dp}{\omega_p^2} \sim 1/(E_L^3 m).
\eeq
 $\lambda_{2.3}$ is similar except with the three particle phase space, whose $E\gg m$ leading-log asymptotics is
$\Phi_3(E)\sim (\log E/m)/E^2$, giving an extra log. The validity of the $L\to\infty$ approximation is justified for these diagrams in the same way as for
$\lambda_{4.2}$. There is also sensitivity to $P_{\rm ext}$, for the same reason as for $\lambda_{4.2}$.

{\bf Diagrams \hyperref[l44]{4.4} and \hyperref[l24]{2.4.}} \label{d44}
Let $q$ be momentum going around the loop of $\lambda_{4.4}$. Then $E_1=2\omega_q$, $E_2= \omega_q+E$, with $E$ the energy of a two particle state of momentum $q$. In particular $|q|\gg m$. So this diagram is given by
\begin{align}
\lambda_{4.4}\sim \int \frac{dE_1}{E_1} \int \frac{dE_2}{E_2} \int \frac{dq}{\omega_q^2}\, \delta(E_1-2\omega_q) \Phi(E_2-\omega_q,q)\,
\sim \int \frac{dE_1}{E^3_1} \int \frac{dE_2}{E_2} \Phi(E_2-E_1/2,E_1/2)\,,
\end{align}
where we neglect the particle mass. The invariant mass of the two particle state is 
\beq 
s = (E_2-E_1/2)^2 - (E_1/2)^2 = E_2(E_2-E_1)\,.
\eeq
If we denote $E_2=E_1 x$ and use the approximation $\Phi(s)\sim 1/s$ we get ($E=E_1$)
\beq
\lambda_{4.4}\sim \int_{E_L}^\infty \frac{dE}{E^5} \int_1^\infty \frac{dx}{x^2(x-1)}\,.
\eeq
The integral over $x$ is log-divergent at the lower limit, and must be cut off at $x\sim m^2/E^2$ because of the cutoff $s>4m^2$ which we ignored so far. So 
\beq
\lambda_{4.4}\sim \int_{E_L}^\infty \frac{dE}{E^5} \log(E/m) \sim (\log E_L/m)/E_L^4\,.
\eeq
Although the leading contribution involves two particle states with small invariant mass, their total momentum was large. As a result this diagram will not be particularly sensitive to finite $L$ and $P_{\rm ext}$ effects. $\lambda_{2.4}$ is similar but involves the three particle phase space, with an extra log in the asymptotics.

{\bf Diagrams \hyperref[l45]{4.5}, \hyperref[2|4]{$\mathbf{2|4}$}, \hyperref[4|4]{$\mathbf{4|4}$}} \label{d45}
\beq
\lambda_{4.5}\sim L \int \frac{dE_1}{E_1} \int \frac{dE_2}{E_2} \delta(E_1-E_2) \Phi_4(E_1)\sim L (\log E_L/m)^2/E_L^3\,,
\eeq  
where we used the leading-log four particle phase space asymptotics $\Phi_4(E)\sim (\log E/m)^2/E^2$ 
The overall factor $L$ arises because the diagram is disconnected. The other two diagrams are fully analogous, except three and two particle spectral densities are involved, and the factor $L$ is absorbed into the bilocal operator. 

{\bf Diagrams \hyperref[l46]{4.6} and \hyperref[l62]{6.2}} \label{d46}
\beq
\lambda_{4.6}\sim \frac1
{\omega_P} \int \frac{dE_1}{E_1} \int \frac{dE_2}{E_2} \delta(E_1-E_2-\omega_P) \Phi_3(E_1)\sim (\log E_L/m)/(\omega_P E_L^3)\,,
\eeq  
where we used the leading-log three particle phase space asymptotics, and $P$ is the external momentum flowing in through the central vertex (we assume $\omega_P\ll E_L)$ 
The expressions in \reef{l46} and in the table correspond to $\omega_P\to m$ which neglects the $P_{\rm ext}$ dependence and overestimates the diagram. $\lambda_{6.2}$ is similar but involves the two particle phase space. Notice that the mechanism for $P_{\rm ext}$ dependence of these two diagrams is different and simpler than for 2.2, 2.3, 4.2, 4.3. 

{\bf Diagram \hyperref[l61]{6.1}.} \label{d61}
Denoting by $p$ the momentum going around the loop we have
\beq
 \lambda_{6.1}\sim  \int \frac{dE_1}{E_1}\int\frac{dE_2}{E_2}  \int \frac{dp}{\om_p^3} \,  \delta(E_1-2\om_p)\delta(E_2-2\om_p) \sim   \int \frac{dE_1}{E_1}\int\frac{dE_2}{E_2^4}    \,  \delta(E_1-E_2)\sim \frac{1}{E_L^4} \, , 
  \eeq
 where we neglected $m$ in the second approximation. 
The bottom line is forced to carry a large momentum, which is different from $\lambda_{4.3}$ and $\lambda_{2.3}$ where the main contribution came from soft bottom line momenta. As a result this diagram clearly has no finite $L$ or $P_{\rm ext}$ sensitivity.
   
  {\bf Diagram \hyperref[l21]{2.1}} \label{d21}
  \beq
 \lambda_{2.1}\sim \int \frac{dE_1}{E_1}\int \frac{dE_2}{E_2} \int \frac{dk}{\om_k} \Phi(E_1-\om_k,k) \Phi(E_2-\om_k,k). 
  \eeq 
 While naively one may have expected $1/E_L^4$ asymptotics,  
  there are two regions of phase space which give an enhanced contribution. The first one is that of small $k$, whose contribution is
 \beq
 \lambda_{2.1}\supset const. \int \frac{dE_1}{E_1}\int \frac{dE_2}{E_2} \Phi(E_1) \Phi(E_2) \int \frac{dk}{\om_k}  \sim (\log E_L/m)/E_L^4\,,
 \eeq
 where we cut off the log-divergent $k$ integral at $k\sim E_L$, where the small $k$ approximation breaks down.
 
 The second enhanced region is $2k\sim E_1 \sim E_2$, where the invariant masses of the two particle states
 \beq
 \label{eq:si}
 s_i = (E_i-\omega_k)^2 -k^2 = E_i(E_i-2\omega_k) +m^2
 \eeq
 are small. Consider the half of the integral where $E_2>E_1$. Introduce $E_2 = x E_1$, $x>1$ and $\omega_k = y E_1/2$, $0<y<1$.
Neglecting the $m^2$ in the r.h.s.~of \reef{eq:si} for the moment, and using the approximation $\Phi(s)\sim 1/s$, which will be adequate to pick the leading-log part, 
 we get
\begin{gather}
 \lambda_{2.1}\supset const. \int \frac{dE_1}{E_1^5} \int_0^1 \frac{dy}{y(1-y)} I(y)\,,\label{eq:21appr}\\
I(y) = \int_1^\infty \frac{dx}{x^2(x-y)} = -\frac{\log(1-y)}{y^2}-\frac1{y}\,.
 \end{gather}
 Notice that $I(y)$ has a log singularity as $y\to 1$ but has a finite limit as $y\to 0$. Substituting $I(y)$ into \reef{eq:21appr} and recalling the effective cutoff $m^2/E^2$ for $y$ near 1, we get that the contribution of this region is doubly log-enhanced.

 {\bf Diagram \hyperref[l22]{2.2}} \label{d22}
  \beq
 \lambda_{2.2}\sim \int \frac{dE_1}{E_1}\int \frac{dE_2}{E_2} \int \frac{dp}{\om_p} \Phi(E_2-\om_p,p) \Phi(E_1-E_2+\om_p,p) \,.
  \eeq 
  The invariant masses of the two particle phase spaces are:
  \beq
  s_1 = E^2+2 E\om_p+m^2\quad(E=E_1-E_2),\quad s_2 =E_2^2-2E_2\om_p +m^2\,.
  \eeq
  The dominant region will be $E,p=O(m)\ll E_2\sim E_L$. 
  Contribution from this region is
  \beq
  \label{22inf}
 \lambda_{2.2}\supset const. \int \frac{dE_2}{E_2^2} \Phi(E_2) \times I_3,\qquad I_3=\int dp\,dE\, \frac{\Phi(s_1)}{\om_p} \,.
  \eeq
  We can equivalently write $I$ as $I_3$, where
\beq
\label{eq:I3}
I_N\sim \int \prod_{i=1}^N\frac{dp_i}{\om_{p_i}}\delta(\sum p_i) \sim \int_0^Ldx\, [G(x,0)]^N\,,
\eeq
from which it's clear that the integral converges, and that $I_N\sim 1/m$, leading to the estimate in the table. The finite volume corrections are then suppressed by the same argument as for $\lambda_{4.2}$. There will also be sensitivity to $P_{\rm ext}$ for this diagram.
  
{\bf Diagram \hyperref[l0]{0}.}\label{d0}  We have three groups of two particles, each of the same total momentum $P=q_1+q_2=p_1+p_2=-(k_1+k_2)$. Let $E$ be the energy of the $q_1,q_2$ group, then the other two have energies $E_1-E$ and $E_2-E_1+E$. We have
\beq
 \lambda_{0}\sim \int \frac{dE_1}{E_1} \frac{dE_2}{E_2} dP \,dE\, \Phi(E,P)\Phi(E_1-E,P) \Phi(E_2-E_1+E,P) \,.
  \eeq   
 The dominant region is $P\sim E \sim W=E_2-E_1 \ll E_1\sim E_2\sim E_L$, which gives
\beq
 \lambda_{0}\sim \int_{E_L}^\infty \frac{dE_1}{E_1^4} \times I, \qquad I=\int  dP\,dE\, dW \, \Phi(E,P) \Phi(E+W,P)\,.
 \eeq
We can equivalently write $I$ as \reef{eq:I3} with $N=4$, from which the rest of the argument follows.

\subsection{General lessons}

\label{sec:lessons}
One important lesson of the careful discussion in this section is that one must be cautious applying naive dimensional analysis to predict how the coefficients of the local approximation scale with $E_L$. For the situation at hand, naive dimensional analysis fails as often as it is successful, because other scales with the dimension of energy, $m$ and $L^{-1}$, come in and change the scaling. 

It would be interesting to develop a local approximation procedure appropriate for the renormalization at the cubic order in the context of TCSA, in which $H_0$ describes a CFT. In the $\phi^4$ case the role of the scale $m$ was to regulate IR divergences, and power counting may be simpler in the TCSA case when no IR divergences are present. However, as mentioned in note \ref{note:cubicTCSA}, we do expect bilocal operators to appear in the TCSA case as well. 

\section{Local approximation: checks of accuracy}
\label{convEL}

As explained in section \ref{PI} and appendix \ref{la}, we have the scales $E_L,E_L^\prime,E_L^{\prime\prime}$ which control the accuracy of the local approximation used to compute the ultrahigh pieces of $\Delta H_2$ and $\Delta H_3$. In this appendix we present some numerical checks of how accurate the local approximation is. For illustrative purposes, we pick a rather low $E_T$. 

\begin{figure}[t]
\begin{center}
\includegraphics[scale=.42]{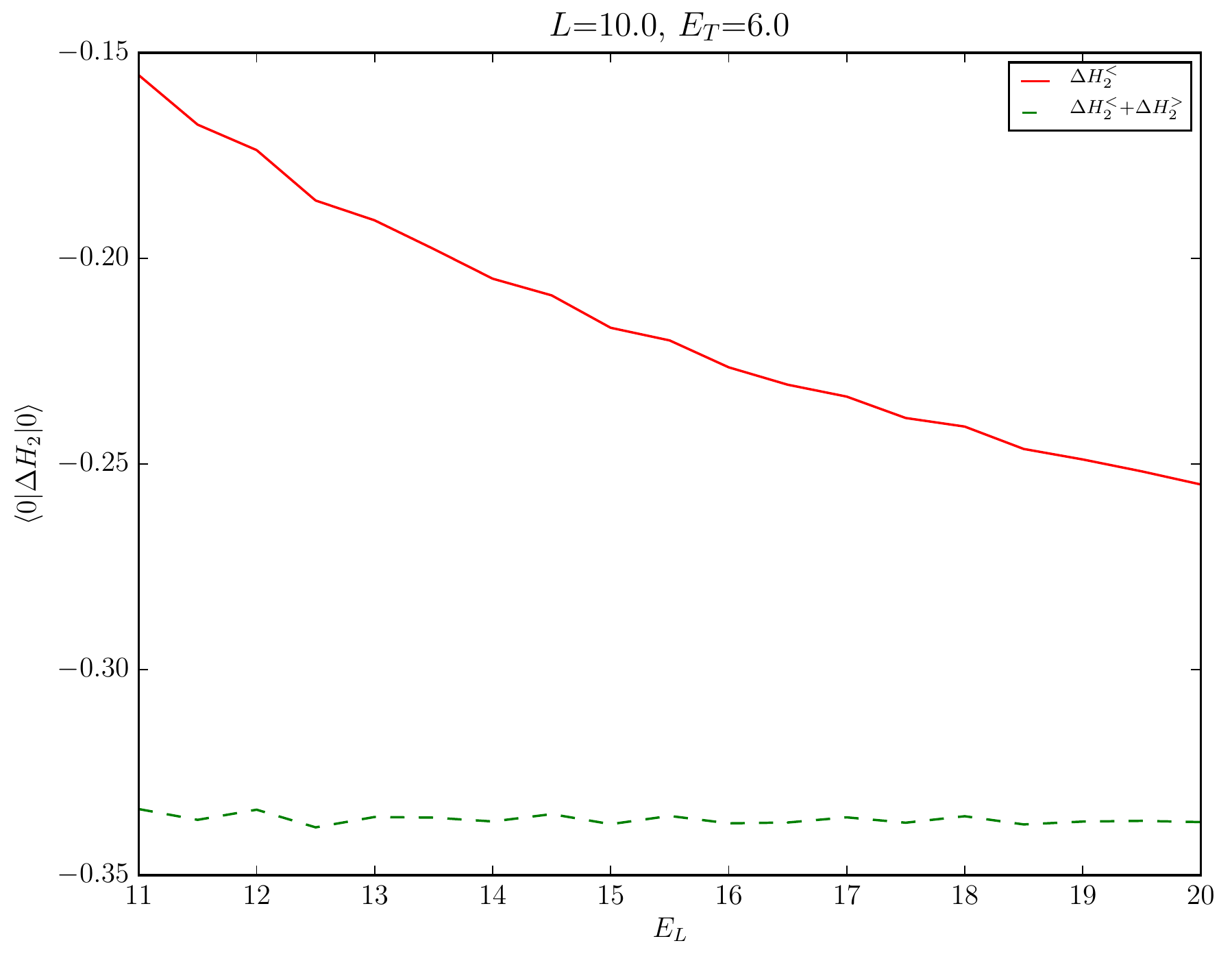}
\includegraphics[scale=.42]{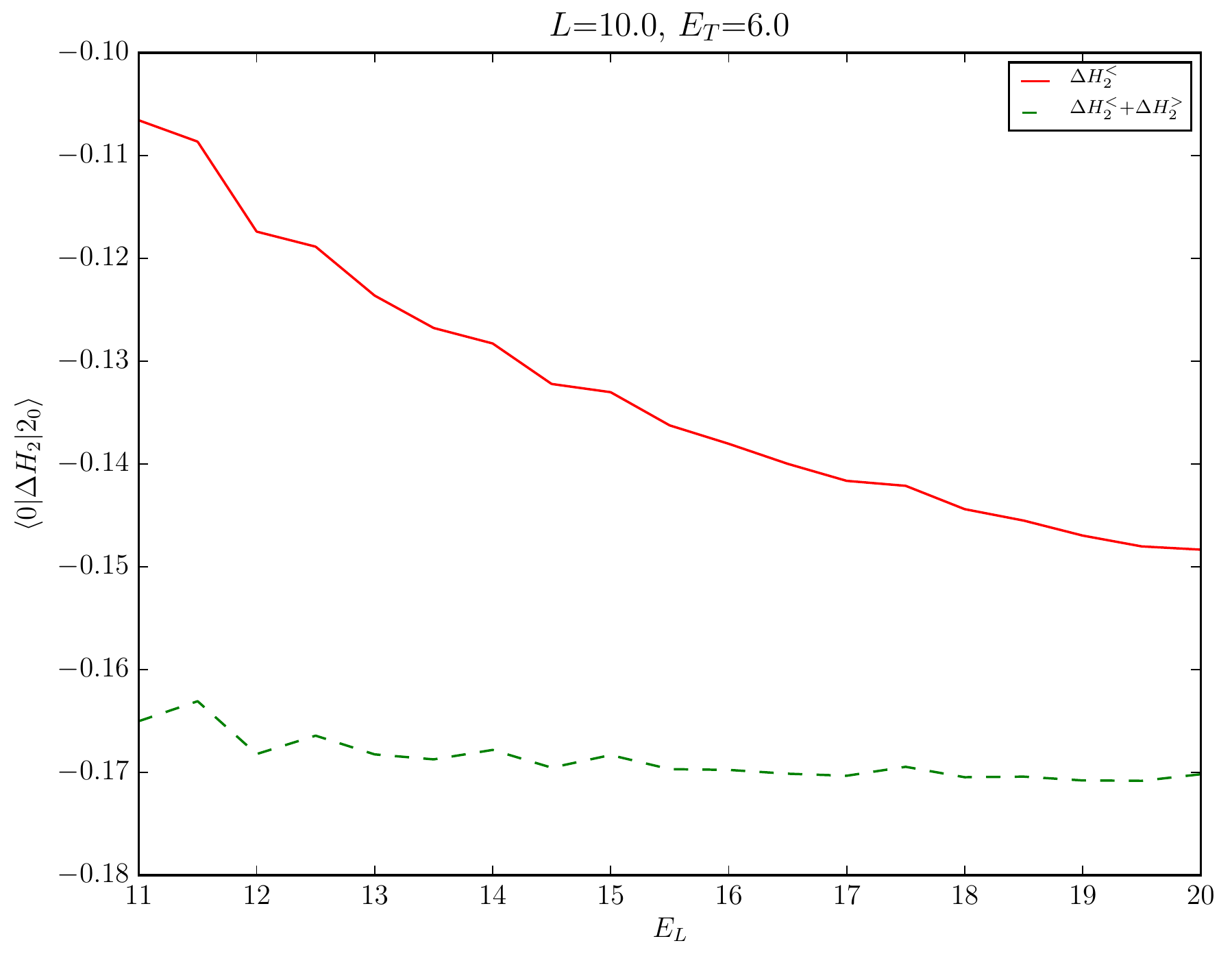}
\caption{Two low-energy matrix entries of $\Delta H_2^<$ and of $\Delta H_2^<+\Delta H_2^>$ as a function of $E_L$.
$\Delta H_2^<$ is computed exactly and $\Delta H_2^>$ in the local approximation.}
\label{panelDH2}
\end{center}
\end{figure}

In Fig.~\ref{panelDH2} we plot two matrix elements of $\Delta H_2$ as function of $E_L$: $\bra{0} \Delta H_2\ket{0}$ on the left and  $\bra{0}\Delta H_2 \ket{2_0}$ on the right, with $\ket{0}$  the free theory vacuum and $\ket{N_0}$ denoting $N$ particles at rest.
These matrix elements are computed as explained in \reef{eq:splitDH2}, i.e. by splitting the calculation as $\Delta H_2 = \Delta H_2^< + \Delta H_2^>$. Recall that $\Delta H^<_2$ includes the contribution from states in the range $(E_T,E_L]$. Instead, $\Delta H_2^>$ includes those in the range $(E_L,\infty)$ and is computed in the local approximation in the $L\rightarrow \infty$ limit, using the expressions in \reef{eq:kappamu}. 

In both plots, the steepest solid line does not include the local diagrams but only $\Delta H_2^{<}$. Instead, the flatter line   includes both $\Delta H^<_2$ and the local approximation to $\Delta H_2^>$, showing little dependence on the arbitrary scale $E_L$. The only local operator in \reef{eq:loc} that  can connect the state $\ket{0}$ with itself is $V_0$. Thus, the left plot tests the coefficient $\kappa_0$. Instead,  the right plot tests $\kappa_2$ since $V_2$ is the only operator in \reef{eq:loc} with non-zero matrix element $\bra{0}V_{N}\ket{2_0}$. In these plots $E_T$ was fixed to 10, but in fact this test does not depend on $E_T$ since shifting $E_T$ just adds a constant to both curves.
We did other similar plots for different matrix entries (in particular testing $\kappa_4$), showing equally good behavior. 

\begin{figure}[h!]
\begin{center}
\includegraphics[scale=.42]{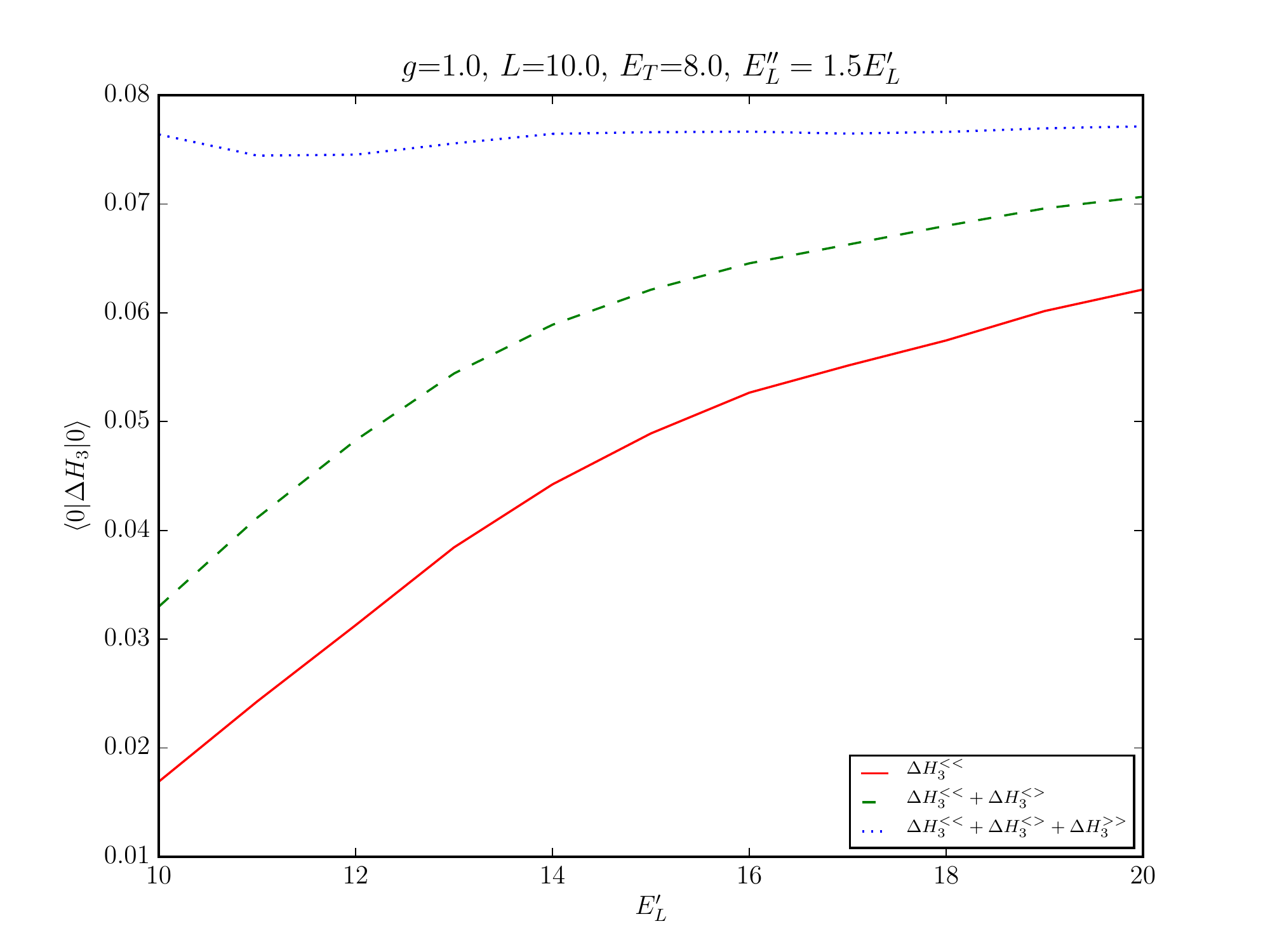} 
\includegraphics[scale=.42]{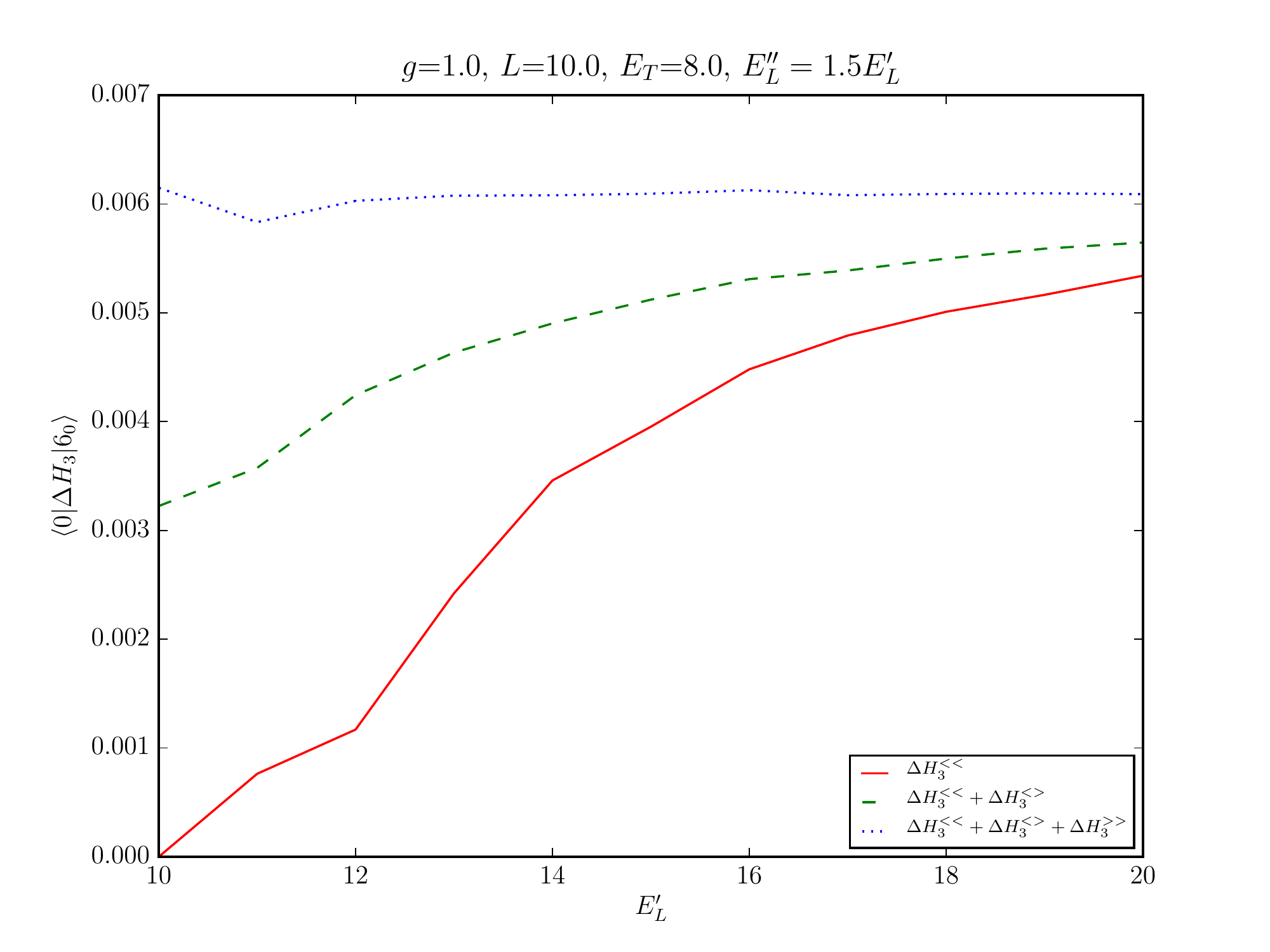}
\caption{   Matrix entries of $\Delta H_3^{<<}$, $\Delta H_3^{<<}+\Delta H_3^{<>}$  and $\Delta H_3^{<<}+\Delta H_3^{<>}+\Delta H_3^{>>}$ as a function of $E_L^\prime$.  }
\label{panelDH3}
\end{center}
\end{figure}

Analogously, in Fig.~\ref{panelDH3}  the matrix elements  $\bra{0}\Delta H_3 \ket{0}$ (left) and  $\bra{0}\Delta H_2 \ket{6_0}$ (right) are plotted as a function of the scale $E_L^\prime$, fixing $E_L''=1.5 E_L'$.
These plots are a numerical test of \reef{d3org}. The steepest solid line of both plots includes only the nonlocal piece $\Delta H_3^{<<}$ of \reef{nonl2}, including the states in the range $E_T<E_k \leq E_L^\prime$.  Instead, the flatter dashed (dotted) lines add to the solid ones  the  operators $\Delta H_3^{<>}$ (and $\Delta H_3^{>>}$)  in \reef{d3org}.   The matrix $\Delta H_3^{<>}$ in   \reef{mix} is computed as explained after \reef{mix2}, i.e.~the contribution of the states between $E_L^\prime$ and $E_L^{\prime\prime}$ is computed exactly doing matrix multiplication while in the range $(E_L^{\prime\prime},\infty)$ we use the local approximation for $\Delta H_2$.  The matrix $\Delta H_3^{>>}$ in \reef{l2} is calculated entirely in the local approximation, taking the $L\rightarrow\infty$ limit. 

 The left plot in Fig.~\ref{panelDH3} is a check of  $\lambda_0$ in \reef{l0}. 
  Instead the right plot tests the $\lambda$ coefficients of those (bi-)local operators in \reef{localexp2} that can connect the vacuum $\ket{0}$ with the six-particle state $\ket{6_0}$. These are the operators  $V_6$ and $\NO{V_2V_4}$. 
 Hence, the right plot is a check of the diagrams $ \hyperref[l61]{\lambda_{6.1}}$, $ \hyperref[l62]{\lambda_{6.2}}$, $\hyperref[2|4]{ \lambda_{2|4}}$. 
  We did similar plots for other matrix elements of $\Delta H_3$ in order to test the rest of the $\lambda$'s, and we obtained similar results to the ones shown in Fig.~\ref{panelDH3}.

\begin{figure}[t]
\begin{center}
\includegraphics[scale=.42]{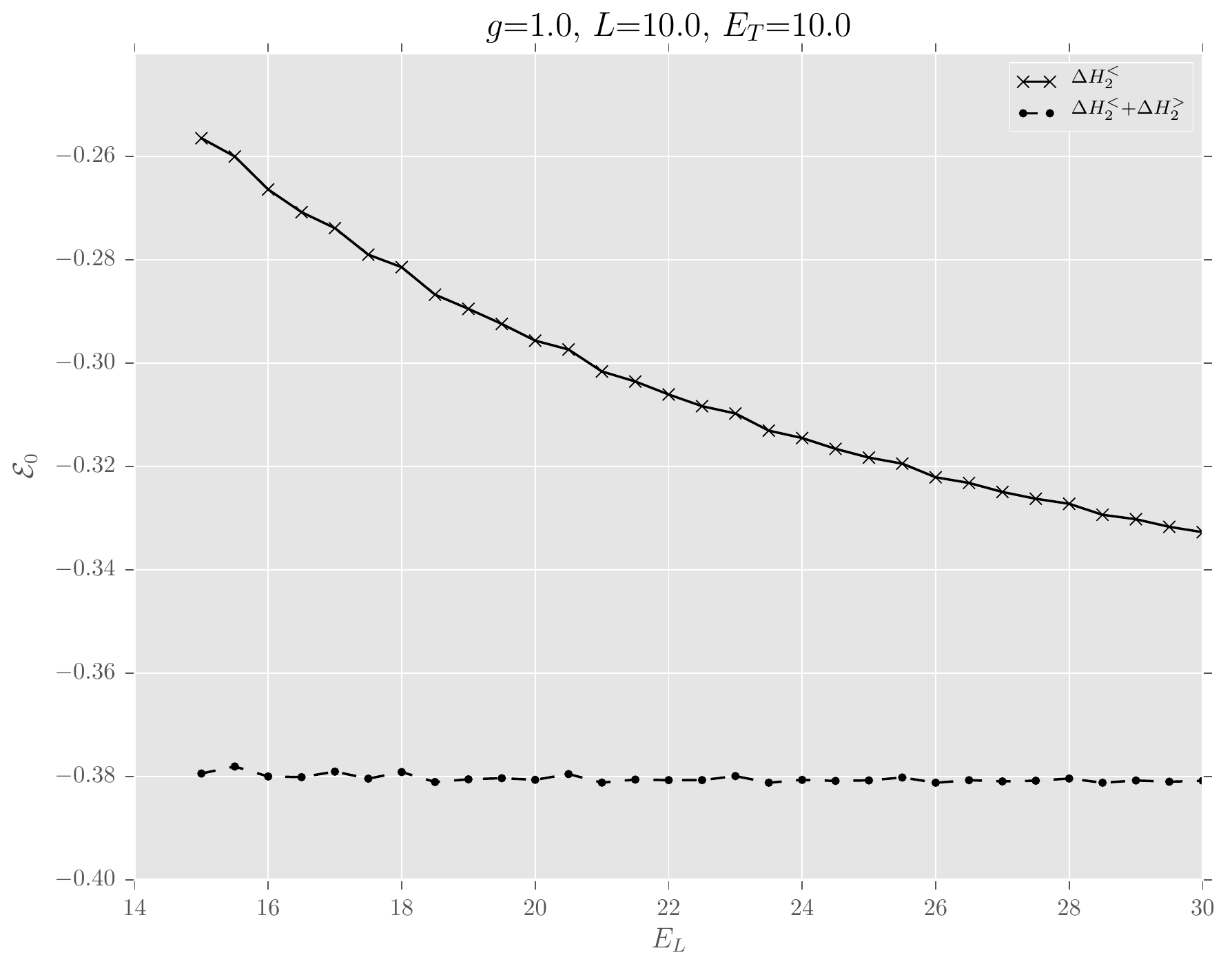} \includegraphics[scale=.42]{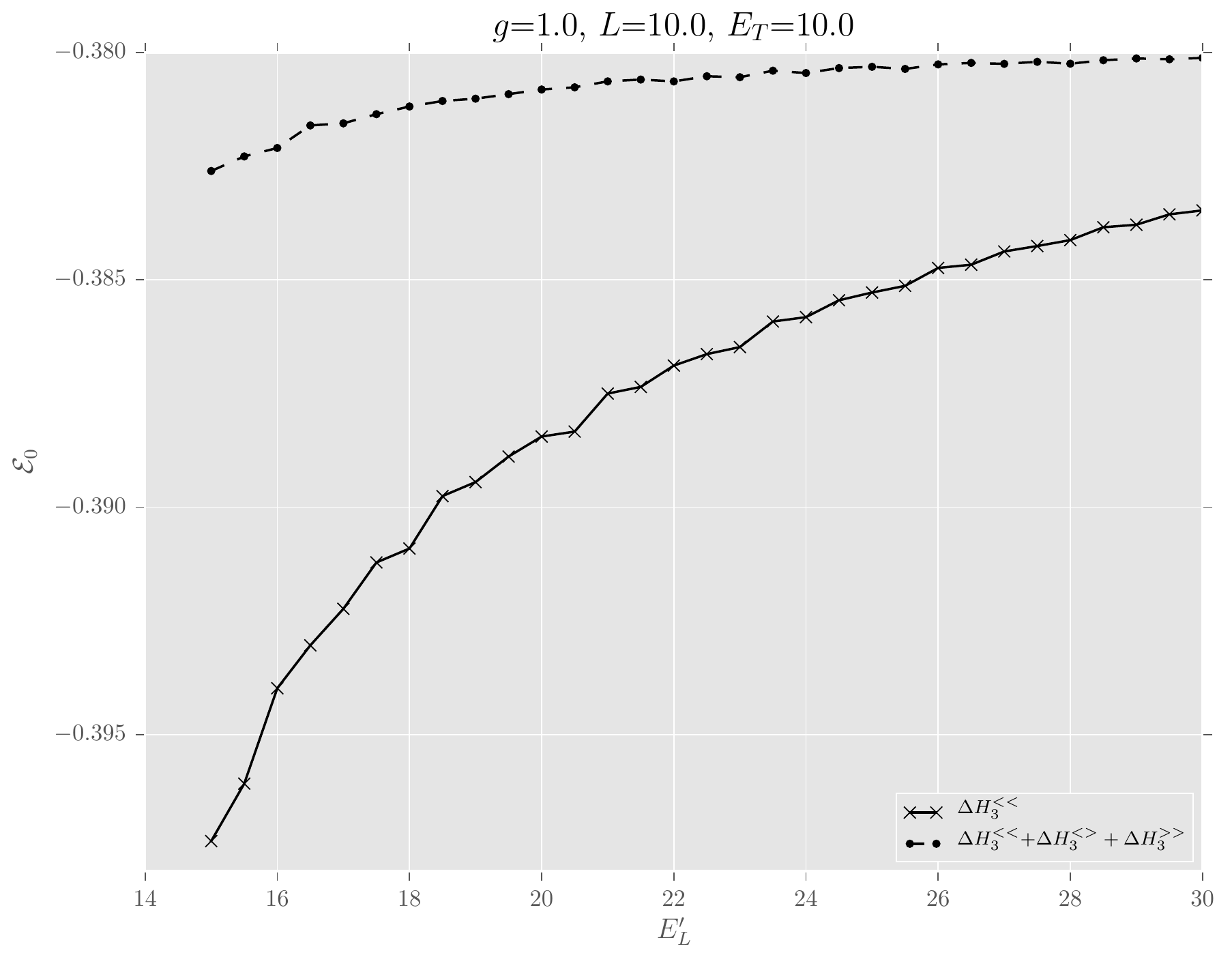}
\caption{ Vacuum energy as a function of $E_L$ (left) and  $E_L^\prime$ (right).}
\label{eigsvsEL}
\end{center}
\end{figure}

Lastly,  in Fig.~\ref{eigsvsEL} we show two plots of the vacuum energy. In the left plot we vary $E_L$ keeping fixed $E_L^{\prime\prime }=1.5 E_L^\prime=2 E_T$, while on the right we vary $E_L'$ keeping $E_L''=1.5 E_L'$ and fixed $E_L=3 E_T$. We use the full $\Delta H_2$ or just its $\Delta H_2^<$ part on the left, and the full $\Delta H_3$ or just its $\Delta H_3^{<<}$ part on the right. The point of these plots is the following. The lines corresponding to the full $\Delta H_2$ and $\Delta H_3$ are quite flat. This is comforting as it shows that there is very little dependence of the spectrum on the unphysical scales $E_L$, $E_L'$, $E_L''$. Note that this is true even for the values of $E_L$, $E_L'$ relatively close to $E_T$. For such $E_L$ we expect our procedure to give a poor approximation for the matrix elements with energies $E_i$ close to $E_T$. However, as we stressed several times, such states have a relatively low impact on the lowest excited states, even for moderately strong couplings $g$. This must be the reason why the spectrum varies so little even for low values of $E_L$ and $E_L^\prime$.  Nevertheless, in the main text we were conservative and took relatively large values of $E_L$, $E_L^\prime$, $E_L^{\prime\prime}$, so that all matrix elements of $\Delta H_{2,3}$ are well approximated. 

As mentioned in section \ref{intanal}, some terms in the local expansion
of $\Delta H_3$ (those marked with $\checkmark$ in Table \ref{tabasym}) are sensitive to the momenta of the external states $P_{\rm ext}$. In our way of approximating those terms, the magnitude of the corresponding matrix entries is overestimated. As the scale $E_L^\prime$ is increased those matrix entries decrease. Perhaps this can be used to explain why the dashed line in the right plot of Fig.~\ref{eigsvsEL} shows some residual growth. Namely, at leading order, the correction to the vacuum due to off-diagonal elements in $\Delta H_3$ is negative, due to the usual level-splitting. Then, as $E_L^\prime$ is increased the value of the vacuum energy should indeed somewhat increase. Although this is a perturbative argument, perhaps there is some truth to it.

In any case, in the future it could be interesting to take better care of $P_{\rm ext}$ dependence, as explained in section \ref{intanal}. This should reduce the residual $E_L'$-dependence of the spectrum.
Perhaps one can then lower further the value of $E_L'$ needed to achieve a given accuracy, saving
a significant amount of computational resources.

\section{Fit procedure} 
\label{sec:fitdetails}

In this appendix we give further details on the fitting procedure. 

\subsection{Infinite cutoff extrapolation}

After we compute the numerical NLO-HT mass and vacuum energy at finite $E_T$, 
we try to extrapolate them to $E_T = \infty$, by fitting the data points 
with a function of the form \reef{eq:etfit}.\footnote{For $g \le 1$, we instead set $\gamma=0$. 
This is motivated by the fact
that by eye the dependence is predominantly linear in $1/E_T^3$, and that the linear fit 
is more robust to the fluctuations of the data around the smooth curve. While this procedure may seem ad hoc, we tested it, and it works well. In the future one can think of more complicated fitting procedures.}
The central value and error bars are computed as follows.
For each $n = 0,1,2,3$, we remove $n$ points in the low $E_T$ part of the data 
sample, specifically $10 \le E_T \le 12.5$, where in total 5 data points 
are present for our choice of discretization of $E_T$. Several subsamples 
are generated, by removing $n$ points according to all possible combinations.
Fitting the model \eqref{eq:etfit} for each subsample, we obtain a series of 
``fit models'' $F_i(E_T)$ and the corresponding  asymptotes $\alpha_i$.
 Then, we take the mean, max and min of the $\alpha_i$'s as the central value, the upper bound and the lower bound estimate.Furthermore, to account for fluctuations for the higher values of $E_T$, we provide
alternative estimates  for the error bars, as follows. We compute the maximum absolute 
difference between the data points and the mean of the $F_i(E_T)$ in the range 
$E_{\rm max} -5 \le E_T \le E_{\rm max}$, where $E_{\rm max}$ is the maximum 
cutoff we attain at a given $L$.  The final error bars are the largest between the two methods.

\subsection{Estimate of the critical coupling}

The critical value of $g$ where the theory undergoes a phase transition is determined from the right plot in Fig.~\ref{fig:specvsG}. The red data points of the plot are fitted with the rational function in \reef{eq:ansatz}, minimizing over $g_1,\, g_2,\, g_3,\, a$ and $g_c$ the ``log-likelyhood function" formed for $N$ data points:
\begin{equation}
\label{eq:rbar}
\chi^2 = \sum_{i=1}^N {(y_i-f(x_i))^2}/{err_i}^2 \, . 
\end{equation}
The central value $g_c=2.76$ reported in Table \ref{table:gc} corresponds to the smallest $\chi^2$, call it $\chi^2(2.76)$.
The uncertainty was determined through the following procedure. 
We fix $g_c$ close to 2.76 and fit the same ansatz \reef{eq:ansatz}, minimizing the log-likelyhood only over $g_1,\, g_2,\, g_3$ and $a$.
The error interval reported in Table \ref{table:gc} corresponds to those $g_c$ for which the root mean square normalized error is within factor 3 of what it is at $g_c=2.76$, i.e.
\beq
\sqrt{\chi^2(g_c)/N}\le 3\sqrt{\chi^2(2.76)/N}\,.
\eeq
We believe that this error determination is conservative.

\section{Algorithmic details}
\label{sec:compdetails}

We will describe here some details of the basis and matrix generation algorithms used 
in this work, highlighting key improvements over the code used in \cite{Lorenzo1}.
It will be important to control both the time and memory complexity of the computation.
The core component of our code is a routine\footnote{The operator $V$ is hermitian and the matrix elements are real in the basis that we consider, so $V_{ij}=V_{ji}$.}
\begin{equation}
F : |i\rangle \to \{ V_{ji}|j\rangle \,\,|\,\, E_j \le \Emax , V_{j i} \ne 0 \}\,,
\label{eq:routine}
\end{equation}
taking as input a state and returning all the states $|j \rangle$ and coefficients 
$V_{ji}$ such that $V_{ji} \ne 0$, in a given energy range.
This routine will be described in more detail in section \ref{sec:routine}. 

\subsection{Basis generation and storage}
\label{sec:basis}

We use two different data structures to represent the Fock states in the Hilbert space. The reason 
to do so will become clear below. Each Fock state is represented in one of the following ways:
\begin{enumerate}
\item as a list of tuples $[(n,Z_n),\ldots ]$ where $n$ represents wavenumber and $Z_n$ occupation number (only $Z_n>0$ are included).
The list is ordered in $n$.  E.g. $[(-1,3),(0,2),(3,1)]$ is a state in this representation. This representation
is convenient to use as input for the routine \eqref{eq:routine}, but it's relatively  expensive in memory.
\item as a fixed-length list of all occupation numbers $[Z_{-n_{\max}}, Z_{-n_{\max}+1}, \ldots, Z_{n_{\max}} ]$, including the zeros. 
E.g. $[0,0,3,2,0,0,1]$ with $n_{\max}=3$ is the above state. 
The state which is part of the output of the routine \eqref{eq:routine} is efficiently computed in this representation.
It can be stored cheaply in memory as a {\it byte sequence} ({\tt bytes} in {\tt python}).
\end{enumerate}

As in \cite{Lorenzo1}, we restrict ourselves to the truncated Hilbert space with total momentum $P=0$. 
Furthermore, we are interested in the part of the Hilbert space which is $\bP$-invariant (where $\bP$ is spatial parity).
Its basis is formed by the states which are either $\bP$-invariant Fock states or have the form
\beq
\label{eq:f1f2}
(|\psi\rangle+ \bP |\psi\rangle)/\sqrt{2}, 
\eeq
where $|\psi\rangle$ is a Fock state such that $|\psi\rangle\ne \bP |\psi\rangle$.
In the latter case the state is represented in the basis by storing either $|\psi\rangle$ or $\bP |\psi\rangle$ (but not both), choosing between the two arbitrarily.
Finally, we work separately in the sectors $\bZ_2=\pm1$ of the field parity $\phi\to-\phi$.
The finite-dimensional Hilbert space that is stored numerically is composed of several parts:
\begin{itemize}
\item ``Low energy" states, i.e. all states with energy $E \le E_T$. This chunk of the 
Hilbert space is stored both in Representation 1 and 2. In this work it typically contains $\sim 10^4$ elements.
\item ``Moderately high'' states with energy $E_T < E \le E_L'$. Typically, in this
work we choose $E_L' \sim 2 E_T$. These states are summed over
in the computations of $\Delta H_3^{<<}$ in in \eqref{nonl2} and of $\Delta H_2^{>}$ 
in \eqref{mix}. Notice that these are {\bf not all} the states in 
the given energy range, as we only need those states $j$ for which there is a nonzero $V$ matrix element connecting them to a low energy state:
\begin{equation}
E_T < E_j \le E_L'\quad\text{ and }\quad \exists V_{ji} \ne 0 \,, \quad E_i\le E_T \,.
\label{eq:hecondition}
\end{equation}
This distinction is important, as the number of all states in the given range grows 
{\it exponentially} with $E_L'$ (for fixed $E_T$), while the number of those respecting the condition 
\eqref{eq:hecondition} only {\it polynomially}.
To generate them, we do the following. As  mentioned, 
we have routine \reef{eq:routine} which, given a 
state $|i \rangle$ in Representation 1 as an input, returns all the states 
$|j \rangle$ such that $V_{ji} \ne 0$ in Representation 2. 
We apply this routine (with $E_{\rm max}=E_L'$) over all the states below $E_T$, 
and save the results in a {\it hashset} ({\tt set} in {\tt python}), which has
constant lookup time. In this way, states are not overcounted.
Finally, the states $j$ so generated are stored in both representations. In this work,
 their number is usually of the order $10^6$.
\item States with energy $E_L' < E \le \max(E_L, E_L'')$, which are summed over 
either in $\Delta H_2^{<}$ in \eqref{nonl1} or in $\Delta H_2^{>}$ in \eqref{mix2}.
In this work we typically choose $E_L \sim E_L'' \sim 3 E_T$. These states 
are generated analogously to the ``moderately high'' states above, but they 
are not saved in the Representation 1 format, because it is not necessary
to act on these states with $V$ anymore. This saves a significant
amount of memory, as there can be around $10^7-10^8$ states in this chunk of 
the Hilbert space.
\end{itemize}

\subsection{Computation of matrix elements}
\label{sec:routine}

We will now describe some details of the routine \eqref{eq:routine}, and of how 
the matrices $\Delta H_2$, $\Delta H_3$ are computed. Suppose we want to find all
the non-vanishing matrix elements $V_{ji}$ between all the states 
$| i\rangle \in \calH_I$, $| j\rangle \in \calH_J$, where $\calH_I$, $\calH_J$
are subsets of the Hilbert space with maximal energies $\Emax^I$ and $\Emax^J$. We assume $\Emax^I\le \Emax^J$ without loss of generality. 
The procedure is described below.

First, the local operator $V$ must be represented efficiently, by decomposing
it into sums of elementary terms. For generality we consider $V = \int \phi^n$,
with $n$ arbitrary. In this way, the code can be used to construct 
both the ``non-local'' and ``local'' parts of $\Delta H_2$, $\Delta H_3$, 
where all the even powers of $n \le 6$ appear.\footnote{We won't describe 
a modification of the algorithm used the compute the ``bilocal'' matrices
$:V_2 V_4:$, $:V_4 V_4:$ appearing in $\Delta H_3$.}
Schematically, $V$ is a sum of products of oscillators
\beq
\label{eq:Vosc}
V \sim \sum_{n_c=0}^{n} \sum_{\{k\}, \{q\}} 
\left( \prod_{i=1}^{n_c} a_{k_i}^\dagger \prod_{i=1}^{n-n_c} a_{q_i}\right), 
\qquad \sum k_i - \sum q_i = 0 \,,
\eeq
where $n_c$ is the number of {\it creation} operators. This sum is infinite, but for given finite $\Emax^I$, $\Emax^J$ only a finite subset will contribute nontrivially to the matrix elements we wish to compute. These relevant terms are selected and stored in memory
as follows: 
 \begin{itemize}
\item For each $n_c$, we cycle over all the states in $\calH_I$, creating a {\tt set}
of all the possible $(n-n_c)$-dimensional tuples $\{q_i\}$ of momenta which are present in at least one state. We will only
need terms in \reef{eq:Vosc} for which $\{q_i\}$ is such a tuple, since all other terms annihilate all states.
\item We iterate over this set of tuples, and for each tuple we generate a list of all 
the possible $n_c$-dimensional tuples of momenta $\{k_i\}$, subject to the constraints
\beq
\label{eq:Vosccreation}
\sum k_i - \sum q_i = 0 \,, \quad \sum_i \omega(q_i) \le \Emax^J\,.
\eeq
This list is then sorted in energy. Clearly we only need terms in \reef{eq:Vosc} for which $\{k_i\}$ is such a tuple, since any other term will either violate the zero momentum condition or raise the energy of the state above the threshold $\Emax^J$ we are interested in.

\item For each $n_c$, we create a {\it hash table} ({\tt dict} in {\tt python})
mapping the tuples of annihilation momenta to the sorted lists of tuples of 
creation momenta. It is useful to use this data structure as it has constant 
lookup time. Also, we construct a similar hash table of the same size, 
containing all prefactors (including the factors $1/\sqrt{2 \omega L}$ 
and the combinatorial factors) for each pair of creation-annihilation sets of 
operators. These coefficients, multiplying the terms in \eqref{eq:Vosc} (not shown in that equation for 
simplicity), are precomputed for efficiency.
\end{itemize}

Next, we cycle over $\calH^J$ and create a lookup hash table ({\tt dict}) of all 
the associations $\{ |j\rangle : j\}$ between the states $|j\rangle$ and the 
row indices of the matrix $V_{ji}$.

Finally, we enter the core routine \eqref{eq:routine}, which makes use of the data
structures defined above, and works as follows:
\begin{itemize}
\item We iterate over $\calH_I$, select a state $| i\rangle$ with energy $E_i$ 
and generate a list of 
all the sets of momenta $\{q_i\}$ than can be annihilated at each value of $n_c$.
\item We iterate over this list, selecting a tuple $\{q_i\}$, and get the corresponding 
list of tuples $\{k_i\}$ previously computed in the hash table.
\item We iterate over the list of $\{k_i\}$. This inner loop is the most expensive 
part of the computation, and it has been optimized using the {\tt cython} extension.
We act on the state $|i\rangle$ with the given sets of 
creation and annihilation operators and generate a new state $|j \rangle$ in 
Representation 2 and partial coefficient $V_{ji}$. The state is looked up 
in the hash table to get the index $j$.
\item We add the partial coefficient to the column $i$ of the matrix, and repeat 
through the previous points, until the column $i$ of $V_{ji}$ is entirely computed.
We add this column to the full matrix in the {\it sparse} format (for maximum efficiency
we use the {\tt coo} format in {\tt scipy.sparse}). 
\end{itemize}
At the end of this cycle one obtains the full matrix $V_{i j}$ over the subspaces
$\calH^I$, $\calH^J$. $V$ is then converted from the {\tt coo} to the {\tt csc} format
to allow for fast algebraic operations.

Some of the tricks describe above reduce the time complexity by orders of magnitude.
We do not report other tricks which speed up the computation by factors of a few. 
For example, many quantities, such as the energies of the 
states, can be precomputed and stored. Also, if $\calH^I = \calH^J$ and $V$ is hermitian, only half of the
terms in \reef{eq:Vosc} related to each other by conjugation can be retained.

\subsection{Evaluation of $\Delta H_2, \Delta H_3$}

The matrices $\Delta H_2$ and $\Delta H_3$ are computed by  summing over basis states 
with energy above $E_T$. As explained in section \ref{PI}, 
they are decomposed into a ``non-local'' part, where the Fock states
are  summed over exactly, and a ``local'' part, where  the sum is approximated analytically.
Here we describe in detail how to evaluate efficiently the non-local part of the matrices.
This step represents the bottleneck of the entire numerical computation.

\subsubsection{$\Delta H_2^{<}$}

The sum \eqref{nonl1} has to be evaluated. To do so, it is most convenient to
to apply the routine \eqref{eq:routine} over the basis states with energy $E \le E_T$,
to construct the matrix $V_{kj}$ in \eqref{nonl1} and its transpose. 
Then, $\Delta H_2^{<}$ is easily evaluated by multiplying those.

\subsubsection{$\Delta H_3^{<<}$}

One has to compute the sum \eqref{nonl2}. The matrices $V_{i k}$ and 
its transpose are constructed as above. Instead, $V_{k k'}$ is sometimes too large 
to be stored in memory, even if it's sparse. If this happens, we divide the 
set of basis states with energy $E_T \le E \le E_L'$ into chunks 
and compute blocks of $V_{k k'}$ one at a time, summing over them sequentially. 

\subsubsection{$\Delta H_3^{<>}$}

The non-local contribution to $\Delta H_3^{<>}$ in \eqref{mix} is evaluated 
analogously to $\Delta H_3^{<<}$. To save resources, it is important to evaluate the matrix elements $V_{k k'}$
cycling over the states with energy $E_T \le E \le E_L'$ and acting on them with $V$, 
rather than cycling over the more numerous states in the range $E_L' \le E \le E_L''$.

%%%%%%%%%%%%%%%%%%%%%%%%%%%
%%%%%%%%                           %%%%%%%%%%%
%%%%%%%%   Bibliography    %%%%%%%%%%%
%%%%%%%%                           %%%%%%%%%%%
%%%%%%%%%%%%%%%%%%%%%%%%%%%

\small

\bibliography{phi4-Biblio}
\bibliographystyle{utphys}

\end{document}